# Review of the Electrochemical Double Layer Theory and Its Applications in Battery Electrode Materials

*Yitao He*[a,b,*]

[a] School of Energy and Environment, Anhui University of Technology, Ma'anshan, Anhui 243002, China

[b] Department of Thin Films and Nanostructures, FZU – Institute of Physics of the Czech Academy of Sciences, Cukrovarnická 10/112, 162 00 Prague 6, Czech Republic

*Corresponding authors: heyitao@ahut.edu.cn; yitao@fzu.cz

**Note:** This paper may generate many significant debates in the fields of electrochemistry and battery research, which is why I have chosen to share it initially on a preprint platform. During traditional anonymous peer review, I often lack the opportunity to engage in open dialogue with reviewers. For this reason, I invite constructive feedback here to address any concerns or differing perspectives directly. Your insights will help refine this work, and I plan to incorporate revisions before submitting it to a peer-reviewed journal at an appropriate time. Thank you for contributing to the improvement of this research. Welcome to share comments via my email.

All the questions and responding answers are listed in the end.

**Abstract**: The interface plays a critical role in electrochemical systems, driving the development of various theories to investigate properties at nanoscale and microscale levels, including the electrictrochemical double layer (EDL) theory and continuum theory. However, the application of EDL theory within the realm of battery electrode materials has not been clearly defined according to material type. The diverse range of materials and their intricate interactions with interface theories often lead to misunderstandings. This review meticulously outlines the origins and principles of the EDL theory. Additionally, it categorizes battery electrode materials into two main groups: semiconductors and conductors. A comprehensive examination is presented on the implementation of EDL and other interface theories, accompanied by a comparative evaluation of the fundamental models and experiments described. The review also explores the potential advancements of EDL and other interface theories, as well as their application in electrode materials. This exhaustive review serves as an invaluable resource for scholars in electrochemistry and materials science, aiding in their understanding of interface theories, their scope of applicability, and the integration of these theories with practical experimental approaches.

# 1 Introduction

The Gouy-Chapman-Stern (GCS) model is a foundational framework for describing the electrochemical double layer (EDL), and it has been widely applied to battery electrode systems.[1] However, the complexity of modern battery materials has revealed critical limitations of the GCS model, such as its inability to accurately capture interfacial energy levels, its neglect of continuity across solid-liquid interfaces, and its oversimplified treatment of interactions among adsorbed species.[2] To address these challenges, researchers have extended the GCS model,[3] and developed new continuum and semiconductor-based theories tailored to specific electrode materials. Despite these efforts, a clear understanding of the interfacial structure and potential distribution remains incomplete, especially for solid-state or heterogeneous systems.[4] This review therefore focuses on recent advances in EDL theory, with particular emphasis on interface models relevant to battery materials, aiming to bridge the gap between classical electrochemical models and emerging interfacial phenomena.

The theory of EDL has a predominant role in history of electrochemistry and energy storage science.[5] The concept of the EDL can be traced back to the early 20th century. In 1859, the physicist Georg Quincke[6] conducted groundbreaking experiments that marked a pivotal advancement in electrokinetic phenomena. His work extended beyond the exploration of electro-osmosis, pumping water through a tube and detecting an electrical potential difference between its ends, unveiling the phenomenon known as streaming potentials. While the magnitude of potential difference varied among these systems, it consistently bore the same sign. His profound contributions extended to the postulation of a space charge outside charged surfaces, crucial for qualitatively explaining both electro-osmosis and streaming potential phenomena. In 1873, James Clerk Maxwell laid the groundwork for the theory of the EDL, proposing a theory that described phenomena at the liquid-solid interface. He concluded that for equilibrium to be achieved in these systems, in accordance with the second law of thermodynamics, opposite (compensating) charges should accumulate near the charged solid surface.[7] Building on this foundation, in 1879, Hermann Ludwig Ferdinand von Helmholtz[8] introduced the concept of the EDL,[9] likening it to a parallel plate capacitor where the inner plate is formed by charges fixed on the solid surface, and the outer plate consists of counterion charges in the adjacent liquid layer. He posited that the thickness of the EDL did not exceed that of two layers of liquid molecules. Further evolving the understanding of the EDL, in 1910, Louis Georges Gouy proposed the existence of a diffuse layer of counterions, resulting from two opposing forces: the thermal motion of EDL ions, which tends to push them away from the charged surface, and the electrostatic attraction of the ions to the charged surface. This hypothesis has since been validated by numerous experiments, significantly advancing the theory of the EDL.[7] Subsequently, in 1947, David C. Grahame's work[10] built upon these earlier contributions and provided a more comprehensive theoretical framework for understanding the EDL, particularly from an electrocapillarity perspective. His work has since become a cornerstone in the study of interfacial electrochemistry. The application of the EDL theory to various fields, including electrochemical batteries, has indeed been a fruitful avenue for research.[11] Understanding the behavior of ions at the electrode-electrolyte interface is crucial in designing and optimizing electrochemical devices such as batteries. Researchers continue to build on the foundation laid by Grahame and others to advance our understanding and improve the performance of these technologies.

Understanding the EDL is crucial for advancing battery technology. The EDL affects battery performance, efficiency, and safety by influencing ion transport and selectivity, and also plays a key role in surface chemistry and morphology of electrodes, enabling the development of materials with increased surface area for higher capacitance and longer cycle life. Furthermore, insights into the EDL dynamics help address challenges such as dendrite formation in alkali-metal-based batteries, enhancing stability and reliability. Understanding the EDL also aids in stabilizing the solid electrolyte interphase (SEI) layer in lithium-ion batteries (LIBs), which is vital for their longevity and safety.[12] Additionally, the EDL impacts thermal management strategies, helping to mitigate heat generation and improve heat dissipation for optimal battery operation and safety.[13]

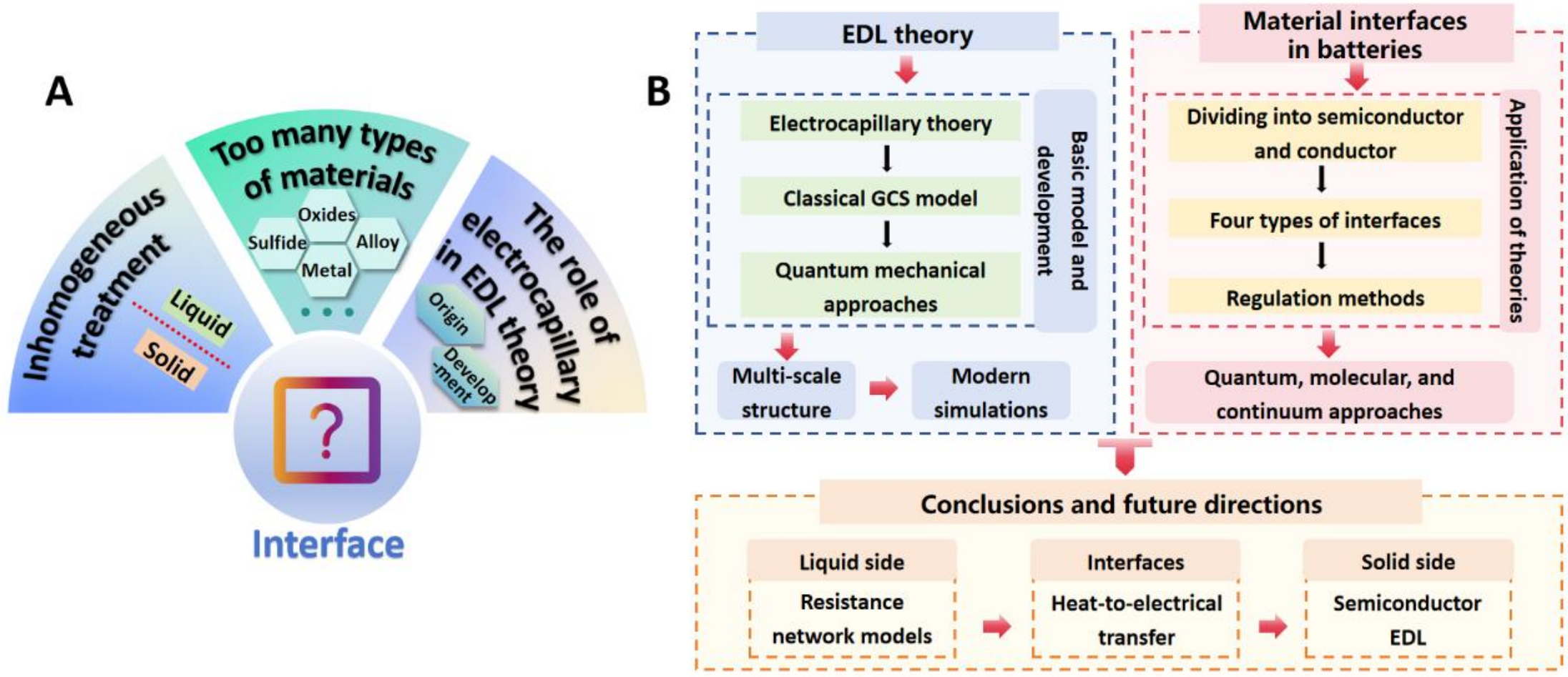

**Figure 1**. (A) The main problems in interface theory in electrochemical systems; (B) The conceptual roadmap for this review.

Moreover, the exploration of EDL structures at the electrode-electrolyte interface reveals a complex tapestry of interactions, intricately tied to the material properties of electrodes. Transition metal oxides, sulfides, carbon-based materials, and lithium-based materials each manifest unique EDL features. The material-dependent desolvation effects further amplify this complexity, where factors like ion size, surface functionalities, and crystalline structures play pivotal roles. For instance, in carbonaceous materials, high surface area, surface groups and tunable electronic properties contribute to diverse compositions within EDL, impacting ion adsorption and desolvation kinetics. Recognizing these material-specific intricacies is crucial for understanding how EDL phenomena influence the desolvation process, subsequently affecting charge storage mechanisms and overall battery performance. The main problems in electrochemical interfaces were listed in **Figure 1** to clearly be visualized.

Simplifying the categorization of electrode materials into semiconductors and conductors is a reasonable approach to streamline the study of the EDL impact across different materials. This categorization helps clarify the complexity of various materials used in batteries, allowing for a more focused investigation into the semiconductor electrochemistry, which is prevalent in many battery systems. Semiconductor electrode materials, which often exhibit unique electronic and electrochemical properties, play a crucial role in the performance of various batteries. Understanding the EDL phenomena at the interface of semiconductor materials provides insights into charge transfer processes, band structure modifications, and the overall electrochemical behavior within these systems. By concentrating on semiconductor electrochemistry, researchers can delve into the nuanced interactions at the semiconductor-electrolyte interface, shedding light on how EDL influences charge storage, ion transport, and the overall performance of semiconductor-based electrodes in batteries. This focused approach allows for a more targeted exploration of the fundamental principles underlying the EDL on semiconductors, potentially contributing to advancements in semiconductor materials-based battery technologies.

In this review, we aim to narrow the focus of EDL research by categorizing electrode materials into semiconductors and conductors, streamlining our exploration primarily towards semiconductor electrochemistry prevalent in many battery systems. By simplifying the material types, we seek to seamlessly bridge foundational EDL research with its specific applications in semiconductor-based battery technologies. Our emphasis lies in elucidating the pivotal role of fundamental science at the forefront of technological innovation within the context of semiconductor electrochemistry. This synthesis systematically traces the evolutionary trajectory from theoretical EDL models to their pragmatic implementation in enhancing battery performance, thereby illuminating intricate pathways through which fundamental electrochemistry catalyzes advancements in energy storage solutions. The scope of our exploration extends to the nuanced interactions within the semiconductor EDL, offering valuable insights into material-dependent phenomena and their profound implications for shaping the desolvation landscape within semiconductor electrode materials. This focused understanding holds the potential to revolutionize the field of material science in batteries, presenting tailored approaches for optimizing energy storage systems across diverse technologies, particularly those employing semiconductor materials. Positioned at the forefront of evolving energy paradigms, this review aspires to be a scholarly testament to the enduring potency of curiosity-driven research, essential for unraveling the mysteries of the natural world and leveraging them for the betterment of humanity.

## 2 Experiments, Theories and Models: Electrocapillary and EDL

### 2.1 Basic theories, models and early experiments

Before delving into the equations of the EDL theory, it is necessary to first introduce the theory of electrocapillarity, which has a close relationship with the battery performances and can be used to determine the properties of electrolytes[14]. The theory of electrocapillarity studies how the surface tension of metals changes when they come into contact with electrolyte solutions. This variation is typically described by an electrocapillary curve, which shows the relationship between interfacial tension and the potential difference applied to the metal. The theory of the EDL provides the foundation for the phenomena observed in electrocapillarity. The shape and characteristics of the electrocapillary curve are influenced by the distribution of charge within the EDL. As the potential changes, the ions within the EDL redistribute, leading to changes in interfacial tension. Key parameters in electrocapillarity theory, such as the electrocapillary maximum and the potential of zero charge[15], are defined based on the concepts of the EDL theory. The EDL theory offers a framework for understanding the physical processes behind the phenomena of electrocapillarity, while electrocapillarity theory provides an experimental method to study and verify the predictions of the EDL theory.

The observations made by William Henry in 1800 regarding the peculiar behavior of mercury in an electrolyte solution under the influence of an applied voltage set the stage for a deeper understanding of electrochemical phenomena, particularly those related to surface tension and electrocapillarity. The phenomenon, where a mercury droplet exhibited movement upon application of voltage or when it came into contact with metals such as iron, hinted at a complex interplay between electrical forces and the physical properties of liquids at interfaces. Building upon these early observations, Gabriel Lippmann[16], in the period between 1871 and 1873, embarked on an exploration of the underlying principles governing these phenomena. Lippmann hypothesized that the observed movement was due to the relationship between surface tension and the chemical potential. To investigate this hypothesis, he ingeniously designed a simple yet effective device known as Lippmann's electrometer, as shown in **Figure 2A**. This device consisted of a glass tube filled with mercury, tapering into a fine capillary with a thin bottom approximately 50 micrometers across. The capillary end was submerged in a sulfuric acid solution, and upon applying a voltage within 0.1 mV, the mercury's movement within the capillary could be meticulously observed with a microscope and measured with remarkable precision.[17] Lippmann's work was not only pivotal in demonstrating the direct relationship between electrical potential and surface tension at the mercury-electrolyte interface but also in mitigating experimental challenges. To reduce the influence of surface contaminants that could skew the observations, Lippmann innovated the dropping mercury electrode. This technique involved the continuous renewal of the mercury surface in contact with the electrolyte, thereby providing a clean interface for each measurement. This equipment later evolved into the polarographic method, a foundational technique in electrochemistry for the quantitative analysis of various substances.

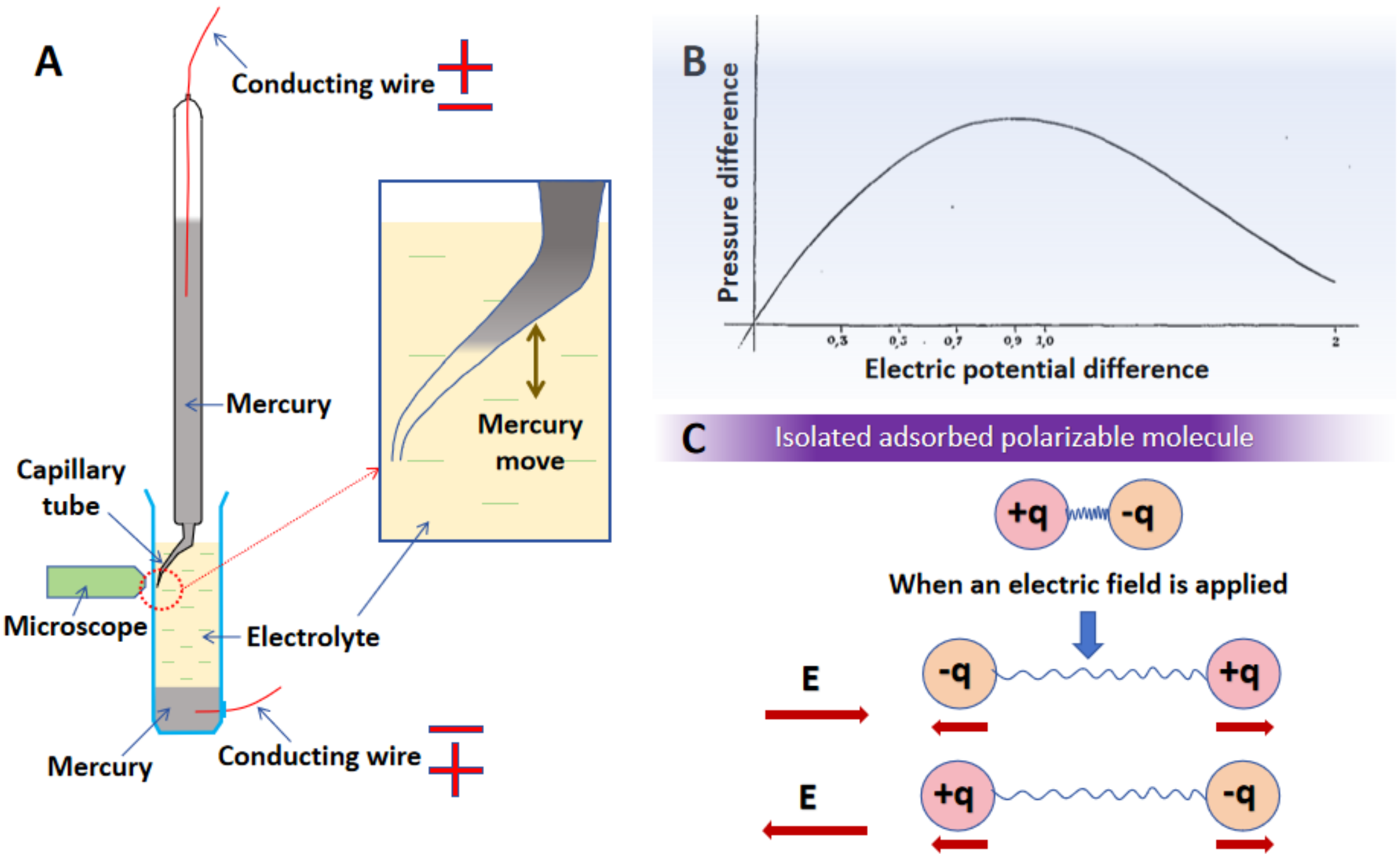

**Figure 2.** (A) Scheme of Lippmann's electrometer measuring the electrocapillary phenomena; (B) Electrocapillary curve[18]; Copyright © 1875 EVISA. (C) An isolated adsorbed polarizable molecule at electrode/electrolyte interface. **William Henry's observations of mercury in electrolytes under voltage were followed by Lippmann's invention of the electrometer. Lippmann's development of the dropping mercury electrode evolved into polarography, which allowed for the quantification of interfacial tension via the Lippmann equation.**

Lippmann's experimental methodology, employing sophisticated microscope and an additional mercury manometer, facilitated a meticulous examination of the interfacial tension between mercury and a sulfuric acid solution. By systematically varying the potential difference across the mercury-solution interface, he observed alterations in the shape of the mercury liquid surface. Because the curvature of the liquid surface is directly related to the interfacial tension. Precise measurements of the curvature and subsequent recording of the corresponding potential differences during the attainment of a new equilibrium position were conducted. Notably, the experimental findings revealed a distinctive electrocapillary phenomenon[18]: the interfacial tension of the mercury liquid surface exhibited a maximum value as the electric potential continued to increase (**Figure 2B**). There was a simple but important equation describing the electrocapillary phenomenon, known as Lippmann equation. R.R. Salem has indeed made a systematic summary.[19] According to the Gibbs adsorption isotherm for multicomponent systems:

$$\mathrm{d}\gamma = -S\,\mathrm{d}T - \sum \Gamma_e\,\mathrm{d}\tilde{\mu}_e - \sum \Gamma_{ss}\,\mathrm{d}\mu_{ss} \quad (1)$$

where $\gamma$ represents surface tension; $S$ is the entropy of neutral substances; $T$ is temperature; $\Gamma_e$ and $\Gamma_{ss}$ are the surface excess concentration of charged and neutral components, respectively; $\mu_e$ and $\mu_{ss}$ are the chemical potential of charged and neutral components, respectively, and the electrochemical potential $\tilde{\mu}_e = \mu_e + z_e\varphi$ ($\varphi$, electric potential). The surface charge density $\sigma$ and the electric potential difference $\Delta\varphi_\beta^\alpha$ of the contacting phases $\alpha$ and $\beta$ in capillary were used to replace the term $\sum \Gamma_e \, \mathrm{d}\tilde{\mu}_e$:

$$\mathrm{d}\gamma = -S\,\mathrm{d}T - \sigma\,\mathrm{d}\Delta\varphi_\beta^\alpha - \sum \Gamma_{ss}\,\mathrm{d}\mu_{ss} \quad (2)$$

The second items in Eqs.(1) and (2) are equal, therefore, by differentiating $\gamma$ with respect to $\Delta\varphi_\beta^\alpha$, we can write the Lippmann equation as follows. When the left term equals zero (or $\sigma = 0$), the surface tension reaches its maximum on the electrocapillary curve.

$$\left(\frac{\partial\gamma}{\partial\Delta\varphi_\beta^\alpha}\right)_{\mu_{ss}} = -\sigma \quad (3)$$

Salem further considered the additional electric energy in the superficial layer by employing an isolated polar molecule model, as depicted in **Figure 2C**. The author posited that in the presence of an electric field $E$, the charges $q$ shift in opposite directions, thereby increasing the distance between them. This led to the formulation of a new equation that describes the impact of the polarizable charge density on surface tension:

$$\left(\frac{\partial\gamma}{\partial\Delta\chi}\right) = -\left(\sigma^{\mathrm{fr}} + \sigma^{\mathrm{bnd}}\right) \quad (4)$$

where $\Delta\chi$ represents the field potential at the point where charge is located; $\sigma^{\mathrm{fr}}$ and $\sigma^{\mathrm{bnd}}$ represent surface densities of the free and bound charges. Eqs. (3) and (4) have similar forms because they were both derived from the basic laws of thermodynamics and the Gibbs adsorption equation. Recently, the work of He's group[20] pioneers the concept of leveraging electrocapillary action to enhance electrode wetting in high-energy LIBs, addressing the challenges posed by large, thick, and highly pressed electrodes that hinder electrolyte filling and wetting. Through theoretical analysis using the Young-Lippmann equation and experimental validation with commercial $LiFePO_4$/graphite pouch cells, they demonstrated that applying an external voltage can significantly accelerate the wetting process, leading to fully wetted electrodes within a shorter timeframe compared to untreated controls. This strategy improves battery manufacturing efficiency and safety by ensuring uniform electrolyte distribution and preventing issues like lithium plating.

Oscar K. Sefler Rice[21] made significant contributions to understanding the mechanism of electrocapillarity by synthesizing prior research on the EDL. According to the derived process of GCS model in the theory of EDL, the Boltzmann distribution equation ($n_{x,i} = n_{0,i} \exp\left(-z_i \omega\varphi / kT\right)$) was used to describe the concentration ($n_{x,i}$) of ion species $i$ at a distance $x$ from the surface and volume charge density; where $n_{0,i}$ represents the concentration in the bulk solution, $z_i$ the valence number including sign, $\omega$ the charge per unit mass of ions, $\varphi$ electric potential, $k$ Boltzmann constant. The total concentration is the sum of positive and negative ion concentrations $n_x = n_{x,+} - n_{x,-}$. Subsequently, the GCS equation can be derived by combining the Boltzmann distribution with Poisson's equation (also known as Gauss's law):

$$Q = \sqrt{8 n_x RT \varepsilon_0 \varepsilon_r} \sinh\left(\frac{|z| \varphi_x F}{2RT}\right) \quad (5)$$

$Q$ surface charge, $\varepsilon_0 \varepsilon_r$ dielectric constant, $\varphi_x$ electric potential drop of diffuse layer, $F$ Faraday constant. The other EDL equations have been introduced in numerous literature and books, and is thus omitted in this review.

Rice proposed the presence of not only a diffuse ion layer near mercury in the solution but also a diffusion layer within the mercury itself. Additionally, it was assumed that there exists a constant-capacitance capacitor at the interface. Employing the theory of EDL, Rice conducted calculations to determine the charge per unit surface. Furthermore, an equation was derived for surface tension by introducing $\mathrm{d}\gamma = -Q\mathrm{d}\varphi$, providing a comprehensive and quantitative framework for exploring the intricate interplay of electrochemical processes at the mercury-solution interface:

$$\Delta\gamma = \frac{2RT}{\omega} \sqrt{\frac{\varepsilon_0 \varepsilon_r n_x RT}{2\pi}} \left( \mathrm{e}^{\frac{\omega\varphi_x}{4RT}} - \mathrm{e}^{-\frac{\omega\varphi_{x,1}}{4RT}} \right)^2 \quad (6)$$

where $\Delta\gamma$ represents the change in surface tension when the electric potential changes from $\varphi_x$ to $\varphi_{x,1}$. The relationship between surface tension and electric potential in this equation also exhibits a parabolic shape and is linked to the diffusion layer electric potential in the EDL theory.

The presence of EDL structure has been experimentally confirmed through measuring double-layer forces. Jacob N. Israelachvili et al.[22] pioneered the development of a surface force apparatus (SFA) capable of measuring electrostatic forces with nanonewton (nN) precision while also providing precise control over the separation distance between surfaces. The equipment picture and scheme of measure mechanism[23] were shown in **Figure 3A and 3B**. In the SFA, two smooth surfaces, typically made of materials like mica or transparent glass, are prepared and mounted on separate piezoelectric transducers. These transducers can precisely control the distance between the surfaces and measure the normal force between them. In electrolyte solutions, surfaces typically carry a charge, leading to the attraction of oppositely charged ions in the solution, resulting in the formation of the symmetrical EDLs. The ionic distribution in the symmetrical EDLs can be seen in **Figure 3C**. The force associated with the two EDLs is generated by the potential difference induced by these counterions in close proximity to the surface. This force escalates quickly as the distance between surfaces decreases, and it frequently becomes substantial over minute distances, especially at the nanometer scale. **Figure 3D** provided a diagram of the relationship between interaction potential energy and distance between two surfaces with the same charge in a 1:1 electrolyte solution under the Derjaguin-Landau-Verwey-Overbeek (DLVO) theory. DLVO represents the combined force of Van der Waals forces and electrostatic double-layer forces. In the figure, there is a swift rise in energy as the separation between the two surfaces narrows, attributable to the EDL repulsion. This force arises from the mutual repulsion of counterions. At elevated electrolyte concentrations, the repulsive force diminishes due to the shielding effect of the abundant ions in the electrolyte, which mitigate the influence of the surface charges. As the surfaces approach each other, an energy barrier emerges, representing a peak in the interaction energy that hinders the particles from coming into closer proximity. Importantly, the surfaces are often crossed at a small angle to create a well-defined contact area, as shown in **Figure 3E**. This geometry helps in controlling the contact region and reduces edge effects that could otherwise complicate the measurements. Thus, the SFA measurements confirmed the presence of counterions at the charged surface and the validity of the basic EDL structure.

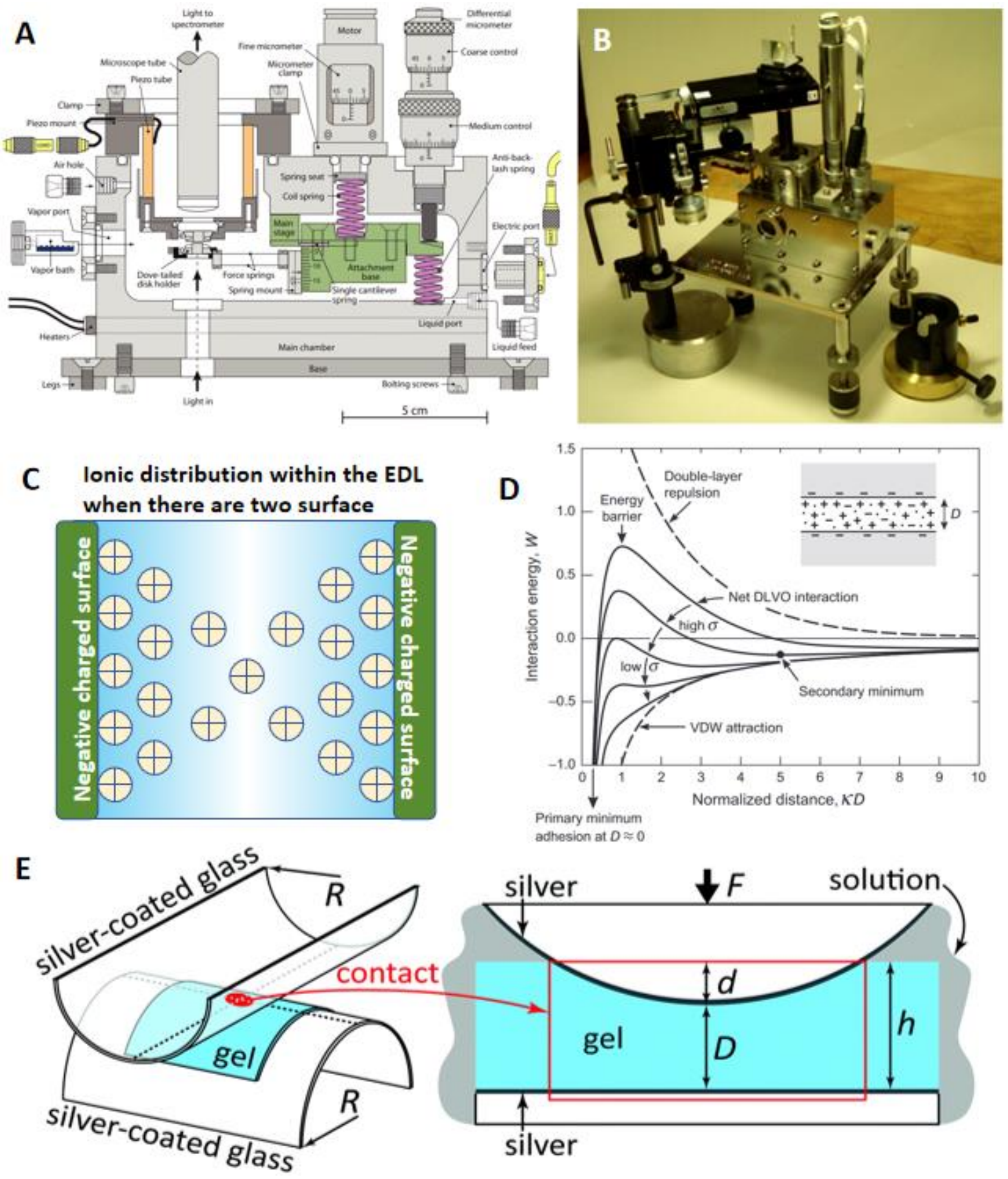

**Figure 3.** (A) The scheme of the center section in SFA 2000 for measuring forces between two molecularly smooth surfaces, and (B) the picture of the apparatus SFA 2000;[23] Copyright © 2010 IOP Publishing Ltd. (C) The scheme of the positive ionic distribution within EDLs when there are two negative charged plane surfaces; (D) Schematic energy versus distance profiles of the DLVO interaction; in the figure, *D* normalized distance between two planar surfaces, *W* interaction energy, VDW van der Waals, σ particle size (radius).[24] Copyright © 2011 Elsevier B.V. (E) Crossed cylinder configuration and schematic of the contact region in the SFA;[25] Copyright © 2020 The Royal Society of Chemistry. **Israelachvili's SFA measurement on electrostatic forces confirmed presence of EDL. DLVO theory explains energy barrier, EDL repulsion, influenced by ionic concentration. Mica or glass surfaces on piezoelectric transducers control separation, reveal force-distance relationships.**

## 2.2 Expansions of EDL theory

While the mentioned theories play the crucial roles in electrochemistry research, they significantly diverge from real-world solution systems.[26] This discrepancy arises from key assumptions within the GCS model. Firstly, the model assumes a one-dimensional distributed electric field,[27] oversimplifying the complex nature of electric fields in real systems. Secondly, it considers the electrode surface as a flat plate, neglecting the intricacies introduced by the actual topography of surfaces. Thirdly, the model treats ions as point charges, overlooking their polarizable and volumetric properties. Additionally, the electrocapillary theory, originating from the liquid-liquid (mercury-solution) interface, encounters challenges when applied to solid-liquid interfaces in practical scenarios.[28] Real-world complexities, such as the influence of crystal lattices, adsorption phenomena, and crystal plane orientation, cannot be disregarded, necessitating a more nuanced and comprehensive consideration of these factors in theoretical frameworks. For example, the $\sigma^{\mathrm{bnd}}$ in Eq. (4) was further explored in Salem's work[19], the number of solvent molecules in a unit volume $N_0$, a mean electronic polarizability of the solvent molecules $\alpha_s$, and the electric field $E$ can be used to represent this term:

$$\sigma^{\mathrm{bnd}} = -N_0 \alpha_s E \quad (7)$$

After uncomplicated derivation, the author got:

$$\Delta\gamma = \frac{1}{2} E^2 \left(1 + \frac{\sigma^{\mathrm{fr}}}{\sigma^{\mathrm{bnd}}}\right)\left(\alpha_s N_0^{\ 2/3}\right) \quad (8)$$

The equation illustrated that the reduction in surface tension observed when an uncharged liquid metal is immersed in a compact environment is dictated by the energy expended in inducing the dipole moment. Furthermore, according to the data of 26 individual solvents, a plot of $\Delta\gamma - \alpha_s N_0^{\ 2/3}$ relationship has been drawn as shown in **Figure 4A**. The point corresponding to water solvent deviated from the general trend in the plot due to the strong hydrogen bonding.

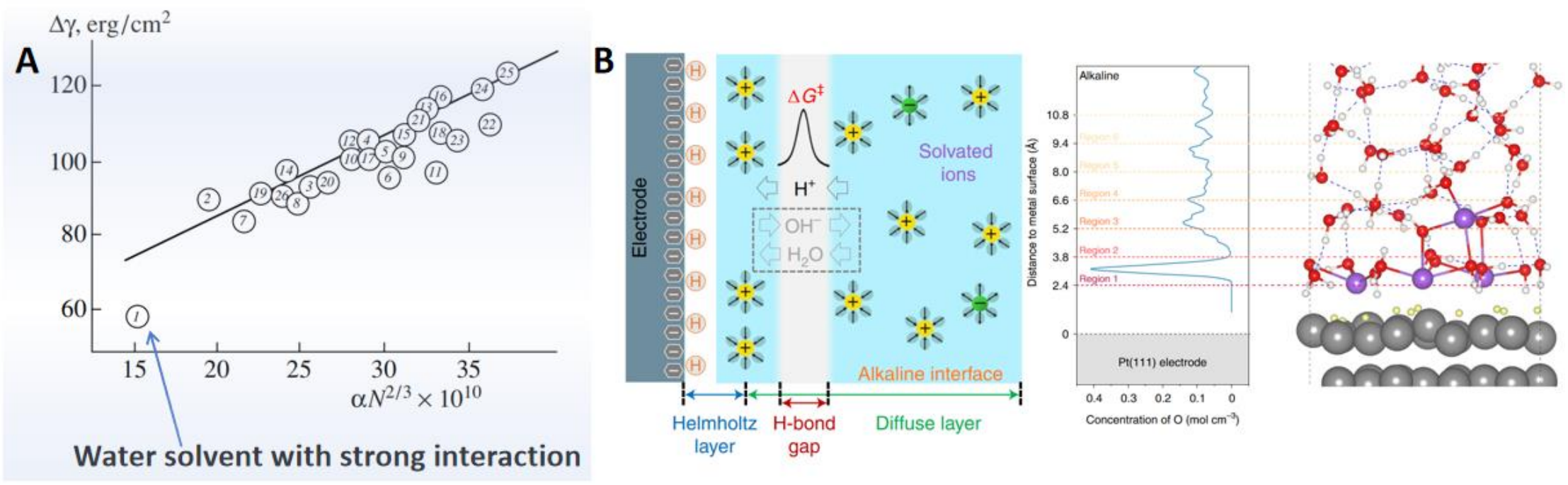

**Figure 4.** (A) The surface tension of mercury Δγ vs. the polarizability coefficient of the solvents[19]: (1) water; (2) methanol; (3) ethanol; (4) n-propanol; (5) n-butanol; (6) tert-butanol; (7) formic acid; (8) acetic acid (9) butyric acid; (10) ethylene glycol; (11) glycerin; (12) diethyl ether; (13) chloroform; (14) formamide; (15) dimethyl formamide; (16) dimethyl acetamide; (17) n-methyl formamide; (18) n-methyl propionamide; (19) acetonitrile; (20) acetone; (21) dimethyl sulfoxide; (22) sulfolane; (23) diethylene glycol; (24) pyridine; (25) benzene; and (26) ammonia; Copyright © 2005 Springer Nature. (B) Schematic representation of proton transfer from the bulk to the interface via hydrogen bond networks in the alkaline system, and schematic illustration of the interfacial region division in the alkaline system, highlighting the gap

region of hydrogen bond networks.[29] Copyright © 2022 Nature. **Chen's group revealed pH effects on Pt electrodes during hydrogen electrocatalysis through studying hydrogen bond network in EDL. AIMD and SEIRAS showed water proton transfer dynamics, influencing hydrogen transfer channels and congestion near Helmholtz plane.**

Therefore, researchers have started to consider the EDL from the perspective of complex interaction networks. Chen's group[29] conducted an investigation into the dominance of hydrogen bond network connectivity in the EDL during hydrogen electrocatalysis on Pt electrodes. This exploration involved a combination of ab initio molecular dynamics (AIMD) simulations and in situ surface-enhanced infrared absorption spectroscopy (SEIRAS) measurements. The study revealed kinetic pH effects, where water molecules could transfer protons to the interface through the hydrogen bond network in the EDL. In an alkaline EDL, the presence of a gap region in the hydrogen bonding network disrupted the continuity, potentially causing a reduction in hydrogen transfer channels and leading to significant congestion near the Helmholtz plane. **Figure 4B** illustrated how the distribution of water molecules and the connectivity of the hydrogen bonding network change within the gap region with increasing distance from the electrode surface. This research on solvent molecule interaction within the EDL offered a novel perspective for designing and optimizing efficient hydrogen electrocatalysts under alkaline conditions. There are also some similar works about simulations of EDL structures recently.[30]

Deriving equations for the EDL considering strong interactions poses a considerable challenge and has proven to be a complex task. A common interface in the electrochemical systems consists of a solid side and a liquid side. Numerous researchers aspire to initiate their investigations from the liquid side, employing quantum mechanical methodologies to derive formulations for the EDL that incorporate the intricacies of quantum effects. In 2007, Alexei A. Kornyshev[31] conducted an in-depth exploration of the structure and characteristics of the double layer formed by ionic liquids at charged interfaces. The author proposed a novel analytical formula to elucidate the potential-dependent behavior of EDL capacitance at planar metal-ionic liquid interfaces. This theoretical framework relied on mean field properties, drawing from the Poisson-Boltzmann lattice gas model, and incorporated a correction accounting for the finite volume occupied by ions. This acknowledged the spatial limitations on ion arrangement, refuting the assumption of infinite ion density. In contrast to the GCS model's reliance on the Boltzmann distribution, Kornyshev employed a Fermi-like distribution:

$$n_{x,\pm} = n_0 \frac{\exp\left(\mp e\varphi_x / kT\right)}{1-\lambda+\lambda\cosh\left(e\varphi_x / kT\right)} \quad (9)$$

where all the symbols retain the same meaning as before. Importantly, the newly introduced parameter $\lambda = \bar{N}/N$ ($\bar{N}$ and $N$ are the total number of ions in the bulk and the total number of sites available for ions, respectively) represents lattice-saturation effect of steric effect on the surface. Incorporating this volume correction led to the derivation of a novel formula for EDL capacitance:

$$C = C_0 \frac{\cosh(u_0/2)}{1+2\lambda\sinh^2(u_0/2)}\sqrt{\frac{2\lambda\sinh^2(u_0/2)}{\ln\left[1+2\lambda\sinh^2(u_0/2)\right]}} \quad (10)$$

where $C$ differential capacitance of EDL; $C_0$ The linear (or initial) capacitance of EDL, that is, the capacitance when the EDL capacitance is close to a constant value at a low potential, can be represented by the Debye capacitance; $u_0$ electric potential vs. bulk. This equation aligned with the GCS model under the limiting condition of $\lambda \to 0$, yet diverged in behavior under other conditions, notably at high lattice-saturation effect. Unlike the exponential increase predicted by the GCS model, the capacitance in this scenario decreased in proportion to the square root of the potential, marking a significant deviation from traditional expectations. In fact, the EDL equation considering lattice-saturation effect has been studied by other researchers[32], and could also be called as Bikerman-Poisson-Boltzmann (BPB) model[33]. However, while the derivation of the Fermi-like distribution was rooted in solid state physics and quantum mechanics, particularly within the domain of semiconductors, the final EDL formulation primarily stemmed from classical thermodynamics.

Furthermore, some modifications have been made by integrating quantum mechanical methods and perspectives based on ion diffusion. For example, Jiang et al.[34] employed time-dependent density functional theory (TDDFT) alongside the conventional Poisson-Nernst-Planck (PNP) equations to examine ion diffusion within electrochemical systems. The PNP equations, which adapt Fick's law to account for electromigration, have seen extensive investigation and application. TDDFT, on the other hand, incorporates considerations for spatial and electrostatic interactions, and the core equation can be expressed as:

$$\frac{\partial \rho_i(\mathbf{r},t)}{\partial t} = \nabla \cdot \left\{ D_i \rho_i(\mathbf{r},t) \nabla \left[ \beta \mu_i(\mathbf{r},t) + \beta V_i(\mathbf{r}) \right] \right\} \quad (11)$$

where $\rho_i(\mathbf{r},t)$ density of ion $i$ at position $\mathbf{r}$ and time $t$, $D_i$ ion diffusivity at infinite dilution, $\beta = 1/kT$ Boltzmann factor, $\mu_i(\mathbf{r},t)$ local chemical potential, $V_i(\mathbf{r})$ external potential to the ionic species. **Figure 5A** illustrates the local charge density, total ion density, and local average potential as predicted by the TDDFT, PNP equation, and modified PNP equation (MPNP, accounting for lattice saturation effects) during the intermediate stages of the charging process. This comparison was conducted at a specific charging time ($t/\tau_D = 5$, where $\tau_D$ represents the ratio of Debye length to diffusion coefficient) and a specific electrode potential (ψ0 = 0.25 V). TDDFT exhibits superior capability in capturing intricate behaviors within electrochemical systems, including wave-like changes in ion density that may not be observable in traditional PNP models.

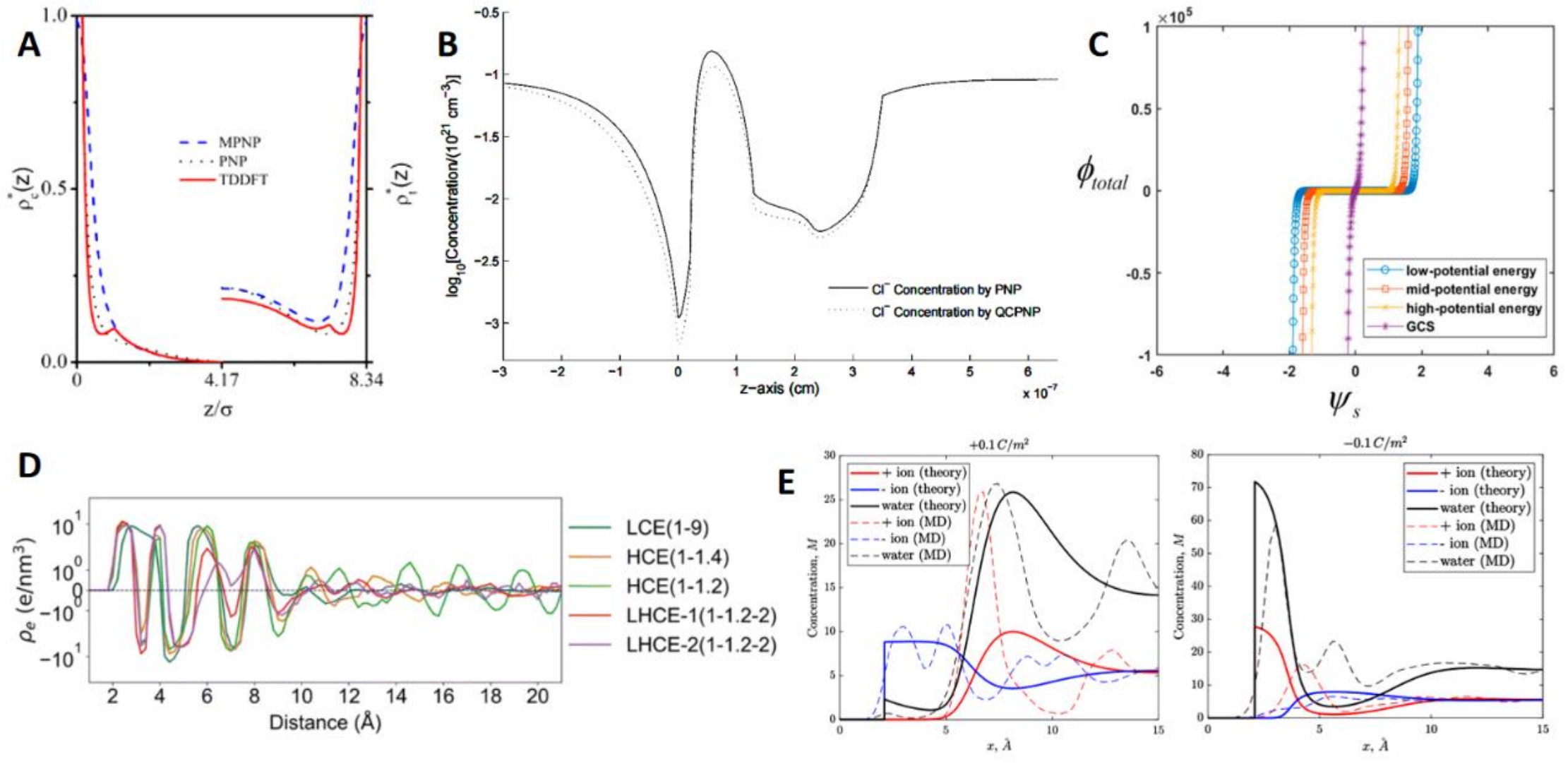

**Figure 5.** (A) Comparison of $\rho_c^*$ the total local charge density, $\rho_t^*$ the total ionic density profile, and the local mean electrostatic potential predicted by TDDFT, PNP and MPNP equations (The horizontal axis represents the position of ions relative to the electrode surface);[34] Copyright © 2014 IOP Publishing Ltd. (B) Chloride concentrations calculated by PNP (solid line) and QCPNP (dotted line) models;[35] Copyright © 2015 De Gruyter publishing. (C) Comparisons of the total potential difference and dispersion layer potential relationship in GCS model with and without quantum tunneling effect of electron;[36] Copyright © 2023 AIP Publishing LLC. (D) Local charge density distributions near charged surfaces from MD simulations for various concentrated electrolytes (Low-, High- and Localized high-concentration electrolytes (LCE, HCE, and LHCE));[37] Copyright © 2025 The Royal Society of Chemistry. (E) Comparison of interfacial concentration profiles for cations, anions, and water molecules under ±0.1 C/m$^2$ surface charge, as obtained from continuum theory and MD simulations.[38] Copyright © 2018 American Chemical Society. **Jiang et al. compared the TDDFT and PNP equations when studying ion diffusion in electrochemical systems, revealing TDDFT's superiority in capturing complex behaviors. Liu introduced QCPNP by incorporating the quantum potential, leading to more precise ion channel simulations. The application of the WKB equation for electron tunneling effects significantly demonstrates the impact of quantum effects on ion transport. The MD simulations illustrates the pronounced concentration-dependent EDL structures. The continuum theory and MD simulations confirm the importance of including solvation and bound water in accurate EDL models.**

There were other works to introduce quantum mechanics to analyze the ionic diffusion on the liquid side. Jinn-Liang Liu[35] introduced the quantum potential modified PNP (QCPNP) model to simulate ion currents in biological ion channels. The quantum potential in this model is derived from the de Broglie-Bohm theory, serving as a first-order modification of classical potential energy. It is employed to characterize moving particles with well-defined trajectories. The quantum potential could be written as:

$$\psi_i^{\pm} = \frac{\mp\hbar^2}{2m_i}\frac{\Delta\sqrt{c_i}}{\sqrt{c_i}} \quad (12)$$

where $\psi_i^{\pm}$ the quantum potential of ion species $i$, $m_i$ ionic mass, $c_i$ ion concentration. It was applied to modify PNP equation and obtained QCPNP:

$$-\nabla\cdot\left(\varepsilon_0\varepsilon_r\nabla\varphi\right) = q_p c_p + \sum_i q_i\hat{c}_i \exp\left(-\beta_i\left(\varphi+\psi_i^{\pm}\right)\right),$$

$$\frac{\mp\hbar^2}{2m_i}\Delta\sqrt{c_i} = \sqrt{c_i}\left[\frac{-1}{\beta_i}\left(\ln\left(\sqrt{c_i}\right)^2 - \ln\hat{c}_i\right) - \varphi\right],$$

$$\nabla\cdot\left[D_i\exp\left(-\beta_i\left(\varphi+\psi_i^{\pm}\right)\right)\nabla\hat{c}_i\right] = 0 \quad (13)$$

where $q_p$ background fixed charge on the channel, $c_p$ background concentration, $q_i$ carrying charge of ion species $i$, $\beta_i = q_i/kT$, $\hat{c}_i$ an extension of the classical Slotboom variable ($c_i = \hat{c}_i\exp\left(-\beta_i\left(\varphi+\psi_i^{\pm}\right)\right)$). The author compared the results of calculating the chloride concentrations by using PNP and QCPNP equations, as shown in **Figure 5B**. The introduction of quantum potential allows quantum effects to be considered in the PNP model. When dealing with ion channel problems, these effects may have a significant impact on ion transport. The QCPNP model can provide a more precise description of ion channel behavior. Therefore, the results of this article can shed some light on how to introduce quantum mechanics into EDL, especially in terms of equation derivation. For example, the Wentzel–Kramers–Brillouin (WKB) formula in quantum electrochemistry can be introduce to describe the quantum tunneling effect of electron within EDL.[36] The EDL capacitance could be expressed as $C_{EDL} = \lambda_e(\phi_{total} - \varphi_s)$, where $\lambda_e$ the amount of charge after removing the electron spillover, $\phi_{total} - \varphi_s$ the difference between total potential difference and potential of dispersion layer. The relationships of $\phi_{total} \sim \varphi_s$ have been shown in the curves, as shown in **Figure 5C**. The platform induced by electron tunneling elevated the potential of the dispersion layer, exerting a more pronounced influence on the region situated between the diffusion layer and the bulk solution.

Meanwhile, for the interface in high concentration electrolyte, Wu and Qi[37] used molecular dynamics (MD) simulations calibrated with density functional theory (DFT)[39] to investigate the structure of the EDL at the interface between graphene electrodes and high-concentration electrolytes. The LiFSI in DME systems were considered. **Figure 5D** presents the charge density profiles of different electrolyte systems under the surface charge density of ±0.6 $e$/nm$^2$. The results show that the high-concentration system when salt:solvent is 1:1.2 displays the most pronounced charge oscillations beyond the EDL region, which contrasts with the classical Debye theory, where screening length decreases with higher ionic strength. These results demonstrate that concentrated electrolytes give rise to more structurally complex EDLs. For the interfaces in aqueous electrolytes, McEldrew et al.[38] proposed a continuum theory for EDL development in water-in-salt electrolytes, incorporating key physical insights such as ion-solvent interactions and polarization effects in the grand canonical free energy functional framework. **Figure 5E** compare the concentration profiles of cations, anions, and water obtained from this theory with those from MD simulations, under surface charge densities of +0.1 C/m$^2$ and -0.1 C/m$^2$, respectively. Despite treating $Li^+$ and its solvation shell as a symmetric species, the model captures key features such as water depletion at positively charged surfaces and interfacial ion layering, highlighting the importance of solvation and electrosorption effects in concentrated electrolyte systems.

In fact, the more widely used quantum mechanical model is Jellium model, which can be used to treat the solid side of EDL. In the Jellium model, which has been described by Wolfgang Schmickler in detail[40], electrons are considered as an electron gas interacting with a uniformly distributed positive charge background (representing the ion cores of the metal). Although the electron density is assumed to be uniform in the bulk, near the metal surface, due to quantum mechanical effects such as the wave-like nature of electrons and quantum overlap, the electron density exhibits non-uniformity and gradually decreases as the distance from the metal surface increases. This phenomenon is known as electron spillover (or electron tunneling). Electron spillover describes how, near the metal surface, the electron density extends beyond the region defined by the positive charge background, showing quantum effects resulting from the interaction between electrons and the positive charge background. This is an important concept in the study of the electronic structure and surface properties of metals. However, this model, while instrumental in advancing the understanding of electron dynamics within metals, encounters significant limitations when applied to semiconductor materials, due to its inherent oversimplification of the metal's positive ion background into a uniform, continuous charge distribution. This approach neglects the discrete, crystalline lattice structure that is pivotal in defining the electronic properties of semiconductors. Semiconductors are characterized by their specific lattice arrangements, which give rise to a distinct band structure, influencing crucial phenomena such as band gap formation and charge carrier mobility. These properties are critically dependent on the quantum mechanical interactions between electrons and the periodic potential of the lattice, aspects that the jellium model cannot accommodate due to its generalized treatment of the ion background. Consequently, while the jellium model provides valuable insights into the general behavior of electrons in metals, its application is markedly constrained in the realm of semiconductor-type electrode materials, where the detailed lattice structure and its effects on electronic properties are of paramount importance.

Moreover, Wu's research[41] has focused on the intricate understanding of EDL, and the study emphasized the importance of quantum mechanical principles and methodologies in elucidating the EDL's microscopic structure and dynamic properties. The concept of quantum capacitance was introduced to consider the effects of electron density fluctuations within the electrode on the overall capacitance. The grand canonical ensemble method was utilized to scrutinize the density profiles of electrons, ions, and solvent molecules within an open system, providing novel perspectives on the EDL's electronic structure and capacitance. Detailed descriptions of the electronic structure at the interface were achieved through the application of Kohn-Sham density functional theory (KS-DFT), which was essential for grasping the non-classical EDL behaviors. The development of Joint density functional theory (JointDFT) marked a significant step in merging quantum mechanical insights with classical theories, offering a new paradigm for capturing the interplay between quantum and classical elements. These quantum mechanical applications deepened our comprehension of the EDL structure and established a robust theoretical basis for the enhancement and innovation of electrochemical systems.

## 3 The EDL on the Electrode Materials in Batteries

The traditional EDL theory, which has primarily been derived from the data related to liquid-liquid interfaces, does not account for the material-specific properties of electrodes. This limitation presents a significant challenge in the field of electrochemistry, especially in the areas of quantum electrochemistry, electrode chemistry, and battery technology. Consequently, advancing EDL theory to accurately model solid-liquid interfaces for a diverse range of electrode materials is pivotal. Addressing this challenge is critical for enhancing our understanding and improving the performance of electrochemical systems. However, it may cause potential confusions when the so many materials were divided into several types according to the components or elements. We notice that most of electrode materials in batteries[42] belongs to the semiconductor, such as some anodes[43] including $In_2O_3$, ZnO, $SnO_2$, $Cu_2O$, NiO, SnO, $Co_3O_4$[44], etc., and most of cathodes including $LiCr_{0.33}V_{0.33}Mn_{0.33}O_2$,[45] $LiFePO_4$,[46] $Li_xNi_{0.8}Co_{0.2}O_2$,[47] $LiCoO_2$,[48] $LiNiO_2$,[49] $LiNiCoMnO_2$,[50] etc. There are some special semiconductor materials, such as silicon-based materials[51] and graphite[52]. The latter is a zero-bandgap semiconductor with a cone-like band structure near the Dirac point.[53] The oxide electrodes in other kind of batteries also belong to semiconductor, such as $Na_xVO_2$,[54] $NaMnO_2$[55]. Meanwhile, some other kind of electrode materials are conductor, such as metal electrodes.

Moreover, semiconductor materials exhibit unique physical characteristics, particularly in terms of their band structure. The band structure of a semiconductor comprises a valence band and a conduction band, with an energy gap between them known as the band gap or forbidden band. Generally, a smaller band gap results in better conductivity for semiconductors at room temperature. At absolute zero, the conduction band of a semiconductor is empty, and the valence band is fully occupied, rendering them non-conductive. As the temperature increases, some electrons gain enough energy to transition to the conduction band, initiating the semiconductor's conductivity. It is important to note that the discussion here pertains to electronic conductivity, not ionic conductivity. Any reference to ionic conductivity will be explicitly specified in the text.

Furthermore, the conductivity of semiconductors can be modulated through a process called doping. Doping involves introducing a small amount of impurity atoms into the semiconductor lattice. These impurity atoms can provide additional electrons (N-type doping) or holes (P-type doping), thereby enhancing the material's conductivity. The interface characteristics of semiconductors significantly influences their electrochemical behavior. Factors such as surface states, surface adsorption, and interface charge distribution can impact the electronic structure and conductivity of semiconductors.

Hence, in order to study and develop the EDL theory from quantum mechanics perspective, electrode materials can be categorized into two main types: semiconductors and conductors. Our attention is directed towards understanding the impacts of some distinct interfaces on the EDL theory—specifically, semiconductor/liquid interfaces, as well as conductor/liquid, semiconductor/semiconductor or conductor interfaces.

## 3.1 Conductor/liquid interfaces

### *3.1.1 Basic model*

Conductors, such as metals, have traditionally been used as collectors or anodes in batteries. Huang et al.[56] developed a grand-canonical model based on the hybrid density-potential functional to describe the distribution of electron and ion densities in the EDL at the metal-liquid interface. At the solid side, metal cation cores are represented as circles with a certain diameter ($a_{\mathrm{mc}}$), arranged periodically ($t$) beneath the metal surface, as shown in **Figure 6A**. A statistical thermodynamic grand potential equation is first established:

$$\Omega = U - TS - \int dV \sum_i \bar{\mu}_i n_i \quad (14)$$

where $\Omega$, grand potential; $U$, internal energy; $S$, entropy; $dV$, a volume unit of the space; $\bar{\mu}_i$, electrochemical potentials of ions and solvent molecules; $n_i$, particle densities. This equation is description of the item in grand canonical ensemble $p = \exp(-\beta\Omega)/Z$ ($Z$, a normalization factor; $\beta = (k_B T)^{-1}$), which is used to represent the possible states thermodynamically. The authors believed that the minimum of the potential corresponded to the microstate with the highest probability of the EDL. By employing the Born-Oppenheimer approximation, they neglected the kinetic energy of the nuclei and decomposed the total internal energy of the EDL into five terms: the kinetic energy of the metal electrons, exchange-correlation energy, electrostatic potential energy, the interaction energy between solvent molecules and the electric field, and the repulsive force between the metal and the solution phase. Using the Euler-Lagrange equation combined with the potential and electron density, they obtained a closed set of control equations for solving the particle density and potential distribution in the EDL. The simulation results were shown in **Figures 6B and 6C**.

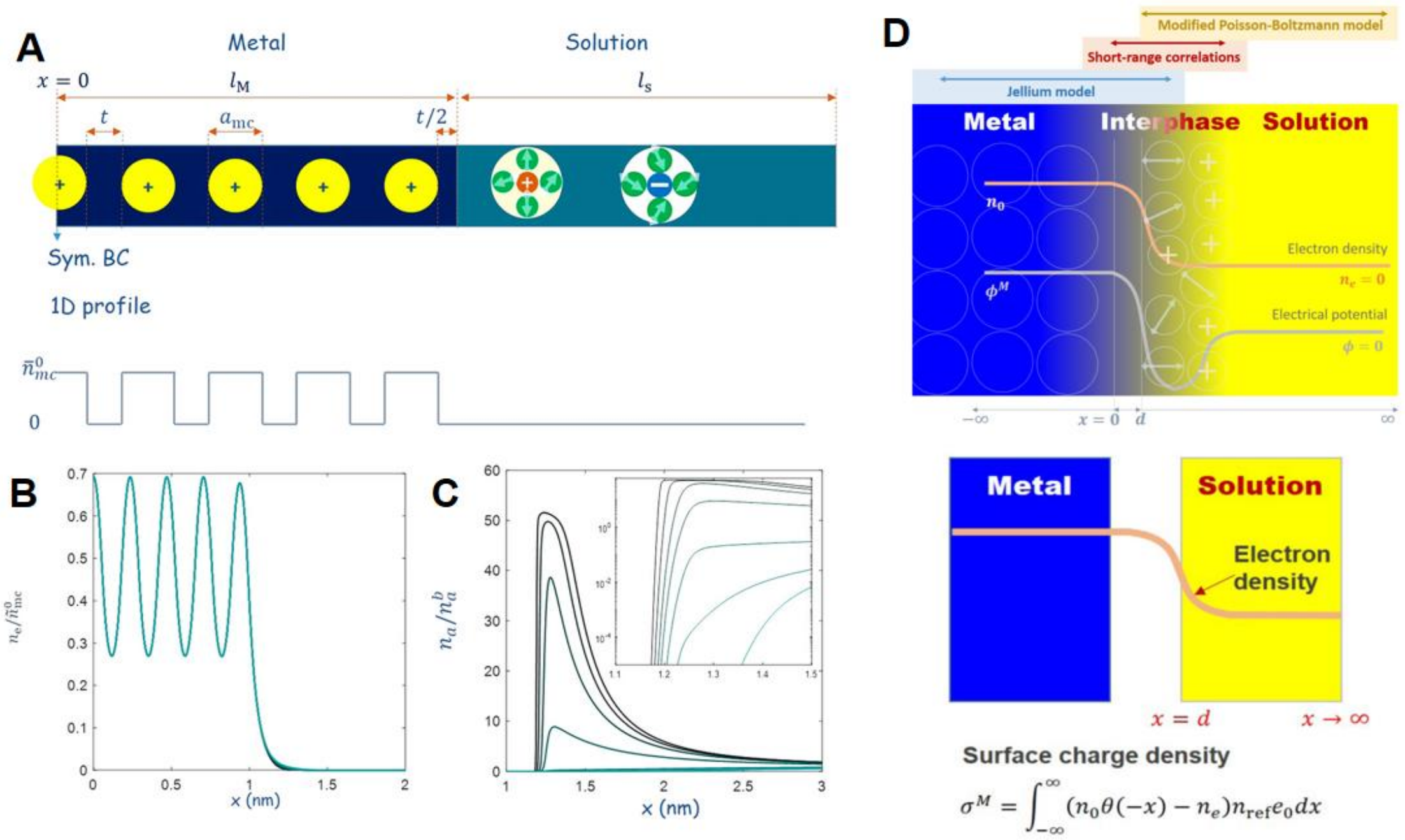

**Figure 6**. (A) Schematic illustration of the distributions of metal ionic cores;[56] (B, C) Corresponding simulation results of electron density and electric potential distributions across the metal–solution interface;[56] Copyright © 2021 American Chemical Society. (D) Scheme of metal-liquid interface model.[57] Copyright © 2020 American Physical Society. **Huang et al. created a grand-canonical model for electron and ion densities in the EDL at metal-liquid interfaces, using a hybrid density-potential functional. The model, incorporating the Born-Oppenheimer approximation, provides equations for particle density and potential distribution in EDL.**

Huang et al.[57] also developed a continuum model that employed the jellium model to describe the metallic electron effect and used a modified Poisson-Boltzmann model to describe the solution side. Short-range correlations between metal electrons and solvents as well as ions at the contact surface were considered, as shown in **Figure 6D**. This work summarized the application of the jellium-Poisson-Boltzmann model in the study of the point of zero charge (PZC) at the metal/solution interface. The model revealed that the entry of free metal electrons into the solution phase led to a deviation between the PZC determined from the $\sigma_M$~$U_M$ curve and the Gouy-Chapman minimum extracted from the $C_{dl}$~$U_M$ curve ($U_M$ represented the dimensionless electrical potential applied onto the metal referenced to the potential in the solution bulk). The quantum effects of metal electrons considered by the model (electron spill-out) resulted in a non-zero charge density $\sigma_M$ at the PZC.

### *3.1.2 Applications and Outlooks*

The widely used metal electrodes are Zn metal and Li metal. For example, Xu et al.[58] elucidated the role of the EDL in controlling competitive reduction reactions, beyond thermodynamic stability, during the formation of the SEI on Li metal anodes. The negative charge on the Li metal surface promotes the formation of a cation-enriched and anion-depleted environment within the EDL. Consequently, only substances that participate in the cation solvation shell can accumulate near the Li metal surface and be preferentially reduced in a continuous dynamic cycle. The authors introduced multivalent cation additives to more effectively incorporate beneficial anionic SEI components into the EDL, leading to the formation of a high-quality SEI. There are other similar works[59] based on this principle.

It can be observed that experimental studies on metal anodes did not delve into deep fundamental theories. However, researchers could leverage the properties of the EDL to adjust strategies to enhance surface stability and suppress dendrite growth. Lv et al.[60] improved the performance of zinc metal batteries by incorporating Zwitterionic ionic liquids (ZIL) into the electrolyte to construct an adaptive EDL. The ZIL additives, which contain both cationic and anionic groups, autonomously form a dynamic electrostatic shielding layer on the electrode surface under the influence of an electric field. The cationic groups (such as imidazolium cations) orient towards the negatively charged zinc anode surface to form the shielding layer, while the anionic groups (such as sulfate) interact with $Zn^{2+}$. The adaptive EDL can adjust its structure in response to local changes in the electric field at the electrode surface, thereby promoting the uniform deposition of $Zn^{2+}$ ions and helping to prevent the formation of zinc dendrites (**Figure 7A**). Furthermore, at the cathode interface, ZIL reduces the direct contact between water molecules and the active material of the cathode by forming a water-lean interface, thus inhibiting the dissolution of the cathode active material and related side reactions.

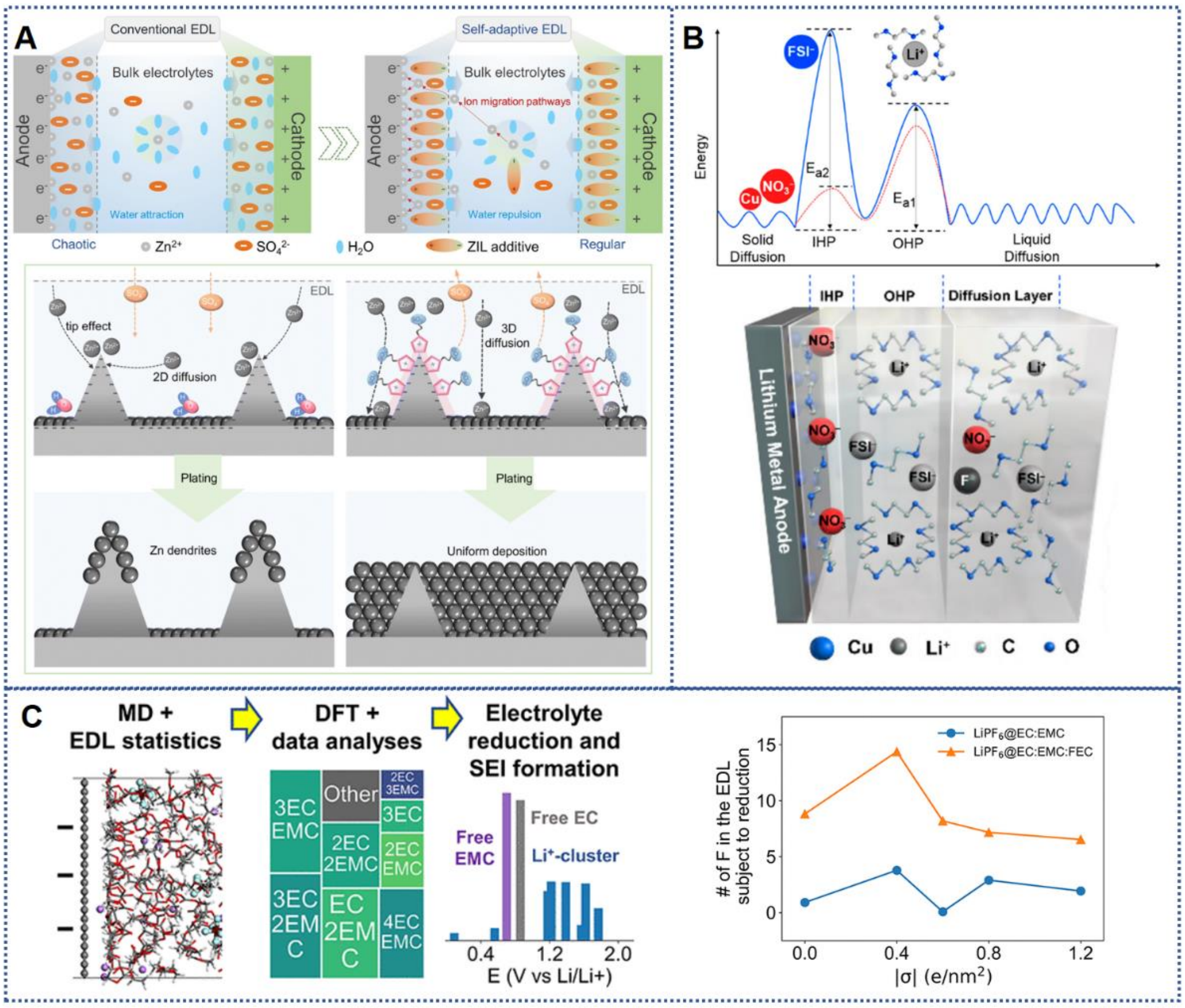

**Figure 7**. (A) Scheme of mechanism for constructing a self-adaptive EDL;[60] Copyright © 2022 The Royal Society of Chemistry. (B) Schematic descriptions of a grim competitive journey of solvated $Li^+$, and correlation between the energy barrier of $Li^+$ and the specific adsorption in the IHP.[61] Copyright © 2019 American Chemical Society. (C) A multiscale computational workflow that links the structure of the EDL to electrolyte reduction and SEI formation; and the MD-calculated number of F atoms within the EDL that are subject to reduction in the specific electrolytes.[62] Copyright © 2023 American Chemical Society. **Lv et al. enhanced zinc battery performance using Zwitterionic ionic liquids with an adaptive EDL. Yan et al. stabilized Li metal anodes by regulating the IHP using $LiNO_3$ and $CuF_2$, forming a stable SEI for improved battery efficiency. Wu et al. developed a novel method to extract information about EDL components by analyzing the EDL structure.**

Yan et al.[61] improved the performance of Li metal anodes by constructing a stable SEI through the regulation of the Inner Helmholtz Plane (IHP). By adding trace amounts of lithium nitrate ($LiNO_3$) and copper(II) fluoride ($CuF_2$) to the electrolyte, $NO_3^-$ was encouraged to adsorb preferentially on the IHP and to be reduced earlier than $FSI^-$ and $F^-$, thereby constructing a stable SEI. The Cu-$NO_3^-$ complex may help to reduce the potential barrier for lithium ion transport between the IHP and the SEI layer, particularly the energy barrier of the de-solvation process (Ea2), thereby enhancing the charging-discharging efficiency of the battery, as shown in **Figure 7B**. In simulations, Wu et al.[62] developed a novel interactive MD-DFT-data-driven model to investigate the EDL structure and electrolyte reduction reactions for $LiPF_6$ in EC:EMC electrolyte. This approach enabled the identification of EDL components and their likelihood of undergoing reduction, particularly focusing on the role of FEC as an additive. The study further revealed how FEC regulates the composition and formation of the SEI layer by increasing the number of fluorine-containing species within the EDL. As shown in **Figure 7C**, through the multi-coupled model, the presence of FEC leads to a higher number of reducible F atoms compared to systems without additives, thereby promoting LiF formation and enhancing interfacial stability of Li metal.

In experimental studies of metallic anode materials, the EDL theory most commonly employed is still the GCS model. The monolayer adsorption within the IHP, the linear distribution of potential in the compact layer, and the concentration changes in the diffuse layer are most often used to explain experimental phenomena. Experimental work tends to be fragmented between the solid and liquid sides, and Groß and Sakong[63] have also pointed out the limitations of the GCS model. These include considering the electrode as a perfect conductor, treating ions as point charges, and assuming that all interactions are of an electrostatic nature. With the rapid development of the electrochemical field in recent years, especially in battery research, the GCS model can no longer meet the requirements for experimental prediction, particularly in aspects such as high-concentration electrolytes, the properties of polycrystalline/single-crystal electrode materials, and the prediction of SEI formation. Therefore, for the metal/liquid interface, in addition to the previously mentioned grand-canonical model and jellium-Poisson-Boltzmann model, Groß has also proposed the necessity of developing AIMD simulations to understand the structure of the EDL at the electrode/electrolyte interface from an atomic level. AIMD is entirely based on quantum mechanical principles, especially DFT, to calculate the behavior of electrons and interactions between atoms. The distribution of changes in the electrostatic potential at the metal/liquid interface can be simulated using atomic arrangements, as shown in **Figure 8A**. However, each method has its own strengths and weaknesses; for example, the grand-canonical model allows for changes in the number of particles, while AIMD typically fixes the number of particles. Nevertheless, electrochemistry is a discipline that pays great attention to changes in the number of particles, especially since changes in concentration are fundamental to establishing electrochemical kinetics. Therefore, AIMD is more suitable for steady-state complex micro-interface systems that require precise electronic structure information.

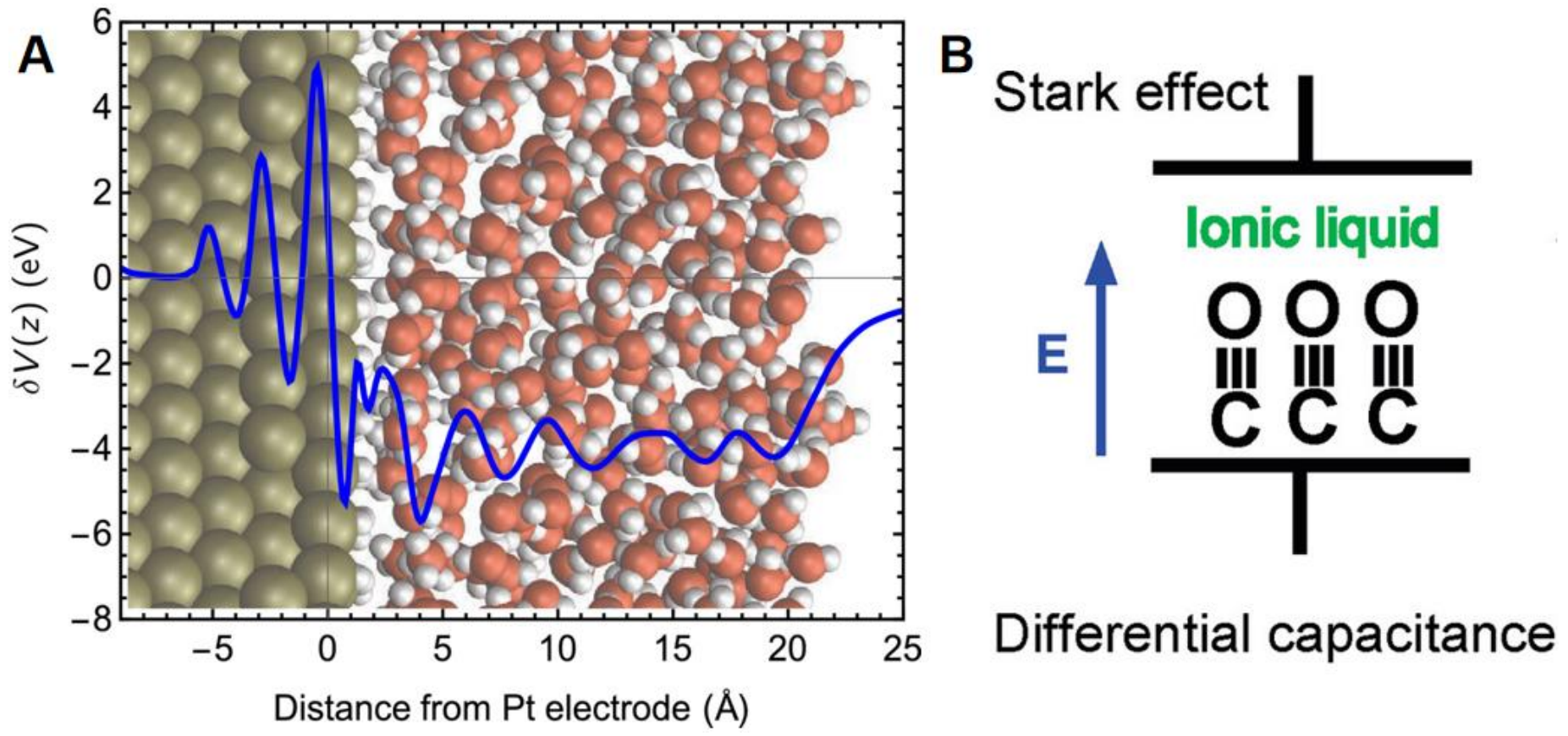

**Figure 8**. (A) A snapshot of the ab initio molecular dynamics trajectory with six layers of water on the hydrogen-covered Pt(111) electrode;[63] Copyright © 2019 Elsevier B.V. (B) Scheme of vibrational Stark shift measurement on the thickness of EDL.[64] Copyright © 2008 American Chemical Society. **The GCS model is commonly used in metallic anode studies but has limitations, which prompts the development of AIMD simulations for atomic-level EDL understanding. The Stark effect can be used to calculate EDL thickness by measuring vibrational frequency changes of adsorbed molecules, offering a method to study molecular polarization within the EDL.**

Furthermore, due to the simplicity of the metal surface conditions, it becomes possible to calculate the thickness of the EDL. When a molecule (such as CO) is adsorbed onto the electrode surface, its specific vibrational modes (such as the C-O stretching vibration) have a definite frequency. If an electric field exists on the electrode surface, this field can affect the electron cloud distribution of the adsorbed molecule, thereby altering the molecule's vibrational frequency (**Figure 8B**). The stronger the electric field, the greater the change in vibrational frequency (i.e., the Stark effect).[64] The proportionality constant between the change in vibrational frequency and the electric field strength is known as the Stark tuning rate (dv/dE), where dv is the change in vibrational frequency, and dE is the change in electric field. Therefore, by measuring the vibrational frequency and the tuning rate, one can calculate the decay length of the electric field in space, indirectly calculating the thickness of the EDL. Following this approach, it is possible to design experimental and theoretical methods to measure the polarization direction of molecules within the EDL.

### 3.2 Semiconductor/liquid interface

The integration of semiconductor physics with quantum mechanics forms a robust theoretical foundation for comprehending electron behavior in solid lattices, notably on the solid side within the EDL.[65] Illustrated by the Kronig-Penney model (**Figure 9A**), a quantum mechanical approach to solving the Schrödinger equation for a one-dimensional periodic potential well, this model yields discrete energy band functions and showcases the quantization of energy levels for electrons in a crystal lattice. In EDL studies, the solid semiconductor characteristics significantly impact the formation and properties of EDL.[66] Quantum mechanical methods can predict and explain EDL features, including charge distribution, potential difference, and the role of interface and surface states.[67] Calculating the Fermi level under specific electrochemical conditions provides insights into the electrochemical stability and reaction kinetics of the EDL. Moreover, interface and surface states, stemming from defects and impurities on the semiconductor surface, have substantial influence on EDL capacitance and conductivity.[68] Quantum mechanical models offer detailed descriptions of the energy distribution and electron occupancy of these states impacting the EDL behavior. Quantum confinement effects, particularly pertinent in nanoscale semiconductor structures, result in band splitting and quantization, affecting electron transport and interfacial reactions within the EDL.[69] Therefore, quantum mechanical descriptions are crucial for capturing the nuanced effects of these phenomena on EDL properties.

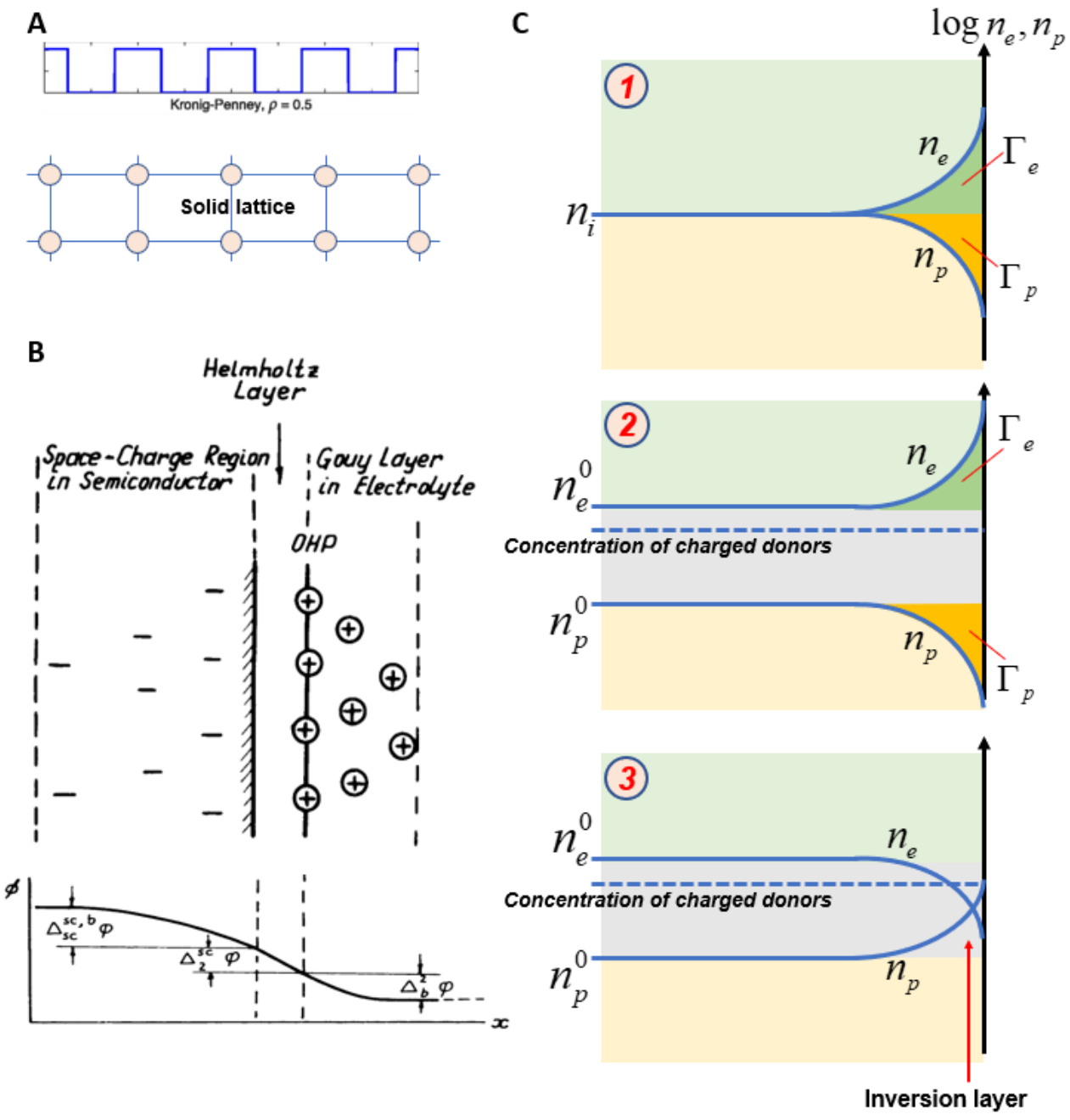

**Figure 9.** (A) The sketch (up) of the periodic potential in Kronig-Penney model[70], and the scheme (down) of solid lattice with the periodic squares; Copyright © 2015 AIP Publishing. (B) Structure of EDL at semiconductor/liquid interface;[71] Copyright © 1980 Springer Science+Business Media New York. (C) Distribution of carrier concentrations at interface: intrinsic semiconductor without doping, extrinsic n-type semiconductor with accumulation layer or inversion layer. **Quantum mechanics, semiconductor physics, and the GCS model are essential for understanding electron behavior in solid lattices and EDLs. These theories predict charge distribution,**

**potential differences, and effects of interface states on EDL properties. Doping and polarization impact conductivity and concentrations, exploring distributions of electrode potential, surface conductivity, and potential drop in semiconductor.**

### *3.2.1 Basic model*

The theory of the EDL at semiconductor solid-liquid interfaces was constructed by Brattain, Garrett, Green, and others based on theoretical and experimental studies of germanium (Ge) electrode. Yu. V. Pleskov[71] has provided a comprehensive review on this topic. In semiconductors, there are two types of charge carriers: holes (positive free charge) and electrons (negative free charge). At the solid-liquid electrolyte interface, dipoles exist, including adsorbed ions and surface groups. Due to the low conductivity of semiconductors, the electric field can penetrate deeply into their interior, necessitating the consideration of a space charge region within the surface layer of the solid side. Concurrently, a diffusion layer as described in the GCS model also exists on the liquid side. The schematic diagram is shown in **Figure 9B**. The electrode is still assumed to be an ideally polarized electrode, meaning no redox reactions have occurred. The distribution of the space charge is described using Poisson's equation:

$$\frac{\mathrm{d}^2\phi(x)}{\mathrm{d}x^2} = -\frac{4\pi}{\varepsilon_{\mathrm{sc}}}\rho(x) \quad (15)$$

where $\phi$ and $\rho$ are the potential and charge density at point $x$, and $\varepsilon_{\mathrm{sc}}$ is the semiconductor dielectric constant. The author further got the expression of charge density:

$$\rho(x) = -\frac{e^2 n_e^0}{kT}\phi(x) \quad (16)$$

where $n_e^0$ is the concentrations of mobile charge in the electrically neutral bulk of semiconductor; $e$, electron charge. This equation was obtained from Boltzmann distribution when the potential differences in the space region is very small compared with $kT/e$. After Combination of these two equations, based on the length of the space charge within the semiconductor and the potential drop, an equation for the electric field strength at the semiconductor surface can be formulated:

$$\xi_{\mathrm{sc}} = -\left.\frac{\mathrm{d}\phi}{\mathrm{d}x}\right|_{x=0} = \frac{\Delta_{\mathrm{sc}}^{\mathrm{sc,b}}\phi}{\kappa_{\mathrm{sc}}^{-1}} \quad (17)$$

where $\xi_{\mathrm{sc}}$ the electric field strength at the semiconductor surface, $\kappa_{\mathrm{sc}}^{-1}$ the Debye length ($\kappa_{\mathrm{sc}}^{-1} = (\varepsilon_{\mathrm{sc}} kT / 4\pi e^2 n_e^0)^{1/2}$), $\Delta_{\mathrm{sc}}^{\mathrm{sc,b}}\phi$ the potential drop in the space charge region in the semiconductor and $\Delta_{\mathrm{sc}}^{\mathrm{sc,b}}\phi = \phi_{\mathrm{b}}^{\mathrm{sc}} - \phi_{x=0}^{\mathrm{sc}}$ ( $\phi_{\mathrm{b}}^{\mathrm{sc}}$ the potential in the bulk of semiconductor, and $\phi_{x=0}^{\mathrm{sc}}$ the potential on the semiconductor surface).

Furthermore, the Helmholtz layer between the diffusion layer in the solution and the space charge in the solid is also considered, which is also known as the "dielectric intermediate layer." The potential drop of the diffusion layer in solution $\Delta_{\mathrm{b}}^{2}\phi$ is:

$$\Delta_{\mathrm{b}}^{2}\phi = \xi_{\mathrm{el}}\kappa^{-1} \quad (18)$$

where $\xi_{\mathrm{el}}$ the electric field strength in Gouy diffusion layer, $\kappa^{-1}$ the Debye length of diffusion layer in solution. The potential drop in the Helmholtz layer can be written as:

$$\Delta_{2}^{\mathrm{sc}}\phi = \xi_{\mathrm{H}}d_{\mathrm{H}} \quad (19)$$

where $\xi_{\mathrm{H}}$ the electric field strength in Gouy diffusion layer, $d_{\mathrm{H}}$ the thickness of Helmholtz layer. Therefore, the sum of the potential drops caused by the three regions $\Delta_{\mathrm{el}}^{\mathrm{sc}}\phi$ is:

$$\Delta_{\mathrm{el}}^{\mathrm{sc}}\phi = \Delta_{\mathrm{sc}}^{\mathrm{sc,b}}\phi + \Delta_{2}^{\mathrm{sc}}\phi + \Delta_{\mathrm{b}}^{2}\phi \quad (20)$$

If the electrolyte is of high concentration, the effect of the solution diffusion layer is not considered. Up to this point, our discussion has only added a space charge layer within the solid inner layer compared to the GCS model. Continue to consider the impact of the concentration of charge carriers (i.e., holes or electrons) within this layer. In semiconductor electrochemistry, the parameter of surface conductivity $K_{\mathrm{s}}$ is adopted:

$$K_{\mathrm{s}} = e(\lambda_e\Gamma_e + \lambda_p\Gamma_p) \quad (21)$$

where $\lambda_e$ and $\lambda_p$ represent electron and hole mobilities; $\Gamma_e$ and $\Gamma_p$ are their surface excesses. Garrett and Brattain[72] The expressions for these two surface excesses are obtained by solving the Poisson's equation:

$$\Gamma_e = -\frac{1}{2}n_e^0\kappa_{\mathrm{sc}}^{-1}\lambda^{-1}e^{-N}\int_o^Y \frac{e^y - 1}{F(y,\lambda,P,N)}dy \quad (22)$$

$$\Gamma_p = -\frac{1}{2}n_p^0\kappa_{\mathrm{sc}}^{-1}\lambda e^{P}\int_o^Y \frac{e^{-y} - 1}{F(y,\lambda,P,N)}dy \quad (23)$$

Among them,

$$F(y,\lambda,P,N) = \mp\left[\lambda e^{P}(e^{-y}-1) + \lambda^{-1}e^{-N}(e^{y}-1) + (\lambda-\lambda^{-1})y\right]^{1/2} \quad (24)$$

where $y = (e/kT)\Delta_{\mathrm{sc}}^{\mathrm{sc,b}}\phi$ ; $\lambda = (n_p^0/n_e^0)^{1/2}$ ; $P = (e/kT)(\varphi_p - \varphi_0)$ ;

$N = (e/kT)(\varphi_n - \varphi_0)$. The $\varphi_p$ and $\varphi_n$ represent the "quasi-Fermi levels" for holes and electrons, respectively. There is $\varphi_p = \varphi_n = \varphi_0$ in the case of thermodynamic equilibrium. $Y$ is the value of $y$ at a dividing surface with the zero surface excess of the certain component. The $n_p^0$ and $n_e^0$ are the concentrations of electron and hole, respectively, in the intrinsic semiconductor. The $\lambda$ is a parameter characterizing the degree of semiconductor doping. If there is no doping, then $n_p^0 = n_e^0 = n_i$.

According to the model described, the distribution of concentrations can be ascertained. **Figure 9C** depicts the distribution of carrier concentrations at the semiconductor interface with and without donor doping. In the absence of doping, the intrinsic semiconductor maintains electrical neutrality, with electron and hole concentrations being equal. During cathodic polarization of the electrode, electrons congregate at the interface, leading to a reduction in hole concentration (**Figure 9C1**). Introducing a significant amount of donor doping elevates the bulk electron concentration, and the scenario during cathodic polarization mirrors that of the undoped scenario (**Figure 9C2**). However, during anodic polarization, an inversion layer may emerge at the interface where the concentration of holes surpasses that of electrons (**Figure 9C3**). These shifts in concentration are induced by changes in the surface excess and consequently modify the surface conductivity. Consequently, it is imperative to delve deeper into the interplay among the electrode potential, surface conductivity, and the potential drop in semiconductor.

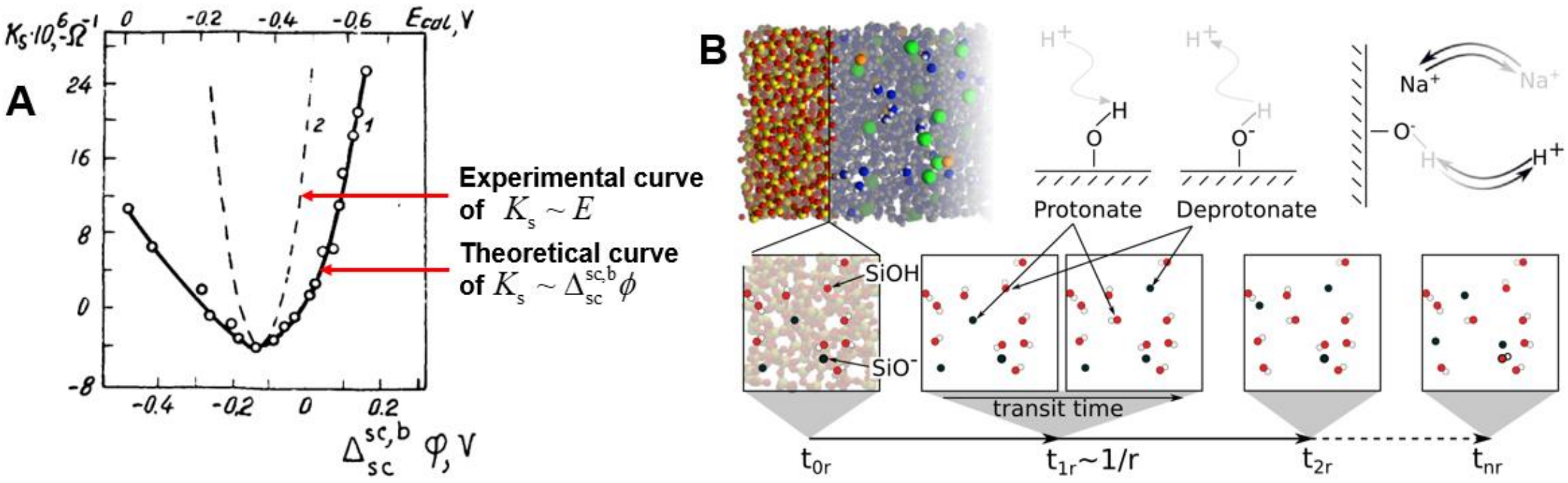

**Figure 10**. (A) Comparison of experimental relationship between surface conductivity and potential, and the theoretical relationship between surface conductivity on germanium and potential drop in semiconductor;[71] Copyright © 1980 Springer Science+Business Media New York. (B) Schematic view of protonation and deprotonation reactions at solid-liquid interfaces, and simulation protocol flow diagram;[73] $t_r$ represents the time interval between consecutive protonolysis reactions, while r denotes the rate of protonolysis. Copyright © 2022 American Physical Society. **Experimental and theoretical curves of the Ge electrode in KBr solution show the former is broader, suggesting a greater impact of surface conductivity on potential change than the potential drop across the Helmholtz layer. Rapid potential changes can prevent Helmholtz layer relaxation. The surface conductivity in EDL remains a hot topic, and proton**

**transfer and diffusion affecting ion behavior at the solid-liquid interface within the EDL were observed.**

When using a high-concentration electrolyte and neglecting the effects of the solution diffusion layer, the change in electrode potential $\Delta E$ can be expressed as:

$$\Delta E = \Delta\Delta_2^{\mathrm{sc}}\phi + \Delta\Delta_{\mathrm{sc}}^{\mathrm{sc,b}}\phi \quad (25)$$

**Figure 10A** exhibits the experimental $K_{\mathrm{s}} \sim E$ curve and theoretical $K_{\mathrm{s}} \sim \Delta_{\mathrm{sc}}^{\mathrm{sc,b}}\phi$ curve of Germanium electrode in KBr solution. The experimental curve is broader than the theoretical curve, indicating that the change in electrode potential caused by surface conductivity, denoted as $\Delta E$, is greater than the change in potential drop, denoted as $\Delta\Delta_{\mathrm{sc}}^{\mathrm{sc,b}}\phi$. Therefore, when $E$ changes, it is necessary to consider the potential drop across the Helmholtz layer. Germanium electrode require several seconds to undergo the relaxation process. Consequently, Brattain and Boddy[74] have experimentally confirmed that rapid changes in potential can prevent the relaxation process of the Helmholtz layer from occurring, at which $\Delta E = \Delta\Delta_{\mathrm{sc}}^{\mathrm{sc,b}}\phi$. The experimental curve and the theoretical curve become equally narrow.

The surface conductivity within the EDL remains a hot topic in recent research. For example, Döpke et al.[73] have observed the impact of proton transfer and diffusion at the solid-liquid interface on the surface conductivity within the EDL. **Figure 10B** illustrates the protonation and deprotonation reactions, as well as how proton dissociation reactions are periodically introduced in molecular dynamics simulations. As the surface proton exchange rate increases, ion adsorption within the EDL diminishes, leading to a reduction in the residence time of ions near the interface. The water structure is affected indirectly, solely through electrostatic coupling with the ions. This indicates that surface reaction kinetics can directly influence the distribution and behavior of ions within the EDL.

### *3.2.2 Applications and Outlooks*

Butler et al.[75] in their comprehensive review of energy material interfaces in various types of devices, have clearly indicated that electrochemical energy storage devices focus on the thermodynamic and kinetic properties at the interface, including the impact of polarization on the electrochemical potential at the electrode-electrolyte interface. Photovoltaic devices, on the other hand, focus on electronic characteristics at the interface, such as bandgap and band alignment. Both the battery and photovoltaic fields have applied different approaches to the same interfacial issues, each yielding rich conclusions. However, there seems to be a barrier between the them (**Figure 11A**), which hinders the mutual application of different methods and conclusions.

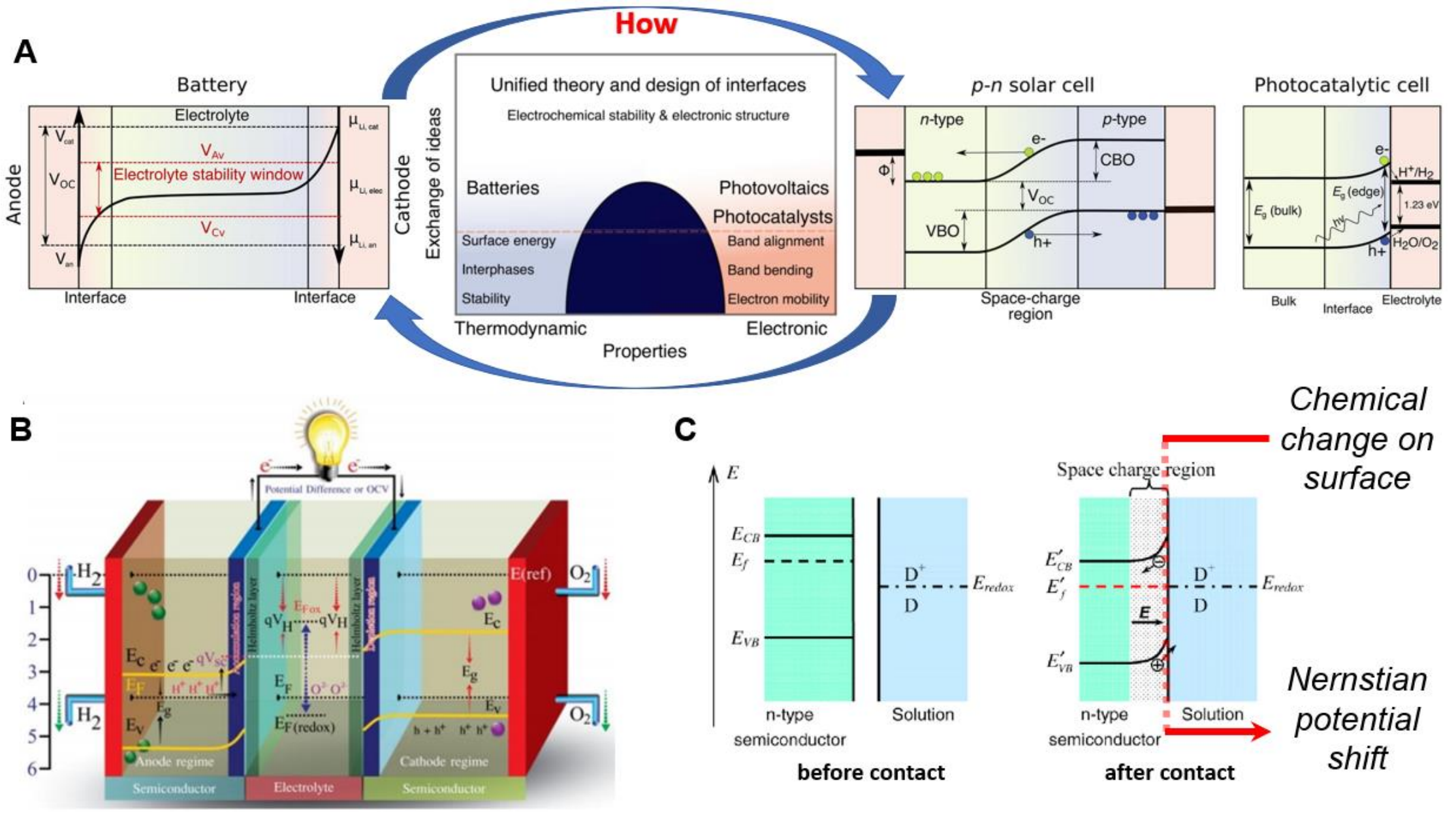

**Figure 11**. (A) The comparison of energy or potential distributions of battery, photovoltaic and photocatalytic devices, and the scheme of a "barrier" that mitigates the exchange of ideas between them;[75] Copyright © 2019 Springer Nature. (B) Schematic of the energy and Fermi levels at semiconductor electrode-electrolyte interface in solid oxide fuel cells;[76] Copyright © 2021 Springer Nature. (C) Scheme of band bending after contact of a semiconductor with a solution containing a molecular redox couple $D^+/D$, and influenced by chemical change. Copyright © 2023 American Chemical Society.[77] **Butler et al. highlighted the importance of interfacial thermodynamic and kinetic properties in energy storage and PV devices. They identified a barrier between these fields that limits the sharing of methods and findings. Electrostatic potential arises from surface dipoles, affecting electron affinity and band structure. Interfacial PCET processes are crucial for understanding proton-coupled electron transfer, necessitating a refined model that considers potential shifts due to chemical changes at the electrode surface.**

In simple terms, semiconductor interfaces focus on electronic energy levels, while electrochemical interfaces are more concerned with the transport and adsorption of ions. The most commonly discussed content in textbooks is the relationship between changes in the electronic energy levels at the electrode and redox reactions.[78] When a voltage is applied causing the electronic energy levels at the electrode to be higher or lower than the matters in the electrolyte, an electron transfer reaction occurs. Although this simple model can be used to explain the standard reduction potential table, it does not yet address the bending of energy bands. In 2019, Zhang et al.[79] have clearly identified LIB electrodes as semiconductors and proposed a semiconductor electrochemical model. The movement of the Fermi level under the influence of an additional electric field and an interface with a high concentration of Li ions was discussed. This model microscopically attributed the electrochemical performance of the battery to the quantization of energy band levels. The perspective based on the semiconductor electrochemical model offers direct strategies for achieving high-performance batteries. These strategies include controlling the applied voltage during battery operation to regulate the Fermi level, increasing the interfacial Li ion concentration, and adjusting the material's energy band structure through doping, etc.

In 2021, Zhu et al.[76] described the semiconductor-electrolyte interface present in SOFCs and the band alignment at this interface due to the equilibration of the Fermi levels. The authors believed that by adjusting the band structure of the semiconductor material, effective control over the transport of ions and electrons can be achieved, thereby enhancing the performance of the fuel cell. The Femi level contains two parts, chemical potential $\mu_e$ and electrostatic potential $e_x$:

$$E_F = \mu_e - e_x \quad (26)$$

The electrostatic potential primarily originated from surface dipoles, which influence the electron affinity and cause the band structure to be associated with the crystal facets. As illustrated in **Figure 11B**, when an electrolyte was interposed between the n-type anode and the p-type cathode, the bands bended at both interfaces to ensure Fermi level alignment. This, in turn, further affected the formation of the Helmholtz layer within the electrolyte and established the voltage across the entire device.

In 2023, James M. Mayer[77] posited that most redox reactions in materials at the interface with protic solutions involve net proton-coupled electron transfer (PCET). It was inappropriate to consider ions or electrons in isolation. From a thermodynamic perspective, the transfer of electrons was typically accompanied by a stoichiometric quantity of protons, akin to the coupled transfer of Li ions in LIBs. PCET induced a Nernstian -59 mV/pH potential shift at the semiconductor/solution interface. Consequently, the traditional model, which only considered the band bending caused by Fermi level equilibration after the semiconductor comes into contact with the solution, should be further refined to account for the potential shifts caused by chemical changes at the electrode surface (**Figure 11C**).

Du et al.[80] described in detail how the onset potentials of OER and ORR change under illumination, thereby affecting the charge and discharge platforms of Li-$O_2$ batteries (**Figure 12A**). The oxidation potential of OER depended on the potential difference between the valence band (VB) of the semiconductor cathode and the $O_2/H_2O$ potential. When light excited electrons to the conduction band (CB), holes were left in the VB, and the electrons traveled through the external circuit to the anode and combined with $Li^+$. Conversely, during discharge, photo-generated electrons were injected into the π2p* anti-bonding orbital of $O_2$, causing its reduction (**Figure 12B**). Thus, illumination resulted in a decrease in OER overpotential and an increase in ORR overpotential, leading to a lower charging voltage platform and a higher discharging voltage platform. This improved battery performance, including energy density and light conversion efficiency.

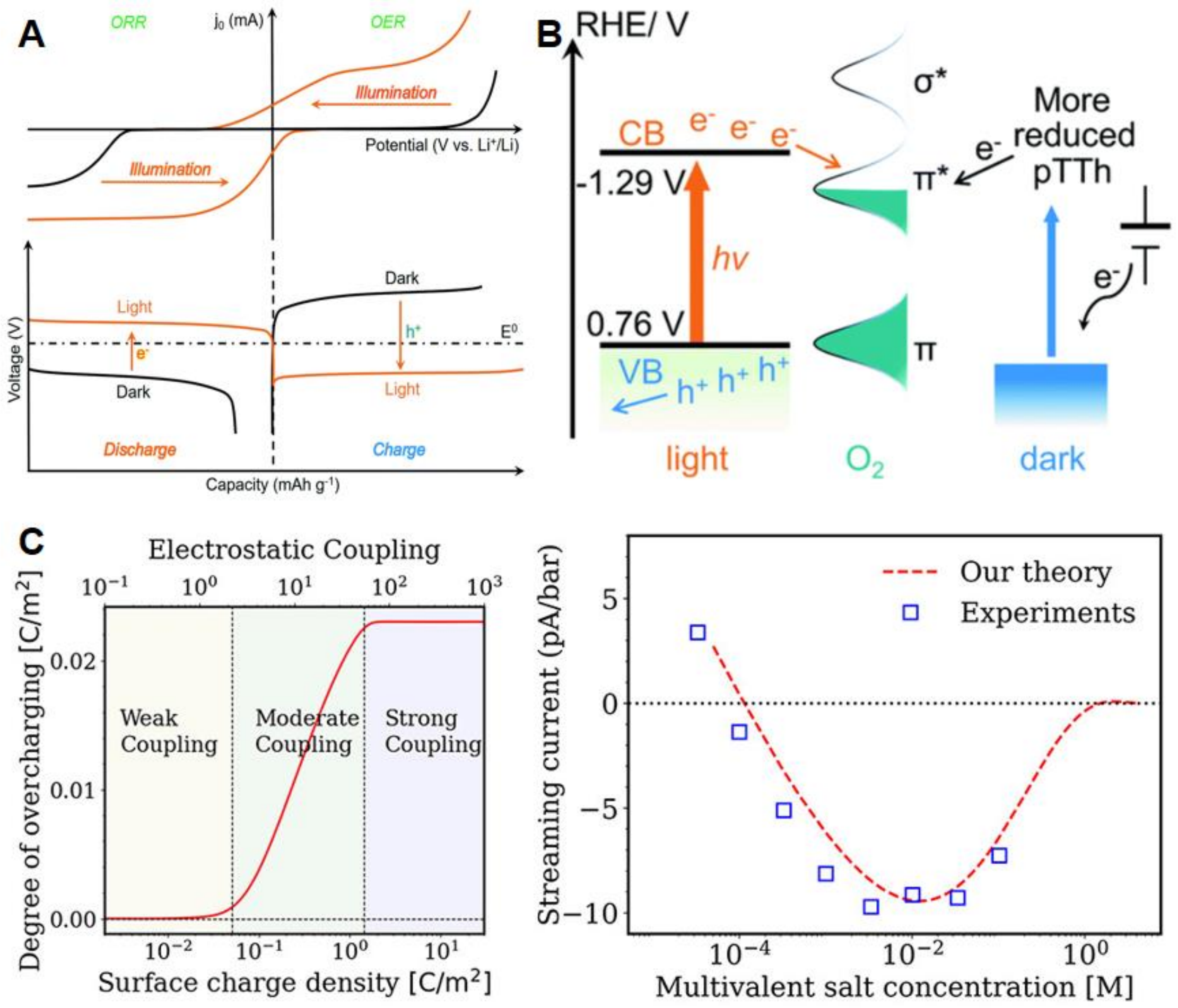

**Figure 12**. (A) Photoinvolved current-potential curves of the ORR/OER and discharge/charge profiles of Li-$O_2$ batteries with and without illumination;[80] Copyright © 2022 The Royal Society of Chemistry. (B) Energy diagram of the ORR on pTTh cathode;[81] Copyright © 2016 Wiley-VCH. (C) Degree of overcharging as a function of surface charge density and electrostatic coupling parameter, and image of charge inversion depending on salt concentration.[82] Copyright © 2023 American Chemical Society. **It was explored that how light affects the onset potentials of OER and ORR, and charge/discharge platforms in Li-$O_2$ batteries. OER oxidation potential depends on the potential difference between the cathode's valence band and the $O_2$/$H_2O$ potential. Illumination reduces OER overpotential and increases ORR overpotential, improving battery performance. Studying the liquid side EDL is challenging due to the complexity of solvated ion microstructure. Agrawal et al. investigated the effects of high counterion adsorption on EDL on liquid side, simultaneously highlighting limitations of classical PB theory. They used a modified Gaussian renormalized fluctuation theory to account for ion correlations and dielectric effects, revealing charge inversion regions and overcharging phenomena.**

It seems more challenging to study the EDL on the liquid side. This is because the microstructure of solvated ions, which are less stable in structure and potential distribution compared to solids, must be considered. However, many studies have aimed to uncover more undisclosed objective facts. For example, in 2024, Agrawal et al.[82] investigated various inversion regions that occur on the liquid side of the EDL under high adsorption of counterions, such as the phenomenon of charge inversion. The limitations of the classical mean-field Poisson-Boltzmann (PB) theory were highlighted, as it does not account for ion correlations, dielectric variation, and the excluded volumes of ions and solvents. The authors adopted a modified Gaussian renormalized fluctuation theory, which allows for the self-consistent consideration of spatially varying ion strength, dielectric constant, and excluded volume effects. If the local concentration of counterions at the charged surface exceeds the amount of surface charge, it can lead to the surface changing from a net negative charge state to a net positive charge state. This excessive accumulation of counterions is also known as overcharging. Surfaces with different charge densities can be divided into three regions, as shown in **Figure 12C**. In regions of low charge density, ion interactions are weak and can be described by the PB equation. However, high charge density attracts a high concentration of counterions, causing overcharging and ion crowding, leading to a nonlinear charge inversion region.

The optimal scenario is to use a single formula to describe both the solid and liquid sides simultaneously. In the field of condensed matter physics, interfaces are viewed as heterogeneous materials and can be described from the perspectives of solid-state physics and thermodynamics. For example, Haymet and Oxtoby[83] derived relationships between the one-body local potential, non-uniform particle density, and free energy by considering the kinetic energy, potential energy, energy due to external potential, and particle density of the substance. If the one-body local potential of a real system is divided into theoretical one-body local potential and effective one-body local potential, then for the liquid structure, the effective one-body local potential can be mathematically expanded to obtain the expression for the thermodynamic potential of the non-uniform system. Furthermore, when dealing with solid-liquid heterogeneous interfaces, if the homogeneous solid is treated as a perturbation to the homogeneous liquid, the relationship between the liquid particle density, solid particle density, and system particle density can be established using Fourier expansion:

$$\rho(\mathbf{r}) = \rho_0\left(1+\eta(\mathbf{r})\right) + \rho_0 \sum_n \mu_n(\mathbf{r}) e^{i\mathbf{k}_n \cdot \mathbf{r}} \quad (27)$$

where $\rho(\mathbf{r})$ is the expectation value of $\rho_m(\mathbf{r})$ (for a system of $N$ and $\mathbf{r}_i$ the position of $i$th particle, $\rho_m(\mathbf{r}) = \sum_{i=1}^{N} \delta(\mathbf{r}-\mathbf{r}_i)_j$ ); $\rho_0$ is the density of the uniform liquid; $\eta(\mathbf{r})$ is the fractional density change on freezing; $\{\mathbf{k}_n\}$ is the set of reciprocal lattice vectors of solid; $\mu(\mathbf{r})$ is the order parameters. The order parameter is the coefficient of the non-uniform density in the Fourier expansion. Since the density expansion of the homogeneous solid phase includes reciprocal lattice vectors, the final effective local potential and thermodynamic potential formulas for the solid-liquid interface are related to the crystal planes of the solid material. Different crystal planes correspond to different particle densities and surface free energies, which in turn affect the energy changes and particle density distribution of the entire interface.

Curtin[84] proposed a method based on density functional theory to study the local particle density distribution at solid-liquid interfaces. In this method, the weighted-density approximation (WDA) is used to approximate the excess Helmholtz free energy of each particle. It should be noticed that WDA is an approach within DFT method used to approximate the exchange-correlation energy functional, which is a key component in DFT calculations. By numerically solving the Euler-Lagrange equation and minimizing the excess Helmholtz free energy under the asymptotic conditions of solid and liquid densities on both sides of the interface, the particle density distribution that results in the lowest energy of the interface system is obtained, as shown in **Figure 13A**. Additionally, this method allows for the calculation of the interfacial free energy, an important physical quantity that describes the energy per unit area of the interface.

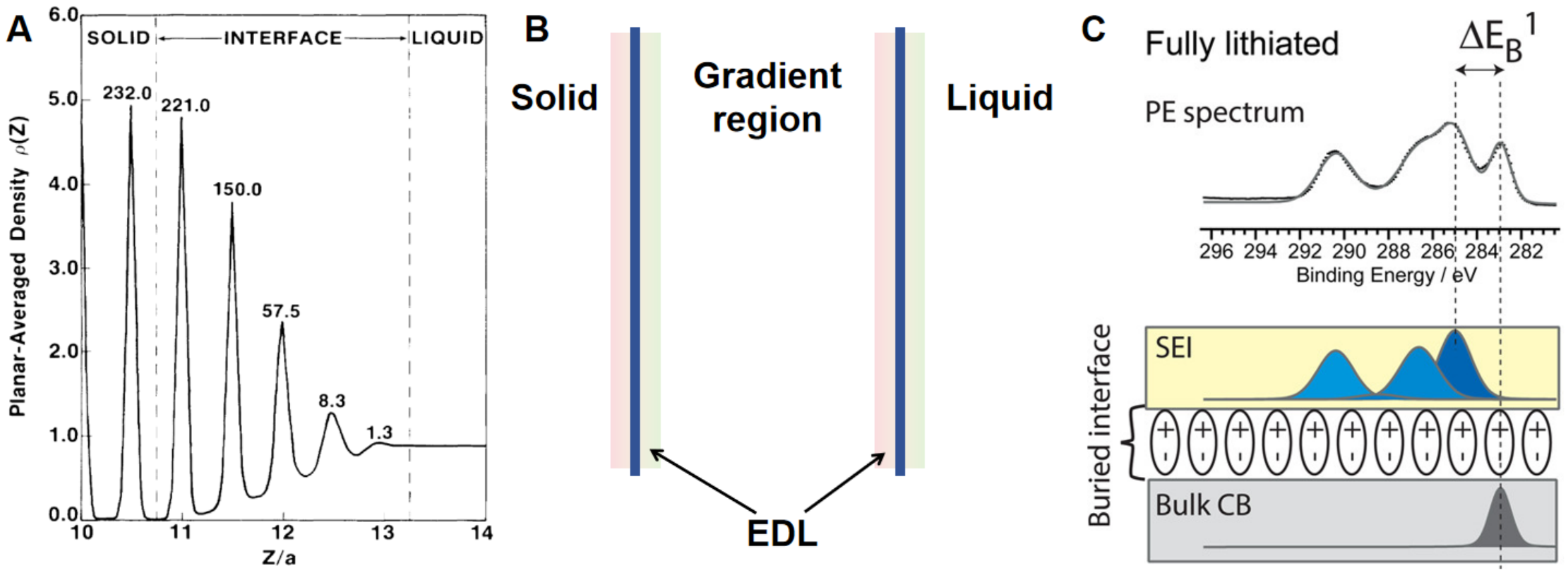

**Figure 13**. (A) Planar-averaged density $\rho$(Z) of the equilibrium solid-liquid interface in Curtin's work;[84] Copyright © 1987 American Physical Society. (B) Scheme of the presence of two EDL at solid-gradient region interface and gradient region-liquid interface, respectively; (C) Scheme of model of fully charged dipole at SEI-anode interface.[85] Copyright © 2016 American Chemical Society. **Curtin's DFT method uses WDA to study particle density distribution at solid-liquid interfaces. It was used to calculate interfacial free energy and provides insights into the lowest energy interface system. The gradient region, a solid-liquid mixed layer, have two distinct solid-solid and solid-liquid EDLs. Maibach et al. discovered a potential gradient at the SEI-bulk electrode interface in LIBs, affecting dipole layer strength and potential drop. Interactions between these two EDLs on the micro-nanoscale are important to deeply understand the electrochemical interface.**

In comparison, although DFT and double layer theory are both used to describe solid-liquid interfaces, they focus on different specific parameters, respectively. DFT primarily involves parameters such as particle density, external potential, and Helmholtz free energy. In contrast, double layer theory focuses on parameters such as the surface charge density of the electrode, electrolyte ion concentration, double layer capacitance, and potential. Additionally, "semiconductor electrochemistry theory" considers parameters such as Fermi level, band structure, band gap, interfacial charge, carrier concentration, and diffusion length. Although QSLT (to avoid confusion, we refer to the solid-liquid interface theory based on DFT as "quantum solid-liquid interface theory (QSLT)") does not directly address electrochemical concepts like ion adsorption, diffusion, or the space charge region in semiconductors, its advantage lies in its ability to describe the solid and liquid on both sides of the interface using a unified formula. The main difference is that double layer or semiconductor electrochemistry theories treat the liquid and solid as two separate parts, whereas QSLT reveals a gradient region from solid to liquid, indicating that the change in particle density is gradual rather than abrupt. In fact, an established fact has proven the necessity of integrating these three theories, namely, the SEI film on the surface of electrode materials in batteries. As clarified in our previous review[86], the SEI on the electrode surface is an interfacial film with a porous outer layer and a dense inner layer. This structure, transitioning from dense to loosely porous, aligns with the conclusions of gradual particle density variation in QSLT. Therefore, what we need to discuss through electrochemical theory is the impact of the gradient length and the rate of change on battery performance.

The existence of the gradient region raises two important questions. One is how to unify the electrons and holes in the semiconductor with the ions in the electrolyte. In fact, they are all charged particles, and by using QSLT, it may be possible to obtain a continuous distribution of charged particles at the interface. Considering the changes in particle density due to the space charge layer at the surface of the semiconductor, a new thermodynamic coupling theory of semiconductor electrochemistry, QSLT, and the double layer can be established. Furthermore, this can extend from thermodynamics to kinetics.

Another key issue of concern in electrochemistry is the exact location of the EDL if a gradient region indeed exists. Strictly speaking, the gradient region is a solid-liquid mixed layer. If the solid-gradient region interface and the gradient region-liquid interface are regarded as two distinct interfaces, then each should have its own EDL. The former would be a solid-solid EDL, and the latter the traditional solid-liquid EDL, as shown in **Figure 13B**. This solid-solid EDL has already been studied; for example, Maibach et al.[85] based on photoelectron spectroscopy (PES) results of different anode materials in LIBs, discovered a potential gradient at the interface between the SEI and the bulk electrode material. An interfacial dipole layer exists between the electrode material and the SEI. During the charge and discharge states of LIBs, the amount of charge accumulated at the interface changes, leading to variations in the strength of the dipole layer and affecting the potential gradient. When the anode material is in a fully lithiated state, a significant amount of charge accumulates at the interface, resulting in a large double-layer potential drop, as shown in **Figure 13C**. Similarly, Ren et al.[87] explored the role of dipole polarization and its contribution to the double dipole layer formation at the interface of $LiNbO_3$@PVDF/Li. The structured alignment of dipoles enhances the local electric field, and strongly influences a uniform $Li^+$ flux for smooth deposition. Furthermore, how the two EDLs, which are close to each other on the micro-nanoscale, interact seems to be an issue that needs to be addressed.

Besides, in previous discussions, the improvement of quantum effects on the macroscopic ion transport equations did not have a significant impact on the results, mainly because quantum effects are more pronounced at the microscopic scale. Whether it is Fick's law or the PNP equations, the focus of thermodynamic analysis is primarily on the ion migration behavior at the liquid-liquid interface. Directly addressing the macroscopic ion migration problem with quantum mechanics seems difficult to yield an ideal result. Furthermore, batteries require a large energy storage density, so the number of ions is enormous for a microscopic system, and the curves obtained from experiments are often the sum of the entire system, where the treatment of the microscopic system does not significantly affect the curves obtained by the macroscopic system.

The same principle applies to the EDL theory, which is based on the macroscopic Poisson-Boltzmann particle distribution equation. Moreover, in actual systems, even for oxide particle powder with a relatively small amount of pore sturcture, the specific surface area of the material is often greater than 10 $m^2$ $g^{-1}$. Undoubtedly, the EDL is also a macroscopic system. Therefore, how to use microscopic theories to study this macroscopic system is a topic worth discussing.

### 3.3 *Semiconductor/semiconductor interfaces*

Except for the most important semiconductor/liquid interface, there are semiconductor/semiconductor interface, semiconductor/conductor interface and conductor/liquid interface. This section will discuss the theories and applications of the semiconductor/semiconductor interface in more detail.

#### *3.3.1 Basic model*

Among various interfaces, semiconductor/semiconductor interface may be more complex, especially for this kind of interface in electrochemical batteries, such as solid-state batteries.[88] In a common condition, the electronic structure at the interface has been made clear, and be usually called junction.[89] The semiconductor/semiconductor junctions have been also classified into two types, homojunctions and heterojunctions. The homojunction means the interface contacting the two same semiconductors with different dopants. The heterojunction means the interface contacting two different semiconductor materials. As shown in **Figure 14A and 14B**, the former has a matched forbidden gap, and meanwhile, the latter has two forbidden gaps.[90] Theories involving junctions have been more extensively described in the field of solar cells, and will not be elaborated on here.

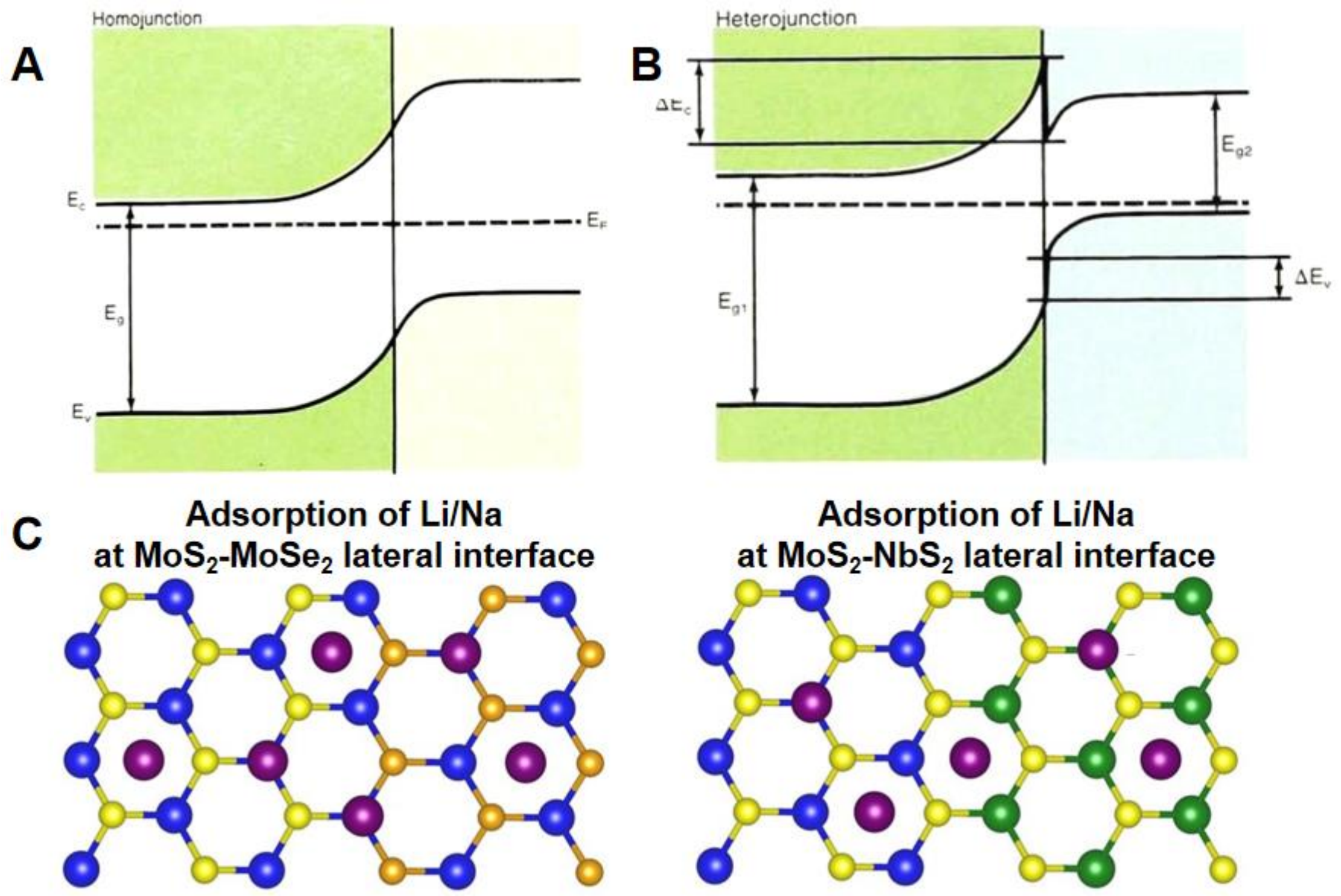

**Figure 14**. Schemes of (A) homojunction and (B) heterojunction ($E_v$, valence band edge; $E_c$, conduction band edge; $E_g$, forbidden gap);[90] Copyright © 1987 AIP Publishing. (C) Various Li/Na adsorption positions at $MoS_2$-$MoSe_2$ and $MoS_2$-$MoSe_2$ interfaces.[91] Copyright © 2022 Elsevier B.V. **Semiconductor/semiconductor interfaces in electrochemical batteries, like solid-state batteries, can be complicated. These interfaces are classified into homojunctions (same semiconductors with different dopants) and heterojunctions (different semiconductor materials). Heterojunctions have two forbidden gaps, while homojunctions have a matched gap. The polarized solid phase interface, often referred to as an EDL, is formed when two solid materials contact. Adhesional bonding occurs due to EDL and molecular forces. DFT calculations suggested that 2D $MoS_2$-$MoSe_2$ and $MoS_2$-$NbS_2$ lateral heterostructures are promising anode materials for Li-/Na-ion batteries, and meanwhile the $MoS_2$-$NbS_2$ structure is particularly advantageous due to its higher adsorption energy and stability. The $MoS_2$-$MoSe_2$ structure undergoes an electronic phase transition during lithiation/sodiation for improving conductivity.**

### *3.3.2 Applications and Outlooks*

This kind of polarized solid phase interface was also called EDL.[92] When two solid materials contact, the adhesional bonding will be formed due to the EDL and the action of molecular forces. Barika and Pal[91] investigated the potential of two-dimensional (2D) $MoS_2$-$MoSe_2$ and $MoS_2$-$NbS_2$ lateral heterostructures (**Figure 14C**) as anode materials for LIBs and sodium-ion batteries (SIBs) using first-principles DFT calculations. These heterostructures not only offered high theoretical specific capacities (518 mAh $g^{-1}$ and 677 mAh $g^{-1}$, respectively) but also exhibited good rate performance and ultra-fast ion diffusion capabilities. In particular, the $MoS_2$-$NbS_2$ heterostructure was considered a superior anode material choice due to its higher adsorption energy and thermodynamic stability. Furthermore, the $MoS_2$-$MoSe_2$ heterostructure undergone an electronic phase transition from semiconductor to metal upon lithiation/sodiation, thereby enhancing electronic conductivity.

Fan et al.[93] conducted a systematic investigation of the electrochemical properties of a series of $MX_2$ (M=Mo, W, Nb, Ta; X=S, Se) during Li/Na intercalation using first-principles calculations. They found that stacking different $MX_2$ with distinct characteristics could form TMDs heterostructures, offering another strategy for tuning electrochemical properties, including voltage, capacity, and electron conductivity. The lattice constants, voltage ranges, and capacities of the semiconductor-semiconductor heterostructure $MoS_2$-$WS_2$ fall between those of their individual components, indicating that heterostructures can integrate the advantages of both materials. Therefore, forming heterostructures by combining $MoS_2$ with higher theoretical capacity and $WS_2$ with lower voltage and minimum Li ion migration barrier can enhance electrode capacity while maintaining low voltage and high Li ion diffusion.

Swift and Qi[94] predicted the potential distribution at the interface in an all-solid-state battery based on a first-principles model. They first obtained the expression for the electrochemical potential of Li ion $\tilde{\mu}_{Li^+}$ through the following system of equations:

$\tilde{\mu}_{Li^+} = \mu_{Li} - \tilde{\mu}_{e^-}$;

$\tilde{\mu}_{e^-} = \mu_{e^-} - e\phi$;

$\mu_{e^-} = -\psi + \psi^a$;

$\psi = IP - E_F$. (28)

then, the electrochemical potential of Li ion is:

$$\tilde{\mu}_{Li^+} = \mu_{Li} - (E_F - IP + \psi^a - e\phi) \quad (29)$$

where $\mu_{Li}$, chemical potential of lithium ion; $IP$, ionization potential; $\psi$, work function (superscript $a$ indicates the anode); $\phi$, electrostatic (Galvani) potential; $E_F$, Fermi energy level; $\tilde{\mu}_{e^-}$ and $\mu_{e^-}$, electrochemical potential and chemical potential of electron, respectively.

Furthermore, the Fermi level, work function, ionization potential, and other parameters were determined through DFT calculations. Based on this model, the authors illustrated the band structure and charge transfer at various interfaces within the Li|LiPON|$Li_xCoO_2$ system (lithium|lithium phosphorus oxynitride|lithium cobalt oxide), as depicted in **Figure 15A**. During lithiation of the positive electrode, Li ions migrate from $LiCoO_2$ to LiPON, leading to the accumulation of positive charges on the $LiCoO_2$ side, thereby increasing the potential and impeding further Li ion transport.

Conversely, during delithiation, Li ions migrate back from LiPON to $Li_{0.5}CoO_2$, reducing the accumulation of positive charges on the $Li_{0.5}CoO_2$ side. Consequently, the EDL formed at the interface serves as a thermodynamic driving force for the transfer of Li ions and electrons.

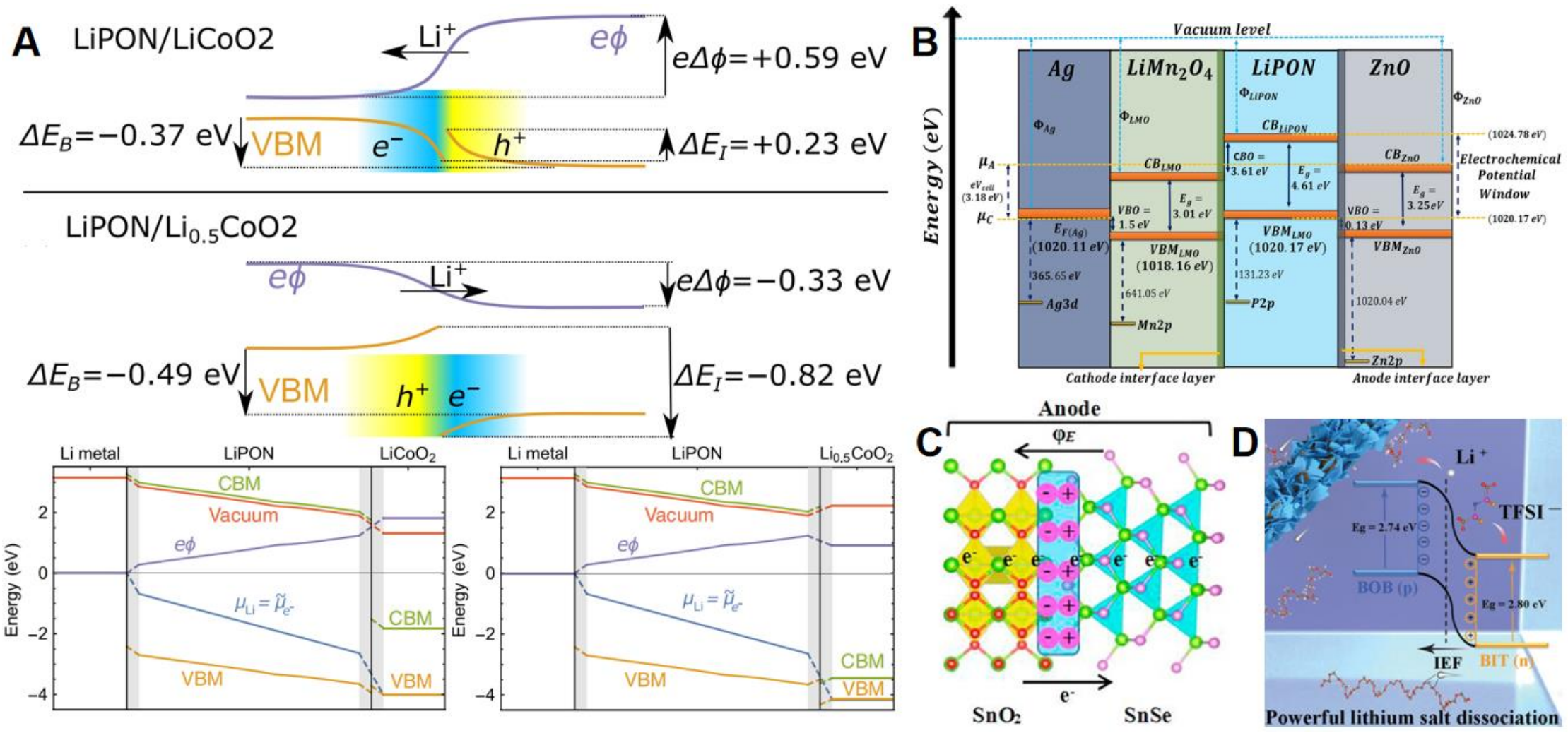

**Figure 15**. (A) Schematic of the electrostatic potential and valence bands at the interfaces between LiPON and $LiCoO_2$ or $Li_{0.5}CoO_2$;[94] Copyright © 2019 American Physical Society. (B) Schematic representation of the electronic energy band diagram for a solid-state Li-ion battery with $Ag/LiMn_2O_4$/LiPON/ZnO structure;[95] Copyright © 2024 Springer Nature. (C) Scheme of semiconductor heterojunction in $SnO_2$/SnSe anode;[96] Copyright © 2018 Elsevier B.V. (D) Schematic diagram of the mechanism for promoting Li salt by dissociator.[97] Copyright © 2023 Wiley-VCH. **DFT calculations determined key parameters for the Li|LiPON|$Li_xCoO_2$ system, illustrating band structure and carriers transfer at interfaces. Lithiation and delithiation processes alter potential and Li ion transport. The EDL at the interface acts as a thermodynamic driving force for Li ion and electron transfer. Martinez et al. designed solid-state batteries with $Ag/LiMn_2O_4$/LiPON/ZnO and reconstructed the energy band diagram. Band misalignment at interfaces can lead to SEI layer formation and affect battery performance. Precise control of potentials enables batteries to possess long lifespan. Additionally, semiconductor heterojunctions in composite electrodes provide built-in potentials and electric fields for charge and discharge processes to improve electrochemical performances.**

Martinez et al.[95] designed solid-state batteries with an $Ag/LiMn_2O_4$/LiPON/ZnO structure and utilized the Kraut method to reconstruct the energy band diagram, focusing on the formation of the EDL at solid-state interfaces and its impact on battery performance. **Figure 15B** illustrates the energy band structures of different material layers in the solid-state battery. The electrochemical window of the solid-state battery is determined by the difference between the bottom of the conduction band and the top of the valence band of the solid LiPON electrolyte. Both the $LiMn_2O_4$/LiPON and LiPON/ZnO heterojunctions exhibit staggered gap structures, characterized by discontinuous band edges and the presence of band gaps. This band misalignment may result in the formation of a SEI layer at the interface. At the $LiMn_2O_4$/LiPON interface, the formation of an electron-insulating SEI layer prevents electrolyte decomposition, contributing to battery stability. However, at the LiPON/ZnO interface, an electron-conducting SEI layer is formed, potentially leading to continuous degradation of the solid electrolyte. Therefore, by precisely controlling the electrode's electrochemical potential and the electrolyte's potential window, solid-state batteries with excellent performance and long lifespan can be designed.

Additionally, semiconductor heterojunctions within composite electrodes can also enhance battery performance. For example, the built-in potential in the $SnSe/SnO_2$ heterostructure provides an additional driving force for electrons and Li ions during the charge and discharge processes (**Figure 15C**).[96] Furthermore, the built-in electric field in one-dimensional ferroelectric ceramic-based heterojunction nanofibers can act as an "acceleration zone", facilitating the rapid transport of lithium ions. This field also assists in the dissociation of Li salts by forming paired electric dipole layers, referred to as "dissociators" (**Figure 15D**).[97]

Semiconductor/semiconductor interfaces are more commonly applied in solar cells because they can easily form p-n heterojunctions and built-in electric fields. These interfaces are fundamental for creating energy conversion devices.[98] Although there is relatively limited literature on the application of semiconductor/semiconductor interfaces in traditional energy storage battery materials, it is foreseeable that with the large-scale application of solid-state batteries in the future, semiconductor/semiconductor interfaces will receive increased attention.

In 2023, Kanno[99] reviewed the development of all-solid-state batteries, including prospects for the application of semiconductor/semiconductor interfaces. The author explicitly noted that intercalation electrodes in Li batteries are typically semiconductors, and within solid-state batteries, the electrochemical reactions at the EDL are described as reactions occurring in the so-called space-charge or depletion regions. Therefore, the study of the EDL at semiconductor/semiconductor interfaces in solid-state batteries is crucial for their development.

Kanno’s group[100] studied the changes in the electronic band structure of all-solid-state thin-film batteries during charge and discharge processes using operando hard X-ray photoelectron spectroscopy (HAXPES). They identified the band edges and Fermi levels of each component in the battery, including the aluminum current collector, $Li_2MnO_3$ cathode, LASGTP solid electrolyte, $Li_3PO_4$ buffer layer, and lithium anode. They also observed band bending at the cathode ($Al/Li_2MnO_3$/LASGTP) and anode (LASGTP/$Li_3PO_4$/Li) interfaces, as shown in **Figure 16A**. During charging, the cathode material $Li_2MnO_3$ transitions from an n-type to a p-type semiconductor, with the Fermi level dropping relative to the valence band maximum (VBM), leading to increased band bending at the interface with the aluminum current collector. During discharging, the Fermi level rises, and the electronic structure reverses. Both interfaces within the current collector-$Li_2MnO_3$-LASGTP structure undergo changes in the direction and degree of band bending, as illustrated in **Figure 16B**.

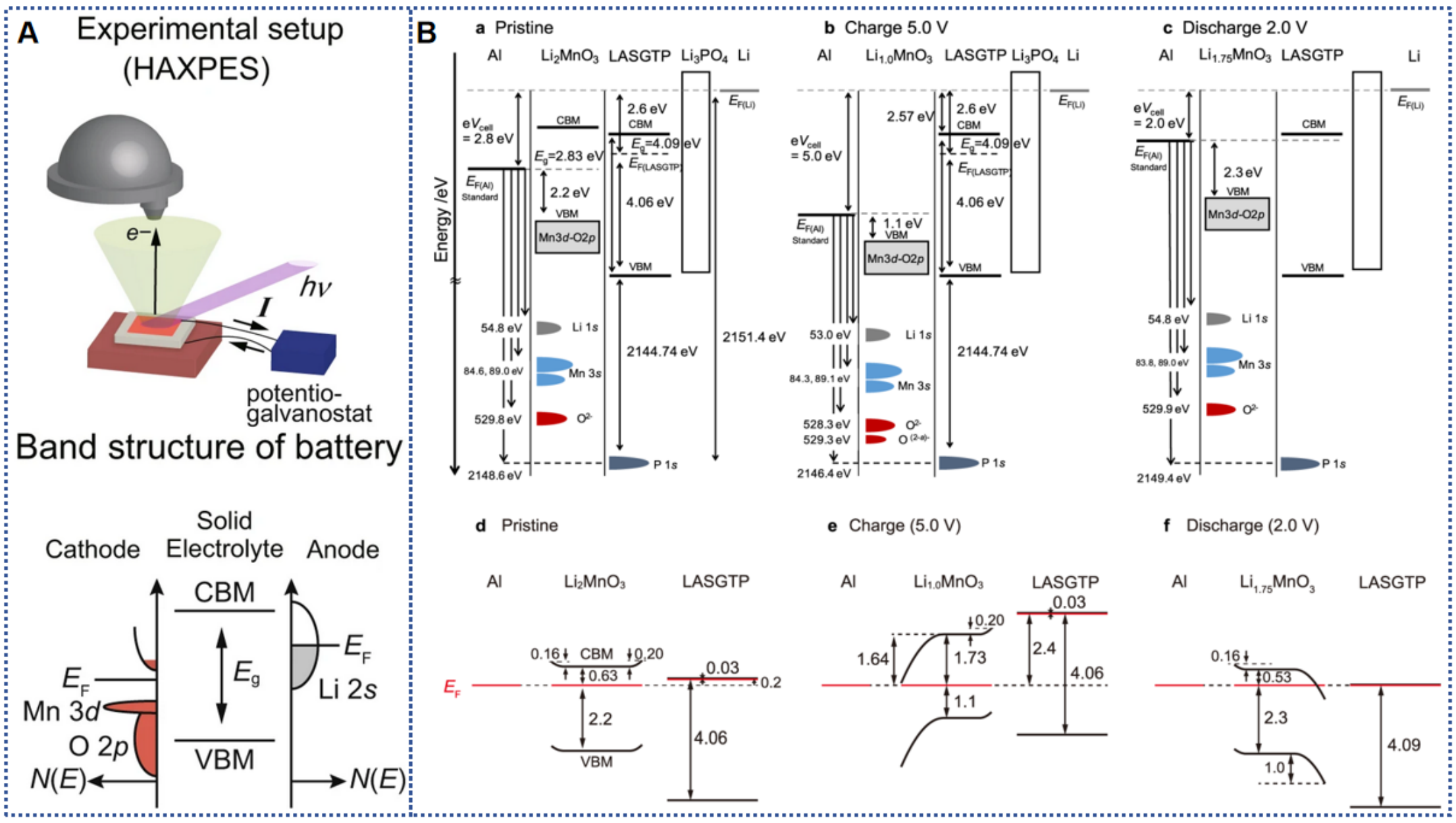

**Figure 16**. (A) Scheme of experimental setup of *Operando* HAXPES technique, and band structure of battery;[100] Copyright © 2022 Springer Nature Limited. (B) Band diagrams of the battery during cycling (CBM, conduction band minimum; VBM, valence band maximum; $E_F$, Fermi level; $E_g$, band gap).[100] Copyright © 2022 Springer Nature Limited. ***Operando* HAXPES was used to study the electronic band structure changes in all-solid-state thin-film batteries during charge and discharge. Band edges and Fermi levels were identified for each component. Band bending occurred at the interfaces on cathode and anode. During charging, $Li_2MnO_3$ transformed from n-type to p-type with dropping of Fermi level and more bending band. Both interface structures within the current collector-$Li_2MnO_3$-LASGTP underwent band bending direction change.**

Therefore, based on the above analysis, it is necessary to establish a comprehensive semiconductor electrochemical theory to describe the EDL at semiconductor/semiconductor interfaces in all-solid-state batteries. The formation and stability of EDL in solid-state batteries significantly influence the electrochemical performances, interfacial resistance, and overall electrochemical kinetics, with implications for optimizing design and enhancing safety in next-generation energy storage systems.[101]

## 3.4 Conductor/semiconductor interfaces

### *3.4.1 Basic model*

Similar to the semiconductor/semiconductor interface, if the metal contact to the semiconductor (electrically connected), the Fermi energy will reach equilibrium, that is, the chemical potential of the electrons has the same level in the two materials.[102] Sparnaay[103] have claimed that the voltage is dependent on the difference in work functions (or contact potential difference Δc.p.) between the metal $W_m$ and the semiconductor $W_{se}$, as show in **Figure 17A**. The work function is the minimum energy required for an electron to escape from the interior of a metal or semiconductor into a vacuum $E_{vac}$. For semiconductors, this is related to the semiconductor's Fermi level $W_{se} = E_{vac} - E_F$.

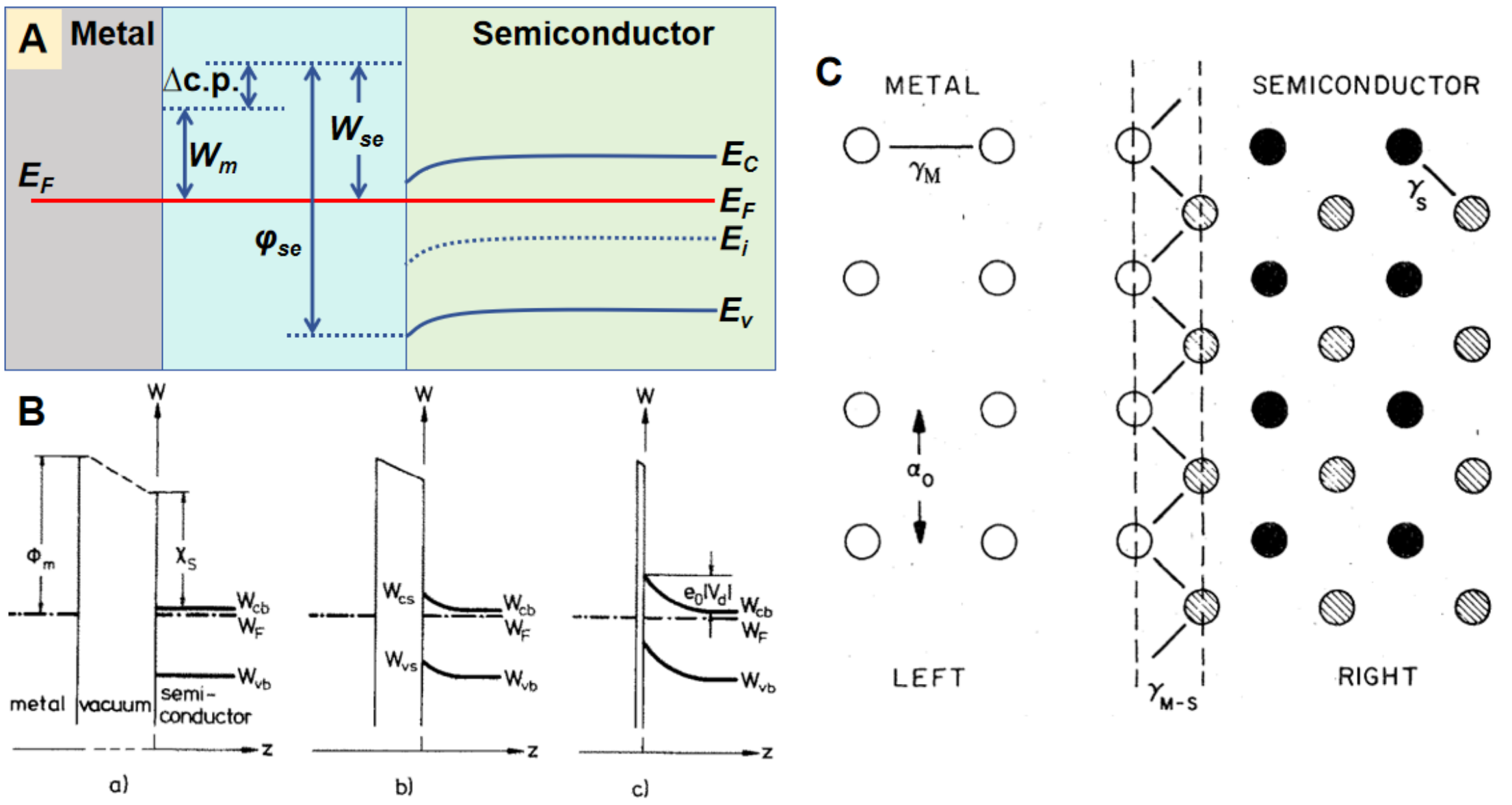

**Figure 17**. (A) Potentials diagrams of metal-semiconductor interface;[103] Copyright © 1967 Elsevier B.V. (B) Development of a Schottky barrier with gradually change in metal-semiconductor distance;[104] Copyright © 1994 Elsevier B.V. (C) Structure of interface between cubic metal and cesium chloride semiconductor.[105] Copyright © 1978 American Physical Society. **Similar to semiconductor/semiconductor interfaces, when a metal contacts a semiconductor, Fermi energy reaches equilibrium due to equal chemical potential of electrons in both materials. Sparnaay noted that voltage depends on the difference in work functions between the metal and semiconductor. Mönch discussed the formation of Schottky barriers at semiconductor/metal interfaces, explaining how does the surface charge accumulation form a charged interface and EDL. The interfacial electric field penetrates the semiconductor to create a space-charge layer but it will do not occur in the metal. The Dyson equation containing Green's functions can be used to obtain the distribution of interface state density and other important parameters.**

Therefore, the mentioned parameters can be used to calculate the escape condition of electron. The energy threshold of electron excitation $\varphi_{se}$ is:

$$\varphi_{se} = W_{se} - E_F + E_V \quad (30)$$

where $E_V$ represents energy of the edge of the valence band. $\varphi_{se}$ is the energy required for an electron to be excited from the valence band to the conduction band.

This provides a fundamental model for calculating various potential energies at the metal/semiconductor interface.

Additionally, McCaldin and McGill[106] mentioned that in honor of the significant contributions of scientists Walter Schottky and Nevill Francis Mott to this field, the potential structure at the metal-semiconductor interface can be referred to as the Schottky-Mott or Schottky barrier. According to Schottky's model, when a metal comes into contact with an n-type semiconductor, electrons flow from the semiconductor to the metal until a stable potential difference is established at the interface. This potential difference results in the formation of a depletion region in the semiconductor, leading to the creation of the Schottky barrier. Nevill Mott later extended and refined Schottky's model, providing a more comprehensive understanding of the electronic structure and barrier formation mechanism at metal/semiconductor contacts, particularly in terms of semiconductor surface states and band bending.

Mönch[104] discussed the formation process of the Schottky barrier and its impact on current transmission. When the contact is created by gradually decreasing the distance between a semiconductor and a metal, an interface with bending of the bands can be formed, as shown in **Figure 17B**. Initially, there is an electric field in the vacuum gap due to the difference in work functions of the two materials. The accumulation of surface charge forms a charged interface of opposite polarity, similar to an EDL structure. The interfacial electric field can penetrate into the surface of the semiconductor but cannot significantly penetrate the metal. Thus, an extended space-charge layer can form on the semiconductor surface. If the surface contains positively charged donors, an upward bending of the bands forms a Schottky barrier, which is due to the migration of mobile electrons.

In the field of condensed matter physics, Lowy and Madhukar[105] considered specific material structures. They used nearest-neighbor tight-binding models to describe metals and semiconductors, obtaining closed-form analytical results for the surface (i.e., solid-vacuum interface) Green's functions. The bulk Green's function is defined as:

$$G_B(E,\vec{k}) = \left[E - E(\vec{k})\right]^{-1} \quad (31)$$

where $\vec{k}$, three-dimensional wave vector; $E(\vec{k})$, the energy of band. In order to treat surface, the 3D wave vector was transformed to the wave vectors parallel to the surface ($k_x$ and $k_y$), and the site component normal to the surface ($k_z$). Then, for a cubic lattice in metal side, the band energy is:

$$E(\vec{k}) = E_0 - 2\gamma_M \left(\cos k_x a + \cos k_y a + \cos k_z a\right) \quad (32)$$

where $\gamma_M$, the nearest-neighbor tight-binding integral; $a$, the lattice constant; $E_0$, the energy of the middle of the band. For the cesium-chloride (CsCl) semiconductor:

$$E_j(\vec{k}) = \frac{1}{2}(\Theta_A + \Theta_B) \pm \frac{1}{2}\left[(\Theta_A - \Theta_B)^2 + 256\gamma_S^2 f_1^2(\vec{\varphi})\right] \quad (33)$$

where $\Theta_A$ and $\Theta_B$, the diagonal energies of atoms A and B, respectively; $\gamma_S$, the nearest-neighbor tight-binding integral referred to bonding strength in CsCl; $f_1(\vec{\varphi})$, the function related to wave vectors. Briefly, the Eqs.(30)-(33) are closely connected through the analysis of electronic excitation and band structure at the metal/semiconductor interface. Eq.(30) defines the energy threshold for electron excitation, which depends on the Fermi level, valence band edge, and work function. To understand these energy levels microscopically, Eq.(31) introduces the Green's function formalism, which is based on the band energy of the materials. The specific form of the band energy in a cubic metal lattice is given by Eq.(32), while Eq.(33) extends the band structure analysis to a CsCl-type semiconductor. Together, these equations provide a consistent framework for linking macroscopic energy parameters to microscopic electronic structures.

According to the two sets of equations above, the Green's function of a certain crystal plane can be constructed. Subsequently, the author can represent individual crystal planes of two crystals on the interface (**Figure 17C**) as a system of equations composed of four Green's functions. By further utilizing the Dyson equation containing Green's functions, the distribution of important parameters such as interface state density can be obtained.

### *3.4.2 Applications and Outlooks*

The relative more widespread interface in the field of batteries is at interlayer of conductor/semiconductor, or metal/semiconductor. This kind of interface is more easer to tune the electronic structure compared to the semiconductor/semiconductor interface, and can be used in rechargeable Zn-air batteries.[107] For example, Li et al.[108] investigated the in situ construction of the cobalt/cobalt selenide (Co/CoSe) Schottky heterojunction and explored the role of the electronic structure at the heterojunction interface in the bifunctional oxygen electrocatalysis of zinc-air batteries. As shown in **Figure 18A**, prior to contact, metallic Co and CoSe each possess their own Fermi levels and band structures. Upon close contact between Co and CoSe, due to the difference in Fermi levels, electrons flow from Co, where the Fermi level is higher, to CoSe, where it is lower, until equilibrium between the two is achieved. Concurrently, electrons from the valence band of CoSe flow into Co, causing the bands of CoSe to bend downwards at the interface. An electron-rich region forms on the surface of CoSe, while a hole-rich region forms on the surface of Co, thereby increasing the reactive sites for the OER and ORR reactions.

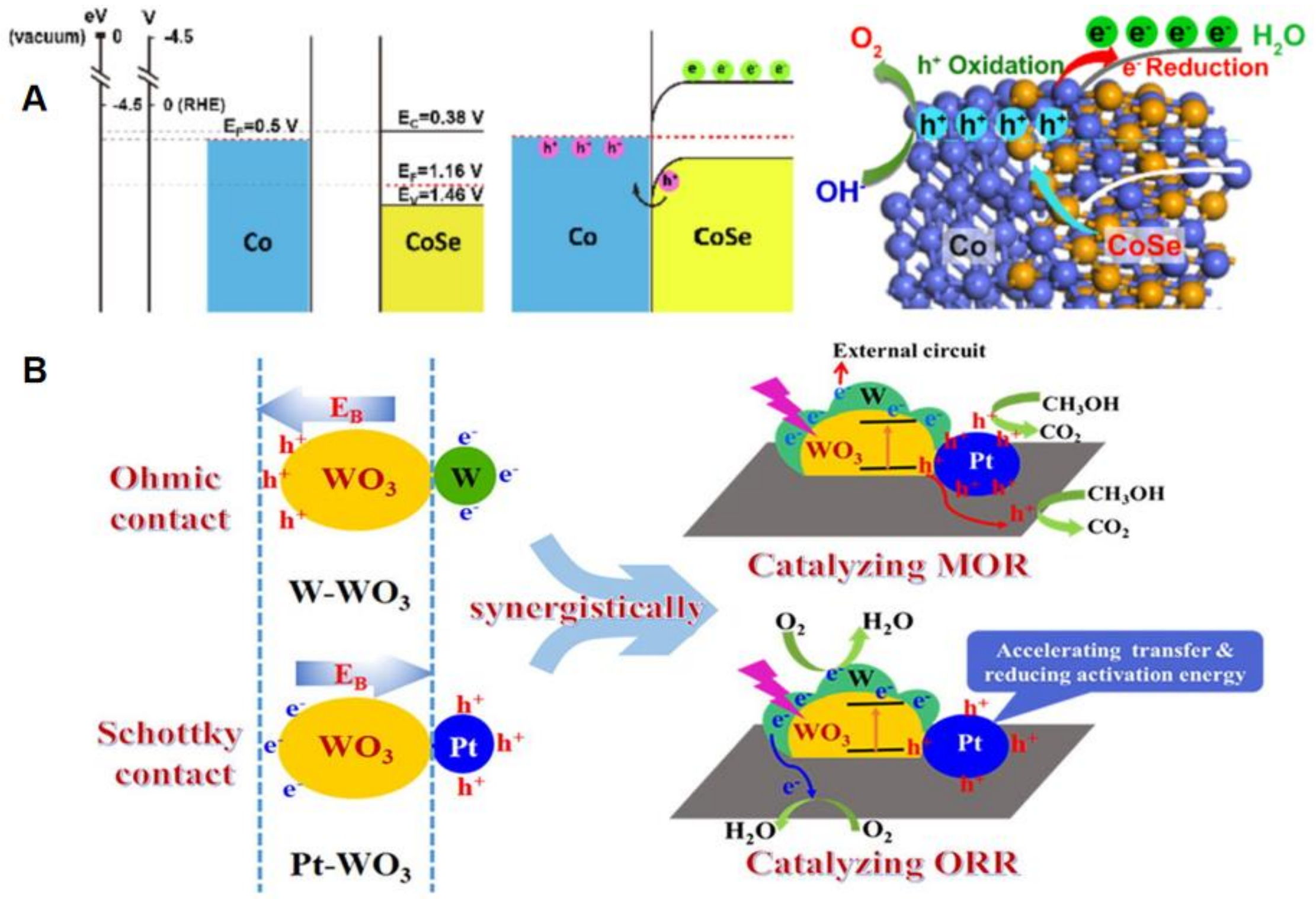

**Figure 18**. (A) The band diagrams of Co/CoSe before/after the Schottky contact, and OER mechanism model for Co/CoSe heterojunction;[108] Copyright © 2022 Elsevier B.V. (B) Scheme of electro-photo catalytic mechanism on Pt-$WO_3$@W composite.[109] Copyright © 2021 Elsevier B.V.

**In batteries, the more common interfaces are conductor/semiconductor or metal/semiconductor, which are easier to be tuned than semiconductor/semiconductor interfaces. Li et al. studied the in situ construction of Co/CoSe Schottky heterojunctions for bifunctional oxygen electrocatalysis in Zn-air batteries. Upon contact, a region rich in electrons on CoSe and another region rich in holes on Co will be formed because electrons flow from Co to CoSe, which increases reactive sites for oxygen reactions. Pt-$WO_3$@W composite materials were also developed for bifunctional electrocatalyst in direct methanol fuel cells. The work function difference between Pt and $WO_3$ causes band bending, facilitating electron transfer in Schottky contact.**

Li et al.[109] constructed a metal/semiconductor oxide heterostructure in the Pt-$WO_3$@W composite material, making it applicable as a bifunctional electrocatalyst for methanol oxidation reaction (MOR) and ORR in direct methanol fuel cells (**Figure 18B**). In ohmic W-$WO_3$ contacts, photogenerated electrons are captured by the W surface, while in Schottky Pt-$WO_3$ contacts, photogenerated holes are captured by the Pt surface; the synergistic effect of these two enhances the catalytic performance. At the Pt-$WO_3$ interface, due to the work function of $WO_3$ being lower than that of Pt, the bands bend upwards, and electrons spontaneously transfer from the semiconductor ($WO_3$) to the metal (Pt). At this time, the electric field direction of the interface EDL is opposite to the direction of band bending, facilitating the transfer of electrons from the semiconductor to the metal; whereas in ohmic contacts, the electric field direction of the EDL is consistent with the direction of band bending, facilitating the transfer of electrons from the metal to the semiconductor.

Liu et al.[110] proposed a multifunctional heterostructure of vanadium oxide and reduced graphene oxide ($VO_2$@rGO), which exhibit semiconductor behavior and metallic properties, respectively. They enhanced the sulfur conversion kinetics in Li-S batteries through a bidirectional Mott−Schottky electrocatalyst enhanced by the interface built-in electric field (BIEF). When the two materials are composited, a difference in work function leads to the flow of electrons from $VO_2$ to rGO upon contact to achieve Fermi level equilibration, forming a space charge region and BIEF. DFT calculations demonstrate the redistribution of electrons at the interface and the plane-average electron difference, as shown in **Figure 19A**. The BIEF improves the chemical adsorption capacity of lithium polysulfides (LiPS), reduces the shuttling effect of LiPS in the battery, and enhances the sulfur utilization rate and the cycling stability of the battery.

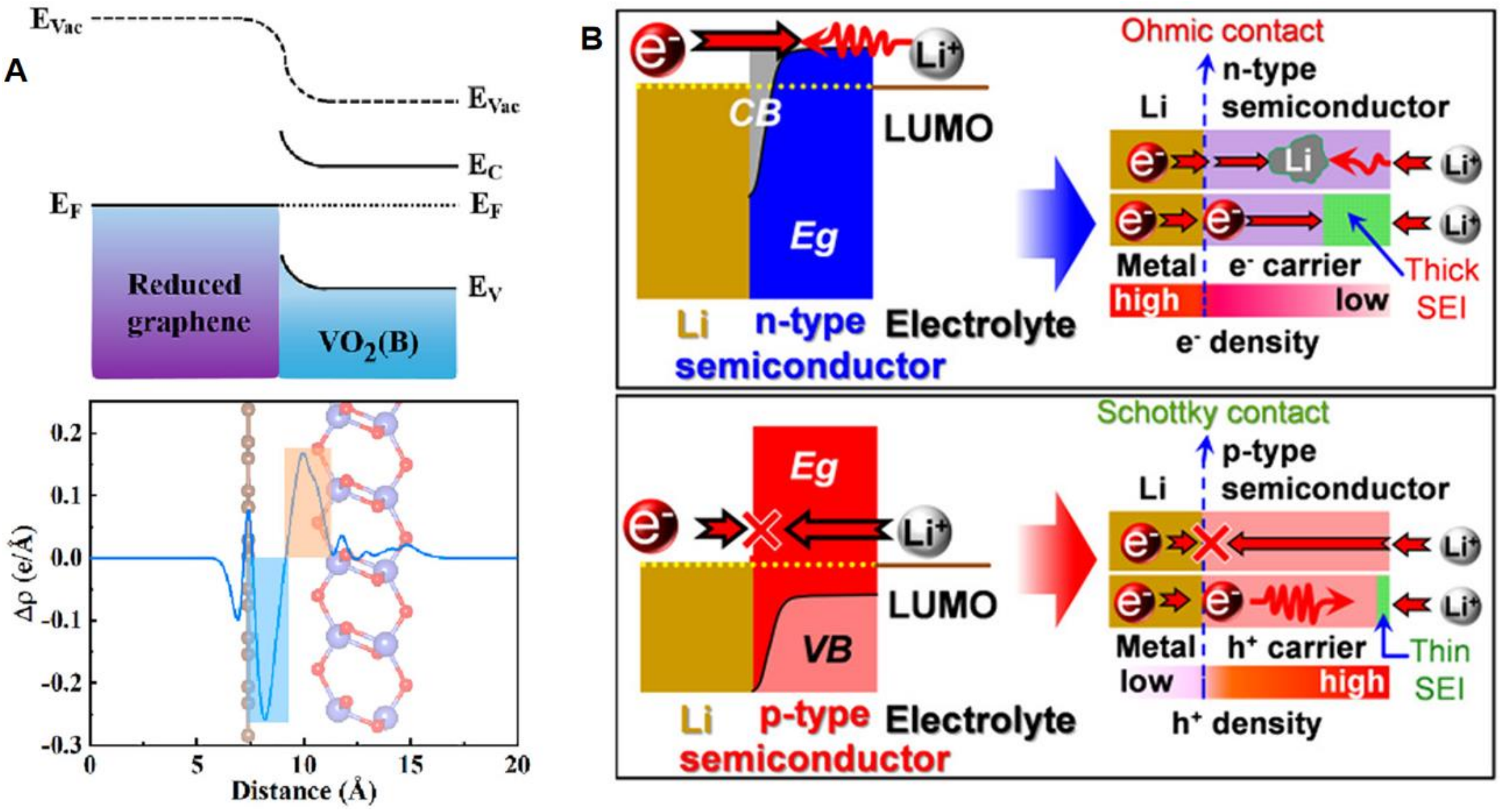

**Figure 19**. (A) Schematic energy diagram of contact between n-type $VO_2$/metallic rGO, and plane-average electron difference of the $VO_2$@rGO heterostructure;[110] Copyright © 2023 American Chemical Society. (B) Schematic deposition behavior of Li-ions in n- and p-PCLs.[111] Copyright © 2021 American Chemical Society. **A $VO_2$@rGO heterostructure was developed to enhance sulfur conversion kinetics for Li-S batteries through formation of the built-in electric field at the interface, where DFT calculations showed electron redistribution and the plane-average electron difference. The electric field improved adsorption of lithium polysulfides, reducing shuttling and enhancing battery stability. Ardhi et al. used n-PCL/p-PCL semiconductor layer on Li metal to suppress dendrite growth. The ohmic n-PCL/Li interface allows electron flow, thickening the SEI. The Schottky p-PCL/Li interface restricts electron transmission, supporting uniform lithium deposition and dendrite suppression.**

Ardhi et al.[111] applied an amorphous polymeric carbon-based semiconductor passivation layer on Li metal electrodes to suppress the growth of Li dendrites. The authors fabricated n-type and p-type semiconductor plasma-polymerized carbon layers (n-PCL and p-PCL), which formed ohmic contact and Schottky contact with Li metal, respectively, as shown in **Figure 19B**. Due to the ohmic contact, electrons can easily pass through the n-PCL/Li interface, leading to the thickening of the SEI. Unlike n-PCL, the Schottky barrier at the p-PCL/Li interface restricts the transmission of electrons from Li metal to p-PCL; only a small number of electrons with energy higher than the barrier can pass through and form the SEI layer, supporting the uniform deposition of Li ions and thus achieving the effect of dendrite suppression.

In addition to modulating the interfacial electronic structure, the EDL also plays a role in enhancing the adhesion between two materials. Derjaguin and Smilga[112] proposed an electronic theory regarding the adhesion strength when dissimilar materials come into contact, detailing the method for calculating the adhesive force when an EDL forms at the metal-semiconductor interface. When a semiconductor comes into contact with a metal, an EDL forms at the interface due to the transfer of electrons from one material to another. The authors provided the expression of the EDL charge density $\sigma$ through derivation of Poisson-Boltzmann equation:

$$\sigma = \sqrt{\frac{2\varepsilon n_1(\infty)}{\pi} kT} \sinh\left(\frac{V_k}{2kT}\right) \quad (34)$$

where $\varepsilon$, dielectric constant; $n_1(\infty)$, electron concentration in the bulk of semiconductor; $V_k$, the potential difference when semiconductor and metal contact, and $V_k = \phi_m - \phi_s$ ($\phi_m$ and $\phi_s$, work functions of metal and semiconductor, respectively). It can be seen that this EDL equation at semiconductor-metal interface has a similar formula to the EDL equation in GCS model because they were derived from Poisson-Boltzmann equation. Higher charge density typically leads to stronger electrostatic attraction between the two solid surfaces, which can enhance adhesion.

On the one hand, although similar theories are used to explain experimental phenomena for electrochemically active materials in both batteries and electrocatalysts, the contact interface between metal current collectors and semiconductor materials in batteries has not received enough attention. Studies have already demonstrated that there is indeed current resistance between the current collector and the electrode.[113] The impact of the EDL at the current collector interface on battery performance is not yet clear. On the other hand, the interface between the metal anode and the solid electrolyte also belongs to the conductor/semiconductor interface. For example, Swift et al.[114] constructed a solid-state EDL model based on the Poisson-Fermi-Dirac (PFD) equation and predicted the band bending and potential distribution at the interface of lithium metal with the $Li_7La_3Zr_2O_{12}$ (LLZO) electrolyte, as well as the Li metal with the inorganic components of the SEI. The PFD equation combines the Poisson equation, Fermi-Dirac statistics, and defect chemistry to simulate the space-charge layer at the interface of the solid electrolyte and electrode materials:

$$\frac{\partial^2\phi}{\partial x^2} = -\frac{N\alpha}{\varepsilon}\left(\frac{1}{B\exp(e\phi/k_BT)+1} - \frac{1}{B\exp(-e\phi/k_BT)+1}\right) \quad (35)$$

where $\phi$, electrostatic potential; $N$, total density of defect sites in solid-state materials; $\alpha$, a certain fixed occupation fraction "saturation parameter" (saturated defect density); $\varepsilon$, dielectric constant; $B$, $B = \alpha\exp(E^f(\phi=0)/k_BT)$ ($E^f$, formation energy of a point defect); $e$, the fundamental charge. The PFD equation provides a framework for modeling charge distributions in solid-state materials with finite site density and electrostatic coupling. It shares conceptual similarities with Kornyshev's Fermi-like distribution (Eq.(9)), which was introduced to account for ion crowding and saturation effects in ionic liquids. Both models improve upon the classical Poisson-Boltzmann model by incorporating Fermi-Dirac statistics to describe occupancy, in contrast to the exponential dependence of ion concentration on potential in the traditional Boltzmann distribution.

**Figure 20A** illustrates the bending situation of the valence band maximum (VBM) and conduction band minimum (CBM) of LLZO obtained by this model, as well as the distribution of positive charge defect concentration. They dominate the formation of the space-charge layer, which is the EDL at the solid interface. The difference between the solid-liquid EDL and the solid-solid EDL is shown in **Figures 20B and 20C**; in the liquid electrolyte, the EDL is composed of solvated ions that are arranged near the electrode surface according to their charge properties. In the solid electrolyte, the EDL manifests as a space-charge layer, composed of point defects (such as ion vacancies or interstitial ions). The potential variation of the latter is closely related to the electronic structure, and band bending must also be considered.

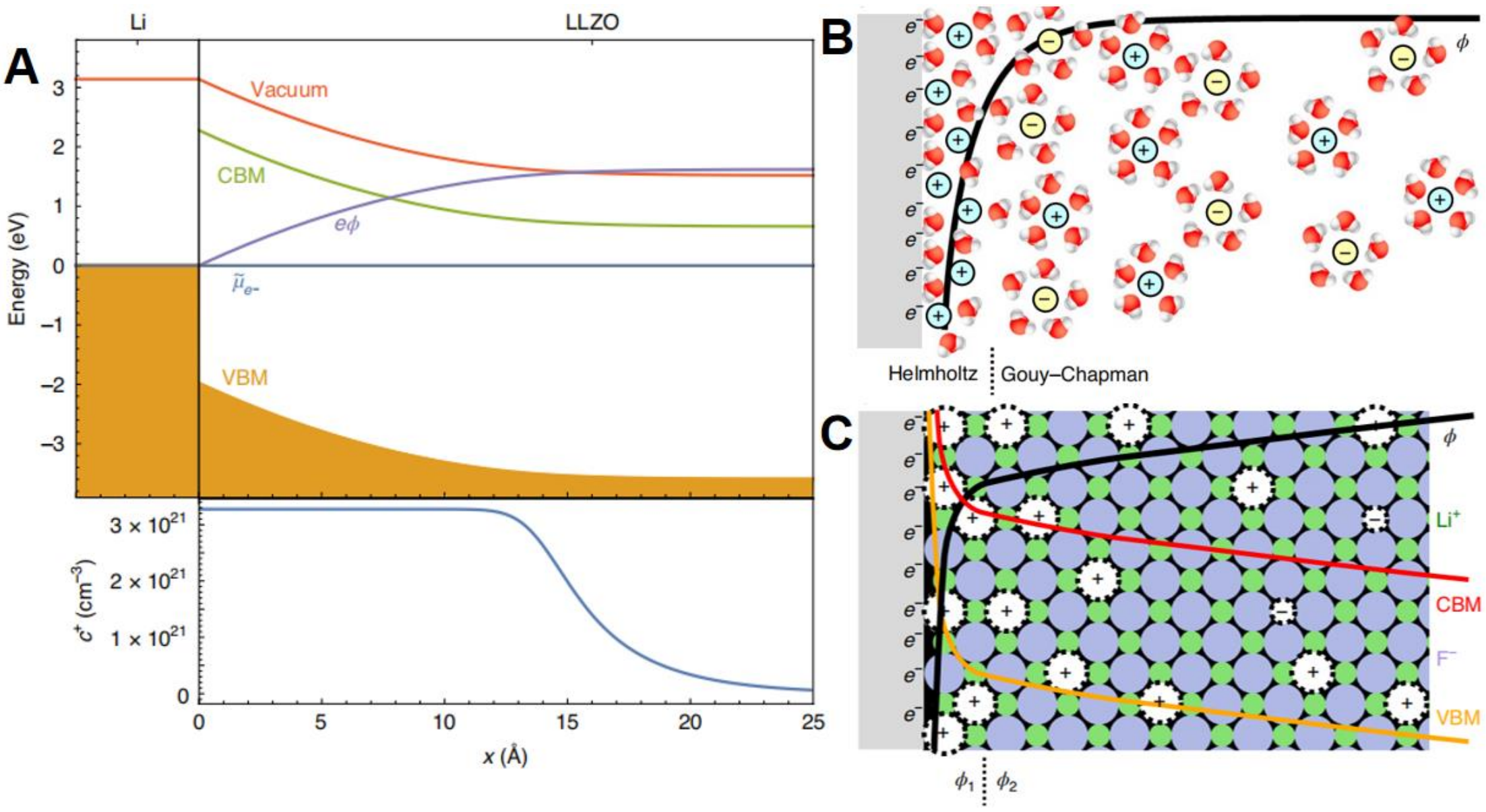

**Figure 20**. (A) Potential distribution and space-charge density at Li/LLZO interface; Comparison of potential distributions at (B) traditional Li/liquid interface and (C) Li/LLZO interface.[114] Copyright © 2021 Springer Nature Limited. **The model shows bending of the VBM and CBM,**

**as well as the distribution of positive charge defects in LLZO, forming a space-charge layer at the solid-solid interface. Solid-liquid EDLs are composed of solvated ions, while solid-solid EDLs, which are space-charge layers, consist of point defects like ion vacancies. The potential variation of EDLs within the solid electrolyte is closely tied to its electronic structure and band bending.**

It can be observed that the various distribution profiles within the EDL are predicated on the Poisson-Boltzmann equation and the Fermi-Dirac distribution (also referred to as the Fermi distribution mentioned earlier). These distribution equations represent a statistical average across all crystallographic planes of a material. Consequently, the theoretical framework does not establish a complete correlation with the specific material composition and crystallographic planes. For a given material, the characteristics of different planes can vary significantly. For example, in Schladt et al.'s work[115], elecrolyte-gated was used to modify the electronic structure of $TiO_2$ single crystals when the ionic liquids acted as the gate material. The electrolyte-gated technique forms an EDL by applying a voltage between the electrolyte and semiconductor, thereby altering the surface electron concentration and conductivity of the semiconductors. It has been found that the metallization behavior of anatase $TiO_2$ is strongly dependent on the crystal face orientation. Specifically, anatase $TiO_2$ with (101) and (001) crystal facets exhibits significant signs of metallization under electrolyte gating, while (110) and (100) crystal facets show no significant effect[115].

The impact of different crystal facets on the solid interface between metal and semiconductor electrode materials may be even greater. For example, in an electrochemical double-layer transistor, the lattice constant near the interface of the $Li_4SiO_4$/niobium (Nb) thin film differs from that of the Nb thin film region far from the interface. This results in an increased state density near the Fermi level at the interface, which promotes an increase of 300 mK in the superconducting critical temperature.[116] Therefore, it is necessary to continue exploring the EDL structure at the semiconductor/metal interface in accordance with the intrinsic crystal face characteristics of solid nanomaterials.

# 4 Critical Challenges and Future Directions in EDL Theory

## 4.1 The paradox of massive voltage disparity at the micro-interface and simulating in-plane EDL distribution

First, let's consider the simplest condition: assuming the thickness of the IHP on a particular electrode surface equals the radius of a water molecule, approximately 1.35 Å, and observing a potential difference of 0.6 V between the electrode and the bulk electrolyte, the electric field strength reaches approximately $4.44\times10^9$ V $m^{-1}$. In this case, we just consider compact layer because it represents the region with the highest electric field strength and the most critical stability challenges. Although the overall EDL thickness can reach approximately 10 nm, the compact layer (or inner Helmholtz plane) directly adjacent to the electrode surface contains the single molecule layer where the electric field intensity is maximized. The extreme field strength exceeds the dielectric breakdown limits for many solid thin films, suggesting that the single molecule layer of liquid electrolyte in the IHP cannot withstand such high electric fields.

Why, then, does this calculated scenario differ from experimental observations? In practice, the EDL on the electrode surface is remarkably stable, particularly within the non-Faradaic potential range. This discrepancy may arise from the distinction between the overall potential difference and the local electric field strength at specific sites on the electrode surface, similar to the difference between total pressure and localized pressure intensity in fluid systems. Consequently, factors like the specific surface area, microstructural irregularities, and atomic-scale surface states greatly influence local electric field variations within the EDL.

If the voltage remains constant but the electric field is measured at different points, variations can occur. Another often overlooked but important issue is how to express the two-dimensional distribution of particle density at the electrode surface within the EDL. Consider a simple scenario: as shown in **Figure 21A**, in a tubular electrolytic cell the size of a single charged molecule, the molecule has only one point of contact on the electrode surface. However, if we slightly increase the diameter of the tube, for example, to the size of two molecular dimensions, where will the charged molecules land on the electrode under the influence of voltage? Furthermore, for a larger cell, such as a pouch cell, how do charged particles choose their landing points on the electrode? More importantly, what role does the EDL play in the process of charged particles selecting their landing points? Although Newman et al.[117] calculated the current density distribution by solving the 1-D Nernst-Planck equation considering reaction kinetics over sixty years ago, and subsequent researchers have since derived distributions using the 2-D Poisson-Nernst-Planck equation[118], it is generally observed that the coupled parabolic and elliptic partial differential equations can yield solutions with unified, smooth gradient curves. This is particularly true for typical electrochemical systems, assuming that appropriate boundary and initial conditions are applied. However, macroscopic models, while useful, may mask the true complexity of the EDL due to atomic-level interactions, non-uniform resistance, and quantum mechanical contributions

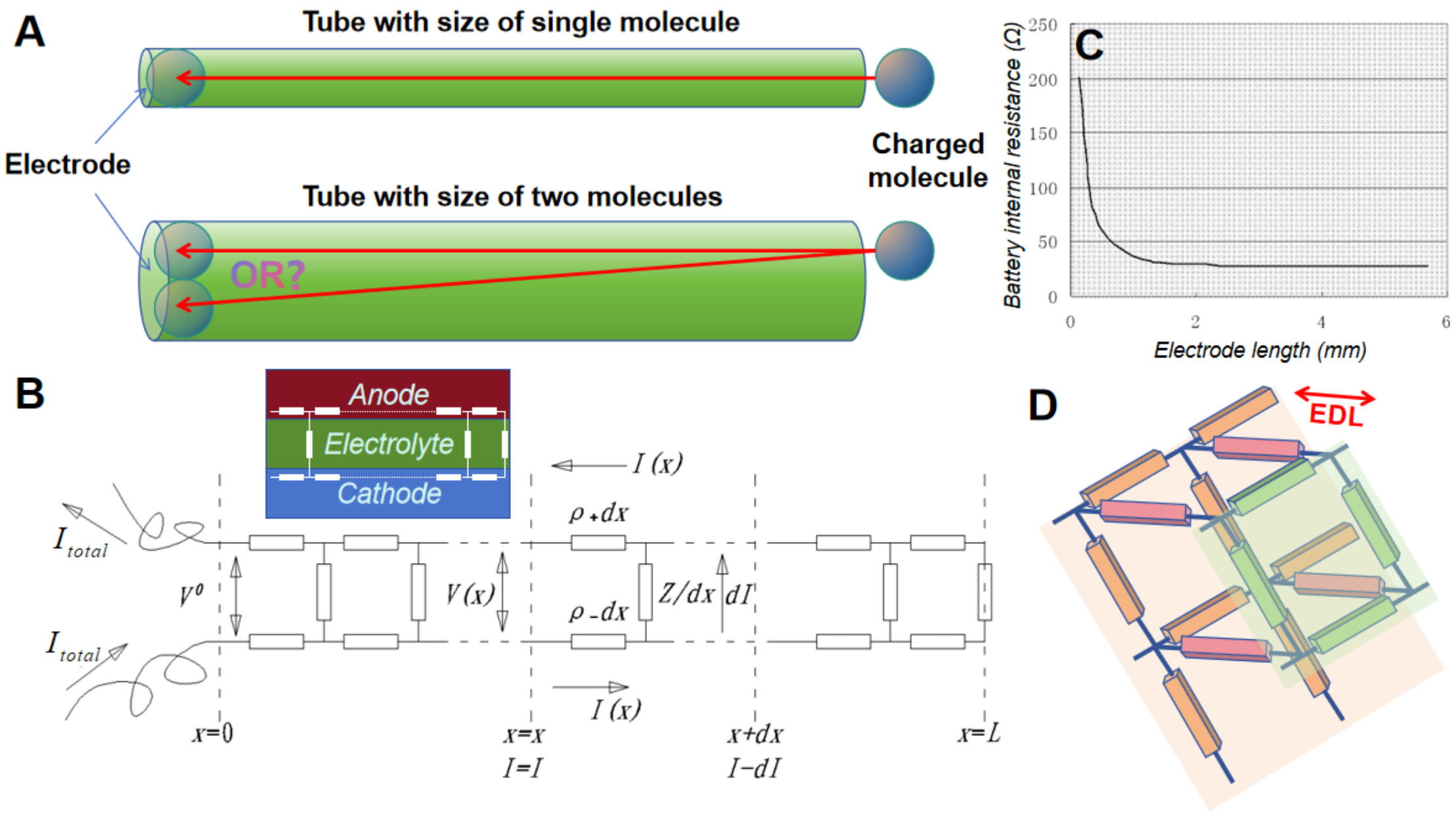

**Figure 21**. (A) Scheme of the question about how to judge landing point of single charged particle; (B) Scheme of cross-sectional battery and infinite resistance grid equivalent circuit diagram; (C) Corresponding internal resistance curve with electrode length;[119] (D) Scheme of 3D infinite resistance grid equivalent circuit diagram of EDL. **The variation of electric field will be present even if voltage is constant. The two-dimensional distribution of particle density at the electrode surface within the EDL is complex. For instance, in a tubular electrolytic cell, charged molecules choose their landing points somewhere on the electrode. In a larger cell like a pouch cell, the process is more intricate. Bian's work used an infinite resistive grid to represent the battery cross-section and calculate resistive distribution. A three-dimensional infinite resistive grid for the EDL could provide a nonlinear resistance distribution, where charged particles may prefer lower resistance areas. This complicated distribution is influenced by factors including voltage, electrode material, and quantum effects.**

Bian’s work[119] can be gained on how to obtain answers in other way. As shown in **Figure 21B**, the authors use an infinite resistive grid composed of countless resistors as the equivalent circuit diagram of the battery cross-section and calculate the resistive distribution along the cross-sectional direction. By solving the resistive differential equation of this equivalent circuit, the curve of internal resistance as a function of electrode length is obtained, as shown in **Figure 21C**. As the electrode lengthens, the internal resistance of the battery will nonlinearly decrease and reach a minimum value. Continuing with this equivalent circuit approach, as shown in **Figure 21D**, constructing a three-dimensional infinite resistive grid for the EDL, if one can successfully formulate and numerically solve the differential equation of this grid, it can be anticipated that a nonlinear variation map of resistance on a two-dimensional plane can be obtained. Charged particles may preferentially land in areas of lower resistance. That is to say, even with a perfect electrode plane and a perfectly formed EDL, the resistance within the EDL will still exhibit nonlinear changes. If further considerations are given to the effects of voltage, electrode material, quantum effects, etc., the distribution of the EDL on a two-dimensional plane would become extremely complex.

## 4.2 The ratio of heat transfer to electrical transfer at the interface

In the field of heat transfer, there has always been a common understanding that materials with high electrical conductivity often also have good thermal conductivity.[120] There must be some correlation between electrical conductivity and thermal conductivity, but this has not been a focus of research in the traditional electrochemical energy storage field. Why does the battery heat up the faster it discharges? By precisely controlling the ratio or balance between heat transfer and electron transfer, we can enhance the Coulombic efficiency of the battery while simultaneously minimizing the risk of thermal runaway. Since electron transfer occurs at the interface, the relationship between heat transfer, electron transfer at the interface, and the mechanism by which the EDL affects this relationship becomes very important.

Actually, in 1835, the Wiedemann-Franz (WF) law was proposed to state that the ratio of charge contribution of the thermal conductivity ( $\kappa_e$ ) to the electrical conductivity ( $\sigma$ ) is proportional to the temperature ( $T$ ):

$$\frac{\kappa_e}{\sigma} = LT \quad (36)$$

where the coefficient $L$ was called as Lorenz number. The WF law is suitable for describing the metals and metal-like conductive materials.[121]

However, there seems to be a lack of detailed discussion and evidence regarding the relationship between heat and electron transfer at the interface. Moreover, if the intention is merely to reduce the heat transfer by using thermal insulation coatings on the electrode surface, it could lead to overheating of the electrode. For example, Gnedenkov et al.[122], in their research on methods to prevent fouling on heat exchange surfaces, found that if a deposit with low thermal conductivity forms on the surface of the heat exchanger, the overall heat flow decreases and the effective heat transfer area is reduced, leading to overheating of the metal and excessive fuel consumption, ultimately resulting in equipment damage. Therefore, it is essential to understand the microscopic mechanisms of charge transfer and heat transfer at the interface in order to continue developing control methods.

In general, there are three modes of electron transfer excitation at the interface in electrochemical systems: photoexcitation, thermal excitation, and electrochemical excitation. Regarding the photoexcitation and thermal excitation of electrons, Hush[123] proposed a unified theoretical framework to describe the outer electron transfer processes of donor-acceptor ion systems in optical and thermal contexts, exploring the correlation between thermal and photoelectron transfer at the metal/solution interface. There was a correlation between the rate of thermal electron exchange and the frequency and bandwidth of the corresponding photointerval transfer absorption. Uniform thermal electron exchange refers to the transfer of electrons between two energy levels or two different positions (such as two ions) under the action of thermal energy. The exchange rate can be described by the Arrhenius equation. Photointerval transfer absorption refers to the process where a system (such as a molecular pair or ion pair) absorbs light energy and electrons transition from one energy level to another.

This transition involves the transfer of electrons between different molecules or ions and is manifested as light absorption at specific frequencies. The frequency and bandwidth of photointerval transfer absorption are related to the energy of the electron transition and the vibrational modes of the system. By establishing the relationship between the two processes, it is possible to predict the characteristics of optical transfer processes by understanding the thermal electron transfer process.

In terms of thermal and electrical excitation of electrons, it has been reported that additional heat can regulate the energy levels of electrons and control the electron transfer at the semiconductor interface ($Sr_3Fe_2O_7$@SrFeOOH) (**Figure 22A**).[124] This also indicates the presence of the impact of heat on electron transfer in electrochemical systems.

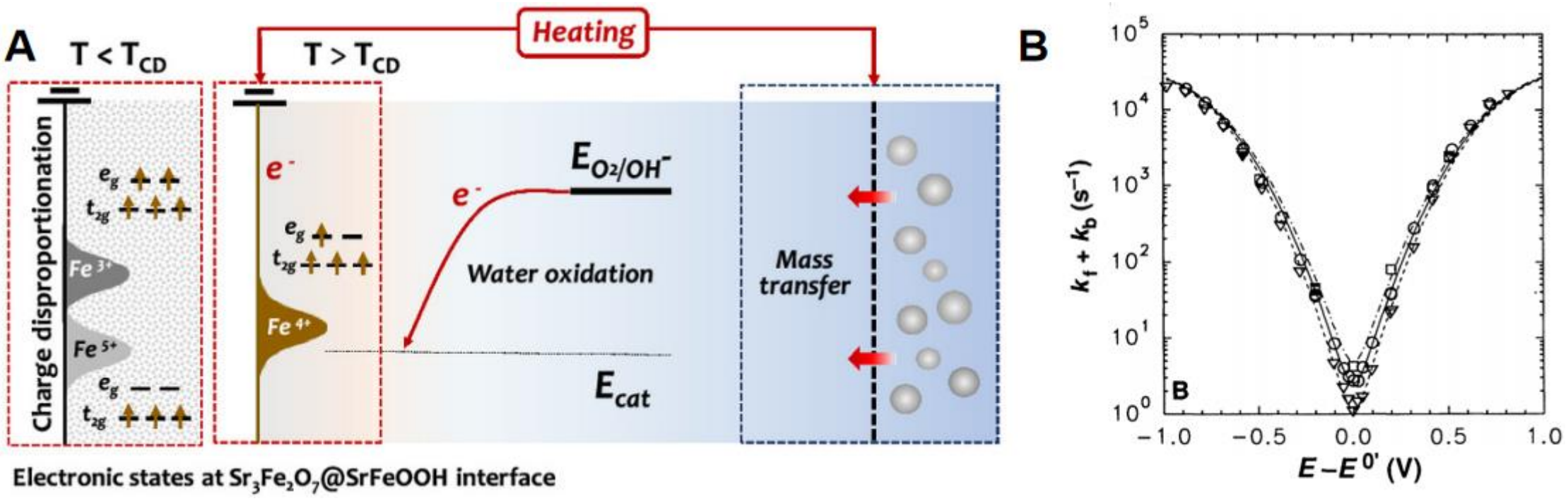

**Figure 22**. (A) Scheme describing the electron transfer at the $Sr_3Fe_2O_7$@SrFeOOH interface under heating;[124] Copyright © 2024 National Academy of Sciences. (B) Semi-log plot of the measured decay rate constants as a function of applied electrode potential at 1°C (triangles), 25°C (circles), and 47°C (squares).[125] Copyright © 1991 the American Association for the Advancement of Science. **Additional heat can regulate electron energy levels and electron transfer at semiconductor interfaces. For the metal/electrolyte interfaces, high temperature increases reaction rates by providing thermal energy for molecular rearrangement.**

Furthermore, if only the effect of an external electric field is considered, the electric field will affect the changes in the potential and kinetic energy of electrons at the interface. For example, Grimes and Adams[126] observed phase transition behavior related to the ratio of electron potential energy to kinetic energy ($\Gamma$) on the surface of liquid helium. When $\Gamma$ is close to or slightly greater than 1, the motion of electrons becomes highly correlated, and the system exhibits liquid-like behavior. When $\Gamma$ is much greater than 1, the potential energy dominates, and electrons form a periodic crystal structure through Coulomb interactions, i.e., a phase transition occurs from a liquid to a crystalline state. By changing the voltage applied to the experimental setup, the binding of electrons on the surface of liquid helium is affected, thereby changing the electron density. Combined with the effect of temperature on kinetic energy, the authors can regulate $\Gamma$. When $\Gamma=137\pm15$, a crystalline two-dimensional electron layer forms on the surface of liquid helium. Although the authors did not mention the EDL and electron transfer, according to basic thermodynamics, kinetic energy can be converted into heat, so there is likely an interaction between the changes in potential energy, kinetic energy caused by the electric field, and the electron transfer process with heat transfer at the interface.

Chidsey[125] investigated the free energy and temperature dependence of electron transfer at the metal/electrolyte interface. The author used a treated gold electrode to measure the electron transfer rate of surface redox reactions in 1 M $HClO_4$ through the potential-step method. By applying the Butler-Volmer equation, the decay rate constant was obtained, and then combined with the Fermi-Dirac distribution equation and the Marcus equation, a diagram of the decay rate constant of electron transfer as a function of potential and temperature was obtained, as shown in **Figure 22B**. The results showed that high temperatures enable more molecules to have sufficient energy to overcome the reorganization energy, thereby increasing the reaction rate. Temperature affects the electron transfer rate by providing additional thermal energy, which makes the molecular or solvent structure more easily rearranged, thus affecting the kinetics of electron transfer.

Bazant[127] extended the Marcus and Butler-Volmer kinetics to concentrated solutions and ionic solids to describe the charge transfer process at the $LiFePO_4$/electrolyte interface. During the charging and discharging process of lithium-ion batteries, $LiFePO_4$ undergoes a phase transition from a Li-rich $LiFePO_4$ phase to a Li-poor $FePO_4$ phase. This phase transition may lead to the formation of different phase regions within the material, i.e., phase separation, which can affect the performance and lifespan of the battery. Increasing the operating temperature of the battery can enhance the diffusion rate of Li ions, thereby reducing the likelihood of phase separation. Furthermore, Feng's group[128] has reviewed the heat generation in EDL and its theory. Therefore, it is also necessary to continue theoretical research on the impact of interfaces containing the EDL on the electron transfer process, as well as the interplay between electron transfer and heat transfer processes from the perspective of electron transition energy levels and electron transfer theory.

## 5 Summaries and Prospective

In this comprehensive review, we have conducted a thorough examination of the EDL theory within electrochemical systems, with particular emphasis on its application and evolution in battery electrode materials. The EDL, a critical interface in these systems, has been extensively studied due to its significant impact on the performance, efficiency, and safety of devices such as batteries and supercapacitors.

*Historical Development and Theoretical Foundations:*

We have provided an in-depth analysis of the GCS model, a foundational theory in electrochemistry and highlighted the importance of surface tension. Our review traces historical evolution of the model, from the seminal work of early researchers to its contemporary modifications. A central aspect of our analysis is the capability of the model to predict ion distribution near charged surfaces and its implications for the capacitance of the EDL. We also highlight the limitations of the GCS model, especially when applied to modern battery materials with complex surface chemistries and structures.

*Classification of Electrode Materials:*

By categorizing electrode materials into semiconductors and conductors, we have offered a more refined perspective on how these materials interact with the EDL. This classification has guided targeted research, fostering a deeper understanding of material-specific EDL phenomena. We focus on the distinct electronic properties of semiconductors and conductors, emphasizing how these characteristics influence the formation and stability of the EDL. This simplified classification provides a clear framework for analyzing the applicability of EDL theories, especially in the context of advanced energy storage technologies like solid-state batteries.

*Interface Theories and Models:*

Conductor/Liquid Interfaces: The study of conductor/liquid interfaces has evolved through various theoretical models to better understand the EDL. A grand-canonical model was developed by using a hybrid density-potential functional, which, alongside the Born-Oppenheimer approximation, provided crucial insights into electron and ion density distributions in the EDL. This was complemented by continuum jellium-Poisson-Boltzmann model, which highlighted quantum effects such as electron spill-out impacting the PZC. Recent advancements include the use of zwitterionic ionic liquids to enhance zinc battery performance and improvements in lithium metal anodes through regulation of the IHP. Despite the advancements, limitations of traditional models have prompted the adoption of AIMD simulations, which offer a more detailed atomic-level understanding of EDL structure. Thus, ongoing developments and refinements in these models continue to advance the field of electrochemistry, providing deeper insights into the complex interactions at conductor or metal/liquid interfaces.

Semiconductor/Liquid Interfaces: The semiconductor/liquid interface is a complex region where semiconductor physics and quantum mechanics intersect, providing a foundation for understanding electron behavior in electrochemical systems. Models such as the Kronig-Penney model and Poisson-Boltzmann equation are instrumental in describing charge distribution and energy band quantization. The interplay between quantum mechanical models and classical electrochemical concepts, such as the GCS model and the Poisson equation, plays a critical role in predicting charge distribution, potential differences, and surface conductivity. Quantum confinement and surface states significantly impact EDL properties, particularly in nanostructured semiconductors, influencing electrochemical stability and reactions. Recent advancements, including the semiconductor electrochemical model and DFT, applied to energy storage and photovoltaic devices, underscore the importance of these interactions in optimizing performance and controlling interfacial phenomena at the atomic level.

Semiconductor/Semiconductor Interfaces: Semiconductor/semiconductor interfaces is crucial for advancing technologies like solid-state batteries. These interfaces, classified into homojunctions and heterojunctions, each exhibit distinct electronic characteristics that influence their application. For example, the $MoS_2$-$MoSe_2$ and $MoS_2$-$NbS_2$ heterostructures have shown promising performance as anode materials due to their high theoretical capacities and stability. Theoretical models such as DFT have been instrumental in predicting the electrochemical properties and potential distribution at these interfaces. Additionally, some work highlighted the role of the EDL in solid-state batteries, where the electrochemical potential and band structure significantly impact battery performance. Similarly, the influence of band misalignment and EDL formation on battery efficiency and stability have been explored. As solid-state batteries become more prevalent, a comprehensive understanding of semiconductor electrochemical theory and EDL formation at these interfaces will be essential for optimizing their design and performance.

Conductor/Semiconductor Interfaces: The conductor or metal/semiconductor interface is governed by the difference in work functions, leading to the formation of a Schottky barrier. This barrier results from the alignment of Fermi levels and band bending, which influences electronic properties and functionality. Key theoretical models, including the use of the Dyson equation with Green's functions, provide insights into interface state density and related parameters. Practical applications, such as in batteries and electrocatalysts, demonstrate the significance of controlling these interfaces to enhance performance and efficiency.

*Future Research Directions:*

Simulation of In-Plane EDL Distribution: The advancement of computational and resistance network models should be recommended to simulate the in-plane distribution of the EDL, considering the non-uniformities of real electrode surfaces. These models would offer a more accurate depiction of charge dynamics at micro and nanoscale levels.

Heat-to-Electrical Transfer at Interfaces: We propose investigating the interaction between heat and electron transfer at interfaces. Understanding this relationship could lead to enhanced battery efficiency and safety, particularly in managing thermal effects during high-rate discharge.

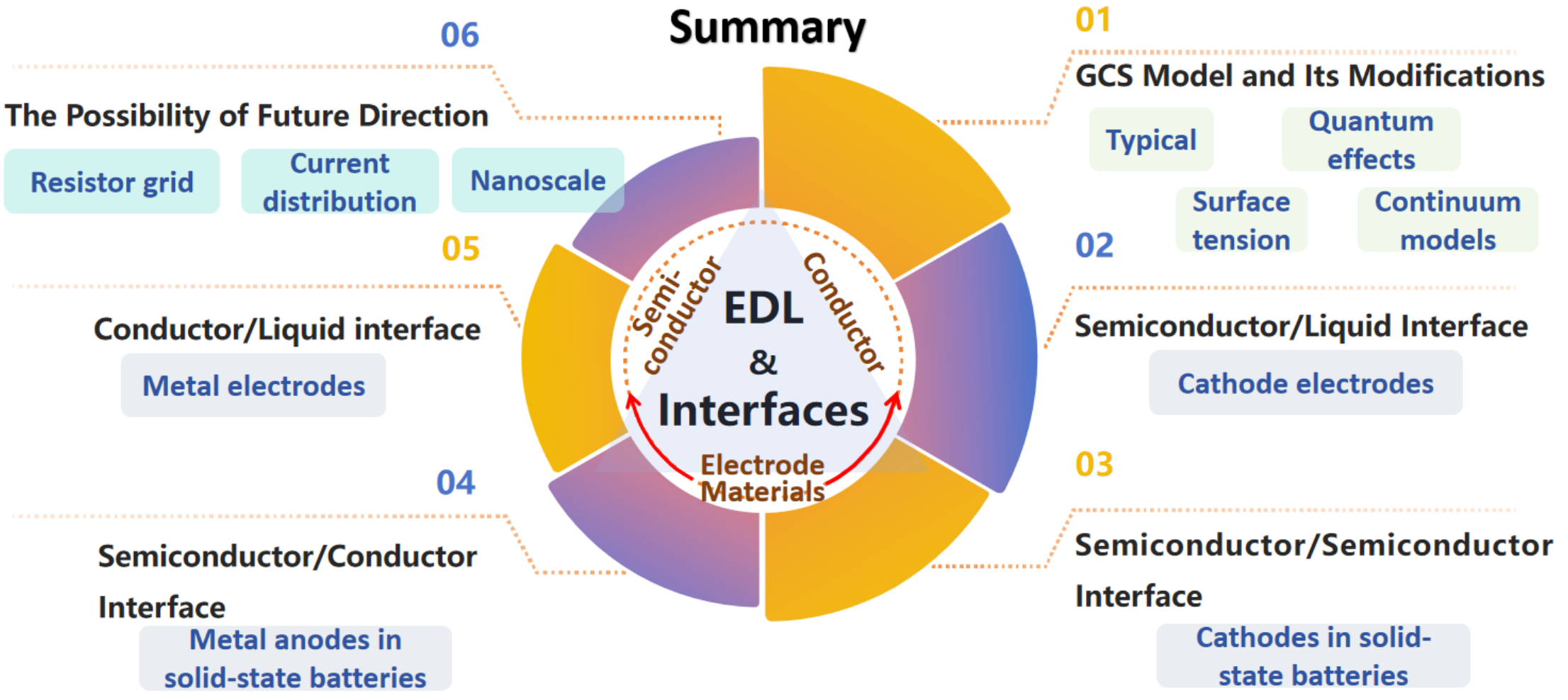

**Figure 23**. Scheme of the summary and key points of this review. **It provides a schematic summary of key topics discussed in the review, highlighting various interfaces in electrode materials, including conductor, semiconductor, and liquid interfaces, alongside potential future directions such as nanoscale modeling and current distribution analysis.**

Summarily, the main points in the review are shown in **Figure 23**, and the future of EDL research lies in developing more advanced computational models that incorporate quantum mechanical effects, surface roughness, material defects, resistance grid or network, and electrolyte composition. These models will be essential for predicting ion and electron behavior at the nanoscale. We advocate for research into new materials with optimized electronic and ionic properties to improve EDL performance in electrochemical devices. Furthermore, collaboration between material scientists, chemists, and physicists is crucial for developing a holistic understanding of EDL phenomena. As the field of electrochemical energy storage advances, EDL research remains a promising avenue for groundbreaking innovations in energy storage technologies.

## Acknowledgments

We would like to show gratitude to the National Natural Science Foundation of China (No. 52202200), the Excellent Young Talents Fund Program of Higher Education Institutions of Anhui Province (No. 2022AH030048), and the European Union and the Czech Ministry of Education, Youth and Sports (MSCA Fellowships CZ FZU III – CZ.02.01.01/00/22_010/0008598).

## References

[1] a)Y. L. Zhu, W. Li, L. Zhang, W. H. Fang, Q. Q. Ruan, J. Li, F. J. Zhang, H. T. Zhang, T. Quan, S. J. Zhang, *Energ Environ Sci* **2023**, 16, 2825; b)L. Jiang, D. M. Li, X. Xie, D. D. Ji, L. W. Li, L. Li, Z. X. He, B. A. Lu, S. Q. Liang, J. Zhou, *Energy Storage Mater* **2023**, 62, 102932; c)S. A. Berlinger, V. Kupers, P. J. Dudenas, D. Schinski, L. Flagg, Z. D. Lamberty, B. D. McCloskey, M. Winter, J. Frechette, *P Natl Acad Sci* **2024**, 121, e2404669121; d)R. Patil, D. K. Das, S. Dutta, *Chem-Eur J* **2023**, 29, e202301117.

[2] a)M. C. Hatzell, M. Raju, V. J. Watson, A. G. Stack, A. C. van Duin, B. E. Logan, *Environ Sci Technol* **2014**, 48, 14041; b)K. J. Jeong, S. Jeong, S. Lee, C. Y. Son, *Adv Mater* **2023**, 35, e2204272; c)P. A. Bonnaud, H. Shiba, *J Phys Chem C* **2023**, 127, 22917; d)G. Zhang, M. Schreier, *Natl Sci Rev* **2024**.

[3] a)F. Q. Yang, *Phys Lett A* **2019**, 383, 2353; b)S. Zhou, *J Mol Liq* **2022**, 364, 119943; c)S. Lamperski, L. B. Bhuiyan, D. Henderson, *J Chem Phys* **2018**, 149, 084706; d)N. Jarvey, F. Henrique, A. Gupta, *J Electrochem Soc* **2022**, 169, 093506; e)L. Chen, H. K. Zhang, R. H. Li, S. Q. Zhang, T. Zhou, B. C. Ma, C. N. Zhu, X. Z. Xiao, T. Deng, L. X. Chen, X. L. Fan, *Chem* **2024**, 10, 1196.

[4] a)T. Binninger, *Electrochim Acta* **2023**, 444, 142016; b)J. D. Elliott, A. A. Papaderakis, R. A. W. Dryfe, P. Carbone, *J Mater Chem C* **2022**, 10, 15225; c)P. Li, Y. Jiao, J. Huang, S. Chen, *JACS Au* **2023**, 3, 2640; d)D. Wang, L. M. Liu, S. J. Zhao, Z. Y. Hu, H. Liu, *J Phys Chem C* **2016**, 120, 4779; e)D. G. Ladha, *Mater Today Chem* **2019**, 11, 94; f)Q. He, B. Yu, Z. H. Li, Y. Zhao, *Energy Environ Mater* **2019**, 2, 264; g)Q. F. Li, C. G. Duan, X. G. Wan, J. L. Kuo, *J Phys Chem C* **2015**, 119, 8662; h)H. Gerischer, *Electrochim Acta* **1990**, 35, 1677; i)F. Yu, T. Huang, P. P. Zhang, Y. P. Tao, F. Z. Cui, Q. J. Xie, S. Z. Yao, F. X. Wang, *Energy Storage Mater* **2019**, 22, 235; j)T. R. Zhang, D. X. Li, Z. L. Tao, J. Chen, *Prog Nat Sci* **2013**, 23, 256.

[5] a)C. Lin, L. He, P. Xiong, H. Lin, W. Lai, X. Yang, F. Xiao, X. L. Sun, Q. Qian, S. Liu, Q. Chen, S. Kaskel, L. Zeng, *ACS Nano* **2023**, 17, 23181; b)D. Wang, Q. Zhang, H. Huang, B. Yang, H. Dong, J.

Zhang, *J Energy Storage* **2022**, 47, 103528; c)E. Johnson, S. Haussener, *J Phys Chem C* **2024**, 128, 10450; d)R. E. Warburton, A. V. Soudackov, S. Hammes-Schiffer, *Chem Rev* **2022**, 122, 10599; e)Y. W. Liu, R. He, Q. F. Zhang, S. L. Chen, *J Phys Chem C* **2010**, 114, 10812.
[6] S. Wall, *Curr Opin Colloid In* **2010**, 15, 119.
[7] I. I. Sudnitsyn, A. V. Smagin, A. P. Shvarov, *Eurasian Soil Sci* **2012**, 45, 452.
[8] H. Helmholtz, *Ann der Physik* **1879**, 243, 337.
[9] M. Khademi, D. P. J. Barz, *Langmuir* **2020**, 36, 4250.
[10] D. C. Grahame, *Chem Rev* **1947**, 41, 441.
[11] S. Sikiru, T. T. Dele-Afolabi, M. Yeganeh Ghotbi, Z. U. Rehman, *J Power Sources* **2024**, 596, 234056.
[12] Z. Li, L. Wang, X. Huang, X. He, *Small* **2024**, 20, e2305429.
[13] T. Zhao, S. Zhou, Z. Xu, S. Zhao, *Journal of Power Sources* **2023**, 559, 232596.
[14] J. Lai, H. Zhang, K. Xu, F. Shi, *J Am Chem Soc* **2024**, 146, 22257.
[15] A. G. Volkov, *Langmuir* **1996**, 12, 3315.
[16] G. Lippmann, *The London, Edinburgh, and Dublin Philosophical Magazine and Journal of Science* **2009**, 47, 281.
[17] A. Sella, https://www.chemistryworld.com/opinion/lippmanns-electrometer/8877.article **2015**.
[18] G. Lippmann, *Ann Chim Phys* **1875**, 494.
[19] R. R. Salem, *Prot Met+* **2005**, 41, 84.
[20] H. Cui, Y. Song, D. Ren, L. Wang, X. He, *Joule* **2024**, 8, 29.
[21] O. K. Rice, *J Phys Chem* **1926**, 30, 1501.
[22] J. N. Israelachvili, P. M. McGuiggan, *Science* **1988**, 241, 795.
[23] J. Israelachvili, Y. Min, M. Akbulut, A. Alig, G. Carver, W. Greene, K. Kristiansen, E. Meyer, N. Pesika, K. Rosenberg, H. Zeng, *Rep Prog Phys* **2010**, 73, 036601.
[24] J. N. Israelachvili, *Intermolecular and Surface Forces*, Elsevier Science, **2011**.
[25] G. D. Degen, Y. T. Chen, A. L. Chau, L. K. Mansson, A. A. Pitenis, *Soft Matter* **2020**, 16, 8096.
[26] a)S. Maheshwari, Y. Li, N. Agrawal, M. J. Janik, in *Advances in Catalysis*, Vol. 63 (Ed: C. Song), Academic Press, 2018; b)A. B. Anderson, *Curr Opin Electroche* **2017**, 1, 27; c)W. Schmickler, D. Henderson, *Prog Surf Sci* **1986**, 22, 323; d)D. W. M. Arrigan, G. Herzog, *Curr Opin Electroche* **2017**, 1, 66.
[27] Y. K. Huang, X. H. Liu, S. Li, T. Y. Yan, *Chinese Physics B* **2016**, 25, 016801.
[28] J. O. M. Bockris, S. U. M. Khan, *Quantum Electrochemistry*, **1979**.
[29] P. Li, Y. L. Jiang, Y. C. Hu, Y. N. Men, Y. W. Liu, W. B. Cai, S. L. Chen, *Nat Catal* **2022**, 5, 900.
[30] G. Jeanmairet, B. Rotenberg, M. Salanne, *Chem Rev* **2022**, 122, 10860.
[31] A. A. Kornyshev, *J Phys Chem B* **2007**, 111, 5545.
[32] M. Z. Bazant, M. S. Kilic, B. D. Storey, A. Ajdari, *Adv Colloid Interface Sci* **2009**, 152, 48.
[33] a)D. Luan, J. Xiao, *J Phys Chem Lett* **2023**, 14, 685; b)L.-L. Zhang, C.-K. Li, J. Huang, *J Electrochem* **2022**, 28, 2108471.
[34] J. Jiang, D. Cao, D. E. Jiang, J. Wu, *J Phys Condens Matter* **2014**, 26, 284102.
[35] J.-L. Liu, *Computational and Mathematical Biophysics* **2015**, 3, 70.
[36] Y. T. He, X. X. Shen, *AIP Advances* **2023**, 13, 075202.
[37] Q. Wu, Y. Qi, *Energy & Environmental Science* **2025**, 18, 3036.
[38] M. McEldrew, Z. A. H. Goodwin, A. A. Kornyshev, M. Z. Bazant, *The Journal of Physical*

*Chemistry Letters* **2018**, 9, 5840.
[39] C. M. Efaw, Q. Wu, N. Gao, Y. Zhang, H. Zhu, K. Gering, M. F. Hurley, H. Xiong, E. Hu, X. Cao, W. Xu, J.-G. Zhang, E. J. Dufek, J. Xiao, X.-Q. Yang, J. Liu, Y. Qi, B. Li, *Nature Materials* **2023**, 22, 1531.
[40] W. Schmickler, *Chem Rev* **1996**, 96, 3177.
[41] J. Wu, *Chem Rev* **2022**, 122, 10821.
[42] Y. J. Yu, S. J. Hu, *Chinese Chem Lett* **2021**, 32, 3277.
[43] H. He, in *Solution Processed Metal Oxide Thin Films for Electronic Applications*, (Eds: Z. Cui, G. Korotcenkov), Elsevier, 2020.
[44] V. M. Leal, E. L. D. Coelho, M. B. J. G. Freitas, *J Energy Chem* **2023**, 79, 118.
[45] M. Tokur, A. Erdas, D. Nalci, M. O. Guler, H. Akbulut, *Mat Sci Semicon Proc* **2015**, 38, 387.
[46] a)J. Y. Li, W. L. Yao, S. Martin, D. Vaknin, *Solid State Ionics* **2008**, 179, 2016; b)Z. H. Cui, X. Guo, J. Q. Ren, H. T. Xue, F. L. Tang, P. Q. La, H. Li, J. C. Li, X. F. Lu, *Electrochim Acta* **2021**, 388, 138592; c)Z. H. Li, D. M. Zhang, F. X. Yang, *J Mater Sci* **2009**, 44, 2435.
[47] I. Saadoune, C. Delmas, *J Solid State Chem* **1998**, 136, 8.
[48] H. Tukamoto, A. R. West, *J Electrochem Soc* **1997**, 144, 3164.
[49] a)L. Fang, M. Wang, Q. Zhou, H. Xu, W. Hu, H. Li, *Colloid Surface A* **2020**, 600, 124940; b)I. Ullah, A. Majid, M. I. Khan, *J Mater Sci-Mater El* **2020**, 31, 7324.
[50] Y. Li, C. X. Xu, M. Y. Dang, C. P. Yu, Y. L. He, W. B. Liu, H. M. Jin, W. X. Li, M. Y. Zhu, J. J. Zhang, *Ceram Int* **2021**, 47, 21759.
[51] K. Seeger, *Semiconductor Physics*, Springer Vienna, **2013**.
[52] a)J. W. Mcclure, *IBM Journal of Research and Development* **1964**, 8, 255; b)C. He, L. Sun, C. Zhang, J. Zhong, *Phys Chem Chem Phys* **2013**, 15, 680.
[53] a)Y. Zhu, H. Ji, H.-M. Cheng, R. S. Ruoff, *Natl Sci Rev* **2017**, 5, 90; b)M. Baskey, S. K. Saha, *Adv Mater* **2012**, 24, 1589.
[54] D. Hamani, M. Ati, J. M. Tarascon, P. Rozier, *Electrochem Commun* **2011**, 13, 938.
[55] L. Zheng, Z. Wang, M. Wu, B. Xu, C. Ouyang, *J Mater Chem A* **2019**, 7, 6053.
[56] J. Huang, S. Chen, M. Eikerling, *J Chem Theory Comput* **2021**, 17, 2417.
[57] J. Huang, P. Li, S. L. Chen, *Phys Rev B* **2020**, 101, 125422.
[58] R. Xu, X. Shen, X. X. Ma, C. Yan, X. Q. Zhang, X. Chen, J. F. Ding, J. Q. Huang, *Angew Chem Int Edit* **2021**, 60, 4215.
[59] a)Y. He, L. Wang, A. Wang, B. Zhang, H. Pham, J. Park, X. He, *Exploration* **2024**, 4, 20230114; b)A. P. Wang, L. Wang, Y. Z. Wu, Y. F. He, D. S. Ren, Y. Z. Song, B. Zhang, H. Xu, X. M. He, *Adv Energy Mater* **2023**, 13, 2300626.
[60] Y. Q. Lv, M. Zhao, Y. D. Du, Y. Kang, Y. Xiao, S. M. Chen, *Energ Environ Sci* **2022**, 15, 4748.
[61] C. Yan, H. R. Li, X. Chen, X. Q. Zhang, X. B. Cheng, R. Xu, J. Q. Huang, Q. Zhang, *J Am Chem Soc* **2019**, 141, 9422.
[62] Q. Wu, M. T. McDowell, Y. Qi, *Journal of the American Chemical Society* **2023**, 145, 2473.
[63] A. Groß, S. Sakong, *Curr Opin Electroche* **2019**, 14, 1.
[64] S. Baldelli, *Acc Chem Res* **2008**, 41, 421.
[65] a)J. Jiang, M. Z. Dai, J. Sun, B. Zhou, A. X. Lu, Q. Wan, *J Appl Phys* **2011**, 109; b)T. Tsuchiya, M. Takayanagi, K. Mitsuishi, M. Imura, S. Ueda, Y. Koide, T. Higuchi, K. Terabe, *Commun Chem* **2021**, 4, 117; c)T. Kim, C. H. Choi, J. S. Hur, D. Ha, B. J. Kuh, Y. Kim, M. H. Cho, S. Kim, J. K. Jeong, *Adv Mater* **2023**, 35, e2204663; d)K. Xu, M. M. Islam, D. Guzman, A. C. Seabaugh, A. Strachan, S. K.

Fullerton-Shirey, *ACS Appl Mater Interfaces* **2018**, 10, 43166; e)K. Tybrandt, I. V. Zozoulenko, M. Berggren, *Sci Adv* **2017**, 3, eaao3659.
[66] a)M. Takayanagi, T. Tsuchiya, D. Nishioka, M. Imura, Y. Koide, T. Higuchi, K. Terabe, *Mater Today Phys* **2023**, 31, 101006; b)H. Yuan, H. Shimotani, J. Ye, S. Yoon, H. Aliah, A. Tsukazaki, M. Kawasaki, Y. Iwasa, *J Am Chem Soc* **2010**, 132, 18402; c)X. Zhang, Q. Wang, J. W. Huang, K. Meng, P. Chen, L. Zhou, M. Tang, C. R. Zhang, X. T. Dai, X. Y. Bi, C. Y. Qiu, H. J. Zhang, W. W. Zhao, H. T. Yuan, *APL Materials* **2021**, 9; d)J. Li, P. H. Q. Pham, W. Zhou, T. D. Pham, P. J. Burke, *ACS Nano* **2018**, 12, 9763.
[67] a)K. Schwarz, R. Sundararaman, *Surf Sci Rep* **2020**, 75, 100492; b)L. Li, Y. P. Liu, J. B. Le, J. Cheng, *Cell Reports Physical Science* **2022**, 3; c)R. Burt, G. Birkett, X. S. Zhao, *Phys Chem Chem Phys* **2014**, 16, 6519; d)J. Vatamanu, D. Bedrov, O. Borodin, *Molecular Simulation* **2017**, 43, 838.
[68] a)H. Yuan, Q. R. Zhang, T. Zhou, W. B. Wu, H. R. Li, Z. P. Yin, J. M. Ma, T. F. Jiao, *Chem Eng J* **2024**, 485, 149926; b)M. S. Wu, K. H. Lin, *J Phys Chem C* **2010**, 114, 6190; c)Q. Bai, H. Li, L. Zhang, C. Li, Y. Shen, H. Uyama, *ACS Appl Mater Interfaces* **2020**, 12, 55913.
[69] a)K. Shibata, H. Yuan, Y. Iwasa, K. Hirakawa, *Nat Commun* **2013**, 4, 2664; b)B. Chen, Z. F. Zhai, N. Huang, C. Y. Zhang, S. Y. Yu, L. S. Liu, B. Yang, X. Jiang, N. J. Yang, *Adv Energy Mater* **2023**, 13, 2300716.
[70] R. L. Pavelich, F. Marsiglio, *Am J Phys* **2015**, 83, 773.
[71] Y. V. Pleskov, in *Comprehensive Treatise of Electrochemistry*, (Eds: J. O. M. Bockris, B. E. Conway, E. Yeager), Springer US, Boston, MA 1980, Chapter 6.
[72] C. G. B. Garrett, W. H. Brattain, *Physical Review* **1955**, 99, 376.
[73] M. F. Dopke, F. Westerbaan van der Meij, B. Coasne, R. Hartkamp, *Phys Rev Lett* **2022**, 128, 056001.
[74] W. H. Brattain, P. J. Boddy, *J Electrochem Soc* **1962**, 109, 574.
[75] K. T. Butler, G. S. Gautam, P. Canepa, *Npj Comput Mater* **2019**, 5, 19.
[76] B. Zhu, L. D. Fan, N. Mushtaq, R. Raza, M. Sajid, Y. Wu, W. F. Lin, J. S. Kim, P. D. Lund, S. I. Yun, *Electrochem Energy Rev* **2021**, 4, 757.
[77] J. M. Mayer, *J Am Chem Soc* **2023**, 145, 7050.
[78] L. R. Faulkner, *J Chem Educ* **1983**, 60, 262.
[79] W. Zhang, D. Wang, W. T. Zheng, *J Energy Chem* **2020**, 41, 100.
[80] D. Du, Z. Zhu, K. Y. Chan, F. Li, J. Chen, *Chem Soc Rev* **2022**, 51, 1846.
[81] B. Zhang, S. Wang, W. Fan, W. Ma, Z. Liang, J. Shi, S. Liao, C. Li, *Angew Chem Int Ed* **2016**, 55, 14748.
[82] N. R. Agrawal, C. Duan, R. Wang, *J Phys Chem B* **2024**, 128, 303.
[83] A. D. J. Haymet, D. W. Oxtoby, *J Chem Phys* **1981**, 74, 2559.
[84] W. A. Curtin, *Phys Rev Lett* **1987**, 59, 1228.
[85] J. Maibach, F. Lindgren, H. Eriksson, K. Edstrom, M. Hahlin, *J Phys Chem Lett* **2016**, 7, 1775.
[86] Y. T. He, J. K. Wang, L. Wang, X. M. He, *Batteries & Supercaps* **2024**, 7, e202400059.
[87] Z. T. Ren, S. C. Gu, T. Li, L. K. Peng, C. H. Zou, F. Y. Kang, W. Lv, *Journal of Materials Chemistry A* **2024**, 12, 32885.
[88] a)T. Qin, X. Zhao, Y. Sui, D. Wang, W. Chen, Y. Zhang, S. Luo, W. Pan, Z. Guo, D. Y. C. Leung, *Adv Mater* **2024**, 36, e2402644; b)J. Maier, *Angew Chem Int Edit* **2003**, 32, 313; c)R. Wang, H. Liu, Y. Zhang, K. Sun, W. Bao, *Small* **2022**, 18, e2203014; d)Y. Xie, C. Yu, L. Ni, J. Yu, Y. Zhang, J. Qiu, *Adv Mater* **2023**, 35, e2209652.

[89] P. Wang, M. Xue, D. Jiang, Y. Yang, J. Zhang, H. Dong, G. Sun, Y. Yao, W. Luo, Z. Zou, *Nat Commun* **2022**, 13, 2544.
[90] R. S. Bauer, G. Margaritondo, *Phys Today* **1987**, 40, 26.
[91] G. Barik, S. Pal, *Appl Surf Sci* **2022**, 596, 153529.
[92] B. V. Deryagin, Y. P. Toporov, *Russ Chem Bull* **1982**, 31, 1544.
[93] S. X. Fan, X. L. Zou, H. D. Du, L. Gan, C. J. Xu, W. Lv, Y. B. He, Q. H. Yang, F. Y. Kang, J. Li, *J Phys Chem C* **2017**, 121, 13599.
[94] M. W. Swift, Y. Qi, *Phys Rev Lett* **2019**, 122, 167701.
[95] J. F. T. Martinez, F. Ambriz-Vargas, P. L. Rodríguez-Kessler, F. M. Morales, A. M. Gámez, C. Gomez-Yañez, *J Mater Sci: Mater Electron* **2024**, 35, 402.
[96] K. Chen, X. Wang, G. Wang, B. Wang, X. Liu, J. Bai, H. Wang, *Chem Eng J* **2018**, 347, 552.
[97] J. B. Kang, N. P. Deng, D. J. Shi, Y. Feng, Z. Y. Wang, L. Gao, Y. X. Song, Y. X. Zhao, B. W. Cheng, G. Li, W. M. Kang, K. Zhang, *Adv Funct Mater* **2023**, 33, 2307263.
[98] H. Kroemer, *Surf Sci* **1983**, 132, 543.
[99] R. Kanno, *Electrochemistry* **2023**, 91, 102001.
[100] K. Hikima, K. Shimizu, H. Kiuchi, Y. Hinuma, K. Suzuki, M. Hirayama, E. Matsubara, R. Kanno, *Commun Chem* **2022**, 5, 52.
[101] a)W.-Y. A. Lam, H. Zhao, B. Zhang, L. Wang, H. Xu, X. He, *Next Energy* **2024**, 2, 100106; b)Y. Wu, Y. Liu, X. Feng, Z. Ma, X. Xu, D. Ren, X. Han, Y. Li, L. Lu, L. Wang, X. He, M. Ouyang, *Adv Sci* **2024**, 11, e2400600.
[102] a)J. Bardeen, *Physical Review* **1947**, 71, 717; b)P. Peljo, J. A. Manzanares, H. H. Girault, *Langmuir* **2016**, 32, 5765; c)R. T. Tung, *Phys Rev B* **2001**, 64, 205310.
[103] M. J. Sparnaay, *Adv Coll Interface Sci* **1967**, 1, 278.
[104] W. Mönch, *Surf Sci* **1994**, 299-300, 928.
[105] D. N. Lowy, A. Madhukar, *Phys Rev B* **1978**, 17, 3832.
[106] J. O. Mccaldin, T. C. Mcgill, *Annu Rev Mater Sci* **1980**, 10, 65.
[107] H. Li, Y. Liu, L. Huang, J. Xin, T. Zhang, P. Liu, L. Chen, W. Guo, T. Gu, G. Wang, *J Mater Chem A* **2023**, 11, 5179.
[108] K. Q. Li, R. Q. Cheng, Q. Y. Xue, P. Y. Meng, T. S. Zhao, M. Jiang, M. L. Guo, H. X. Li, C. P. Fu, *Chem Eng J* **2022**, 450, 137991.
[109] Z. Li, S. Xu, Y. Shi, X. Zou, H. Wu, S. Lin, *Chem Eng J* **2021**, 414, 128814.
[110] G. Liu, Q. Zeng, Q. Wu, S. Tian, X. Sun, D. Wang, X. Li, W. Wei, T. Wu, Y. Zhang, Y. Sheng, K. Tao, E. Xie, Z. Zhang, *ACS Appl Mater Interfaces* **2023**, 15, 39384.
[111] R. E. A. Ardhi, G. Liu, J. K. Lee, *Acs Energy Lett* **2021**, 6, 1432.
[112] B. V. Derjaguin, V. P. Smilga, *J Appl Phys* **1967**, 38, 4609.
[113] B. A. Mei, O. Munteshari, J. Lau, B. Dunn, L. Pilon, *J Phys Chem C* **2018**, 122, 194.
[114] M. W. Swift, J. W. Swift, Y. Qi, *Nat Comput Sci* **2021**, 1, 212.
[115] T. D. Schladt, T. Graf, N. B. Aetukuri, M. Li, A. Fantini, X. Jiang, M. G. Samant, S. S. Parkin, *ACS Nano* **2013**, 7, 8074.
[116] T. Tsuchiya, S. Moriyama, K. Terabe, M. Aono, *Appl Phys Lett* **2015**, 107.
[117] J. S. Newman, C. W. Tobias, *Journal of The Electrochemical Society* **1962**, 109, 1183.
[118] D. He, K. Pan, *Applied Mathematics and Computation* **2016**, 287-288, 214.
[119] H. Bian, *https://muchong.com/html/201801/2926546.html* **2011**.
[120] a)X. Qian, J. Zhou, G. Chen, *Nat Mater* **2021**, 20, 1188; b)X. Xu, J. Chen, J. Zhou, B. Li, *Adv*

*Mater* **2018**, 30, e1705544; c)S. Wu, T. Yan, Z. Kuai, W. Pan, *Energy Storage Mater* **2020**, 25, 251.
[121] a)M. Silhavik, P. Kumar, P. Levinsky, Z. A. Zafar, J. Hejtmanek, J. Cervenka, *Small Methods* **2024**, e2301536; b)H. Yi, J. Y. Kim, H. Z. Gul, S. Kang, G. Kim, E. Sim, H. Ji, J. Kim, Y. C. Choi, W. S. Kim, S. C. Lim, *Carbon* **2020**, 162, 339; c)S. Yigen, A. R. Champagne, *Nano Lett* **2014**, 14, 289.
[122] S. V. Gnedenkov, S. L. Sinebryukhov, D. V. Mashtalyar, V. S. Egorkin, A. K. Tsvetnikov, A. N. Minaev, *Prot Met+* **2007**, 43, 667.
[123] N. S. Hush, *Electrochim Acta* **1968**, 13, 1005.
[124] M. Lu, Y. Du, S. Yan, T. Yu, Z. Zou, *P Natl Acad Sci* **2024**, 121, e2316054120.
[125] C. E. Chidsey, *Science* **1991**, 251, 919.
[126] C. C. Grimes, G. Adams, *Phys Rev Lett* **1979**, 42, 795.
[127] M. Z. Bazant, *Acc Chem Res* **2013**, 46, 1144.
[128] L. Zeng, X. Tan, N. Huang, G. Feng, *Curr Opin Electroche* **2024**, 46, 101503.

**The questions and answers:**

**Q1**: (anonymous) Is EDL thickness really 1.35 Angstrom? There are numerous reports on such thickness, if they are in the order of ~10 nm, the question on dielectric strength/breakdown is not necessary.

**A1**: Sure, as we known, the thickness of EDL is appl. 10 nm because it contains compact and diffusion layers. However, in the initial study, the thickness of the EDL is defined as diameter of single adsorbed molecule layer, like a plane physical capacitor. In this case, we consider the influence of voltage on the intermedium within EDL, especially for steady static EDL. And, the voltage is partially caused by the potential difference between diffusion layer and electrode surface. Therefore, it is reasonable to consider only single layer on the surface. The stability of this single molecule layer needs to be concerned under such ultra-high voltage, which exceed the dielectric breakdown limits for many materials.

Even if we consider the entire EDL thickness of around 10 nm instead of a single molecule layer, the electric field strength would not drastically change. The value would be changed from $4.44\times10^{9}$ V $m^{-1}$ to $6.0\times10^{7}$ V $m^{-1}$, which is still an extremely high electric field near the surface. So, the reason why the molecule keep stable needs to be considered.

**Change 1**: A sentence was added to state the point : "In this case, we just consider compact layer because it represents the region with the highest electric field strength and the most critical stability challenges. Although the overall EDL thickness can reach approximately 10 nm, the compact layer (or inner Helmholtz plane) directly adjacent to the electrode surface contains the single molecule layer where the electric field intensity is maximized.".

(When a better answer appears, this part will be updated.)

------------------------------------------------------------

**Q2**: (anonymous) Although the objective is interesting, and could be potentially beneficial to a rather wide community, it is unfortunately not reached at all. This is because the information is not hierarchized, but mostly presented as a sequence of sometimes barely related works. For unexperienced readers or even for experimentalists, who are the main target, it will be very hard to follow since the different physical origins between very different methods are not explained at all. Many equations are introduced, but the link between them is never made. Many paragraphs look like "fillers" that do not convey any useful information, for example, the very first paragraph of the introduction (and many more). Figures such as Fig. 1 that is completely useless. In addition there are many ill-defined concepts, such as the "chemical potential interaction" (page 8), that do not exist: The chemical potential is a thermodynamic quantity, introduced to describe systems at the macroscopic scale, while interactions are occurring at the microscopic scale. There are many more examples like this that I will not detail. In conclusion, I do not recommend publication of this review in any journal.

**A2**: Thank you for your valuable comments.

**1)** For "This is because the information is not hierarchized, but mostly presented as a sequence of sometimes barely related works. For unexperienced readers or even for experimentalists, who are the main target, it will be very hard to follow since the different physical origins between very different methods are not explained at all."

The original idea of this review is to clarify the origin of EDL theory and its recent developments. The conflicts between the different theoretical models and experimental work were introduced. First, the electrochemistry is an interdisciplinary, which is strongly based on experiments and theories simultaneously. We just used physical methods to study the electrochemical phenomenon. For example, researchers get diffusion equation from Heat Transfer (Fourier's law), and used diffusion equation (Fick's law) to explain the kinetics, such as CV curves. So, the any researcher who are interesting in electrochemistry may be the main readers. We can not divide theories and experiments into separate parts.

However, in fact, there are different description methods in the different models. For example, we used Fick's law and B-V equation to describe the electrochemical kinetics, but used P-B (or PBP) model to describe the EDL. Both them are important topics, and even occur in the same region near the electrode surface, but they are described in different way. That is similar for various electrode materials. We need to use some methods to link them. Before that, we need to classify the types of materials and try to cover all existing electrochemical theories. It is hard to achieve but we need to try. So, in this review, we just divided materials into conductors and semiconductors, which is also consistent with reality. Subsequently, many theories or models will be introduced to describe the four (even more) types of interfaces. This review undoubtedly brings many different description methods. Currently, I have no idea how to avoid this point, which may be hard to be done.

**Change 1**: Adding a conceptual roadmap at the beginning to guide readers through the structure. Please see the new Fig.1 in the original position.

**2)** For "Many equations are introduced, but the link between them is never made."

***For Eqs.(1)-(4)***: In the initial stage, the origin of EDL should be introduced, but before that, the electrocapillary can not be abandoned because it can be explained by EDL theory. Therefore, the Eqs.(1), (2), (3), (4) cover the theory of electrocapillary. The Eqs.(1) and (2) are from the Gibbs adsorption isotherm, which might be introduced in "Physical Chemistry". Eq.(3) was got by differentiating, and Eq.(4) was obtained by further considering in the presence of the polarizable charge density. The concepts of bound charge might be introduced in "Dielectric Materials".

***For Eq.(5)***: The Eq.(5) is the main equation in EDL theory, from P-B model. But it should be noticed that there is also a debate. In the derivation process, both the positive and negative ions was considered, and was used to calculate surface charge on solid side. Many researchers refer to all the types of ions should be considered. However, like a thermodynamic system, we should clarify a system first. If only liquid (electrolyte) is considered, both sides on cathode and anode should be combined, and there are two EDLs because the system should keep equilibrium. If only one interface is considered, such as on cathode/liquid interface, we can only consider surface charge (electron) and adsorbed ions and ions in diffusion layer because the interface system can keep equilibrium. So this topic is complex, but currently, this main equation is acceptable for everyone. Eventually, the Eq.(5) can not directly linked to Eq.(1)-(4) because the classical EDL theory didn't explicitly consider surface tension.

***For Eq.(6)***: The Eq.(6) was listed to declare the method in the literature to combine the previous equations. It derived from the Eq.(5) and $\mathrm{d}\gamma = -Q\mathrm{d}\varphi$.

**Change 1**: A original sentence was modified to clarify the relationship between Eq.(5) and Eq.(6): "Furthermore, an equation was derived for surface tension by introducing $\mathrm{d}\gamma = -Q\mathrm{d}\varphi$,".

***For Eqs.(7) and (8)***: The item of surface densities of the bound charges was further explained by Eq.(7), and got the relationship between surface tension and surface charges in Eq.(8) by solving Eq.(4). The math derivation process was ignored because the conclusion has been published in literature. Subsequently, the experimental results and theoretical predict can be compared in next step.

***For Eqs.(9) and (10)***: The Eq.(9) is description of ion distribution, which was derived from Fermi-like distribution. The occupation space for ions was considered. This equation can be regarded as the modification of Boltzmann distribution $n_{x,i} = n_{0,i} \exp\left(-z_i \omega\varphi / kT\right)$ in P-B model, which has been list in original text. It can be seen that, compared to Boltzmann distribution, the Fermi-like distribution contains a new parameter $\lambda = \bar{N}/N$, which introduce the impact of the number of available sites. The detailed math derivation process can be found in literation. Eventually, the Eq.(10), the modification of EDL capacitance equation can be further obtained, So, these equations are also used to link previous ones.

***For Eq.(11)***: The Eq.(11) described a DFT-modified PNP equation, which discussed the EDL properties from quantum mechanical perspective and on liquid side. Actually, different from the EDL model, this equation discussed the ion diffusion properties in electrolyte near the electrode or in a certain channel. There was lack of transitional sentences when the modification perspective changes.

**Change 2**: A transitional sentence has been added to improve the logical flow: "Furthermore, some modifications have been made by integrating quantum mechanical methods and perspectives based on ion diffusion. For example,"

***For Eqs.(12) and (13)***: The new concept of quantum potential in Eq.(12) was introduced in literature, and was used to modify the PNP equations in Eq.(13). This example was used to exhibit a typical equation modification on liquid side. Because there were many different way to modify the diffusion equation through quantum mechanic method, it is highly challenging to link these equation to previous ones.

***For Eq.(14)***: Because the model is more complicated in literature, here we just briefly introduce the main general equation about grand potential, that is Eq.(14). The subsequent steps were declared by main method and figures. This equation is description of the item in grand canonical ensemble $p = \exp(-\beta\Omega) / Z$ . That is used to represent the possible states thermodynamically, and in the literature, it was used to represent the probability of the microstate of the EDL. It is separate theoretical framework and is hard to link previous equations literally and directly. But we should introduce here to lead to the subsequent content. To clarify this point:

**Change 3**: A sentence was added to explain the origin of Eq.(14): "This equation is description of the item in grand canonical ensemble $p = \exp(-\beta\Omega) / Z$ ( $Z$ , a normalization factor; $\beta = (k_B T)^{-1}$ ), which is used to represent the possible states thermodynamically."

***For Eqs.(15)-(20)***: In the section 3.2.1, we discussed the basic model for semiconductor/liquid interfaces. It also originated from Poisson's equation (Eq.(15)) to get the relationship between potential and charge density. Then, the Eq.(16) was obtained from Boltzmann distribution when the potential differences in the space region is very small compared with $kT/e$. Thus, after combining them, the Eq.(17) for electric field strength can be obtained by differetiating potential. If the Helmholtz and diffusion layers on the liquid side is further considered, the potential drop equation is Eq.(20), containing Eqs.(18) and (19). The equation just combines the potential drops at the interface, including solid side and liquid side. They can undoubtedly be linked to the previous model, and the development was based on EDL structure and surface charge space.

**Change 4**: To improve the logical structure, we added: "This equation was obtained from Boltzmann distribution when the potential differences in the space region is very small compared with $kT / e$."

***For Eqs.(21)-(24)***: As we mentioned in original text, "the electrolyte is of high concentration, the effect of the solution diffusion layer is not considered". That means we shorten the interface region in liquid side, but widen the region in solid side simultaneously. The equation for surface conductivity (Eq.(21)) in semiconductor electrochemistry was introduced, containing mobilities and surface excesses of charge carriers. In the literature, the surface excesses were obtained by solving Poisson's equation, Eqs.(22) and (23). And the Eq.(24) explained the expression of item in Eqs.(22) and (23).

***For Eq.(25)***: This equation discussed the change in electrode potential when the charge space layer in solid side, and only Helmholtz layer were considered in liquid side. We listed Eq.(25) to show how to link the electrode potential, potential drop in the space charge region and surface conductivity, which is beneficial to understand Fig.10A.

***For Eq.(26)***: Because in this part the semiconductor-electrolyte interface and semiconductor properties were discussed, the Femi level and the relationship between chemical potential, electrostatic potential and Femi level needs to be shown in equation.

***For Eq.(27)***: A method was introduce to treat interface, and Eq.(27) was used to describe how the particle density varies spatially in a non-uniform system. In such systems, the presence of a solid surface perturbs the otherwise homogeneous liquid structure, leading to spatial variations in particle density. To capture this variation, the particle density is expressed as a combination of a uniform density, a slowly varying perturbation, and a Fourier series expansion term that accounts for periodic fluctuations induced by the structured solid phase. We can use this equation to link the microscopic structure of the solid to the density modulation in the liquid.

***For Eqs.(28) and (29)***: These equations introduce the important concept of the electrochemical potential of Li ions, which serves as a foundation for the subsequent content in this review. This parameter is critical for DFT calculations and plays a central role in analyzing energy levels.

***For Eqs.(30)-(33)***: We discussed the origin of the voltage that arises when a conductor comes into contact with a semiconductor, which depends on the difference in their work functions. The work function of the semiconductor is related to its Fermi level via $W_{se} = E_{vac} - E_F$ . Next, we used this relationship to calculate the energy threshold for electron excitation, as shown in Eq. (30), which incorporates the valence band edge, the Fermi level, and the work function. Following this, Eq.(31) introduces the use of Green's functions, which incorporates the band energy, to compute electronic states in the bulk material. Eq.(32) presents the band structure of a metal with a cubic lattice based on the tight-binding model, while on the semiconductor side, Eq.(33) describes the energy levels in a specific CsCl-type semiconductor.

**Change 5**: To clarify this part, a sentence was added before Eq.(30): "Therefore, the mentioned parameters can be used to calculate the escape condition of electron.".

A short paragraph was added behind Eq.(33): "Briefly, the Eqs.(30)-(33) are closely connected through the analysis of electronic excitation and band structure at the metal/semiconductor interface. Eq.(30) defines the energy threshold for electron excitation, which depends on the Fermi level, valence band edge, and work function. To understand these energy levels microscopically, Eq.(31) introduces the Green's function formalism, which is based on the band energy of the materials. The specific form of the band energy in a cubic metal lattice is given by Eq.(32), while Eq.(33) extends the band structure analysis to a CsCl-type semiconductor. Together, these equations provide a consistent framework for linking macroscopic energy parameters to microscopic electronic structures.".

An equation was inserted into the original sentence: "For semiconductors, this is related to the semiconductor's Fermi level $W_{se} = E_{vac} - E_F$ .".

***For Eq.(34)***: It needs to be provided to show the EDL charge density at the metal-semiconductor interface through derivation of Poisson-Boltzmann equation. That is an example of application of EDL theory in conductor/semiconductor interface.

***For Eq.(35)***: It was listed to show the Poisson-Fermi-Dirac (PFD) equation, which can be compared with previous Fermi-like distribution (Eq.(9), Kornyshev's model) and Boltzmann distribution.

**Change 6**: A comparison was added: "The PFD equation provides a framework for modeling charge distributions in solid-state materials with finite site density and electrostatic coupling. It shares conceptual similarities with Kornyshev's Fermi-like distribution (Eq.(9)), which was introduced to account for ion crowding and saturation effects in ionic liquids. Both models improve upon the classical Poisson-Boltzmann model by incorporating Fermi-Dirac statistics to describe occupancy, in contrast to the exponential dependence of ion concentration on potential in the traditional Boltzmann distribution.".

***For Eq.(36)***: It was listed to support that my point has been partially studied in other research field.

**3)** For "Many paragraphs look like "fillers" that do not convey any useful information, for example, the very first paragraph of the introduction (and many more). Figures such as Fig. 1 that is

completely useless.”

***For Fig.1 and first paragraph***: We believe this figure plays an important role in the review. It is not merely decorative but serves as a visual summary of the fundamental challenges in current interface theory within electrochemical systems. Specifically, it highlights three often-overlooked but interconnected issues:

a. the inhomogeneous treatment of solid and liquid phases;

b. the diversity and complexity of materials involved at interfaces;

c. the underappreciated role of electrocapillary theory in the development of EDL models.

By presenting these core problems at the beginning, the figure helps guide readers through the motivation of the review.

For the first paragraph in Introduction, we believe it is essential to guide the reader and clearly justify the need for this review. Therefore, we have revised the opening paragraph to better highlight the importance of interface theory, clarify the research gap, and emphasize the unique contribution of this work. Please see the revised version in the main text.

***For Fig.2***: A detailed scheme of Lippmann's electrometer was drawn in Fig.2A, which laid the foundation for the development of polarography. This historical device can be difficult to understand for readers unfamiliar with classical electrochemical instrumentation, so we felt a clear visual representation would aid comprehension. Fig.2B presents a typical electrocapillary curve from the literature, showing the characteristic pressure-potential relationship. Meanwhile, Fig.2C illustrates the behavior of an adsorbed polarizable molecule under an electric field, offering a simplified conceptual model to help readers intuitively grasp the role of dipole reorientation at the interface.

Electrocapillarity reflects how the surface tension at the electrode/electrolyte interface varies with applied potential. This phenomenon is inherently governed by the structure and dynamics of the EDL, which controls the distribution of interfacial charge and potential. Therefore, understanding electrocapillarity requires a microscopic understanding of EDL theory.

***For Fig.3***: After introducing the theoretical framework of the EDL, it is essential to present experimental validation of its structure and behavior. Therefore, in Fig.3, we introduce the surface forces apparatus (SFA 2000), which enables high-resolution force measurements between two smooth surfaces and has been used to investigate electrostatic interactions within the EDL. In particular, the DLVO theory, illustrated through the interaction energy profiles in Fig.3D, provides a classical model for interpreting these measurements by combining van der Waals attraction and double-layer repulsion. Subfigures 3A and 3B show the experimental setup, while 3C-3E illustrate how ion distributions and interfacial forces are affected when two charged surfaces are brought into proximity. These figures provide experimental support and physical insight into EDL structure under confinement.

***For Fig.4***: Fig.4A illustrates how strong hydrogen bonding interactions can significantly affect interfacial behavior. This highlights the complexity of modeling such systems due to strong solvent-electrode, solvent-salt or solvent-solvent interactions. To address this, Fig.4B illustrates a representative study from the literature that used modern simulation methods to analyze and resolve this issue.

***For Fig.5***: The original Fig.5 presents several representative studies exploring quantum mechanical modifications to EDL model. As shown in figures, the incorporation of quantum effects results in only slight changes in ion distributions or potential profiles compared to classical models. This subtlety raises important discussions and even debates about how significant quantum effects truly are in practical electrochemical systems (micro to macro), and how to best leverage them to gain physical insights or improve model predictions.

We acknowledge that the current version of the figure could be visually improved and that the number of quantum-modified EDL studies in the literature remains limited. Thus, a few important researches were added.

**Change 1**: Two more examples were added into section 2.2, and Fig.5 has been improved by adding two more figures to support the explanations. Please see the modifications in page 14-16. Two references were added: "Q. Wu, Y. Qi, Energy & Environmental Science 2025, 18, 3036; M. McEldrew, Z. A. H. Goodwin, A. A. Kornyshev, M. Z. Bazant, The Journal of Physical Chemistry Letters 2018, 9, 5840."

***For Fig.6***: Fig.6 is essential in the section on conductor/liquid interfaces because it provides a detailed microscopic view of electron and ion distributions across the metal-solution boundary, where classical EDL models often simplify or ignore. Panels A-C illustrate how electron density and electrostatic potential evolve across the interface based on a microscopic simulation framework, while panel D introduces a modified Poisson-Boltzmann approach that incorporates electron spillover and short-range interactions.

This figure supports the discussion by showing why a purely electrostatic is insufficient for accurately capturing interfacial behavior at conductive surfaces, especially for solid side. By integrating quantum mechanical effects, the models depicted here offer a more realistic treatment of charge distribution, especially in systems involving metal electrodes. Therefore, Figure 6 serves as a necessary bridge between idealized theories and physically meaningful interfacial descriptions, justifying the need for more advanced models in modern EDL studies.

***For Fig.7***: This figure presents representative experimental examples that highlight the functional role of the EDL at metal interfaces.

**Change 2**: A new reference was added and please see corresponding explanation and analysis in page 21: "Q. Wu, M. T. McDowell, Y. Qi, Journal of the American Chemical Society 2023, 145, 2473."

***For Fig.8***: In this part, we highlighted several key issues that may serve as valuable directions for future research.

***For Fig.9***: To better understand the scheme of Kronig-Penney model, a scheme of periodic solid lattice was complemented in Fig.9A. And through Fig.9B, a concept of space charge region at semiconductor/liquid was introduced. Subsequently, the influences of doping concentration on the energy level within space charge region were shown in Fig.9C.

***For Fig.10***: The significant discrepancy between the experimental and theoretical curves in Fig.10A highlights the necessity of considering the potential drop across the Helmholtz layer at semiconductor–solid–liquid interfaces. Furthermore, the example shown in Fig.10B illustrates additional features and functional roles of the EDL at such interfaces, emphasizing aspects that warrant further attention in related studies.

***For Fig.11***: Fig.11A presents a study that directly compares interfacial behavior in battery electrode materials and semiconductor interfaces in solar cells, making it suitable for inclusion in the application-focused section. Figs.11B and 11C both relate to semiconductor/liquid interfaces and demonstrate clear changes in band energy, effectively illustrating the influence of interfacial conditions on electronic structure.

***For Fig.12***: Figs.12A and 12B illustrate how illumination influences the linear sweep voltammetry curves and energy band alignment of ORR and OER in Li-$O_2$ batteries. These electrocatalytic processes are closely tied to Li ion behavior at the cathode interface, underscoring the importance of interfacial charge transfer in battery performance. The shift in onset potentials and voltage profiles under light exposure suggests new opportunities to tune interfacial reactions using photoelectrochemical strategies. This connection allows us to extend the understanding of such interfaces to broader applications across electrochemical energy storage systems.

Fig.12C introduces a new framework for analyzing the interface, classifying it into three regions, including weak, moderate, and strong electrostatic coupling based on surface charge density. This classification provides a fresh perspective distinct from classical continuum models, offering deeper insight into charge inversion behavior and interfacial overcharging effects under varying ionic environments.

***For Fig.13***: In this outlook part, we mainly focused on the gradient or transition region at the solid/liquid interfaces, obviously including the dual-EDL issue. Further development of polarization of dual-EDL was shown in Fig.13C, especially for recent direction about Li metal protection, etc.

***For Fig.14***: Figs.14A and 14B present schematic energy band diagrams of typical homojunctions and heterojunctions, which are commonly encountered at semiconductor/semiconductor interfaces. These models help illustrate the fundamental differences in band alignment between the two different systems.

Figs.14C and 14D provide a practical example of such interfaces in composite electrode materials, showing the Li/Na adsorption behavior at $MoS_2$-$MoSe_2$ and $MoS_2$-$NbS_2$ lateral interfaces. This particular example underscores the role of interfacial band structure and EDL formation in tuning electrode properties.

***For Fig.15***: This figure focuses on semiconductor/semiconductor interfaces in all-solid-state batteries, highlighting how energy band structures evolve near the interface depending on the materials involved.

***For Fig.16***: This figure introduces an advanced experimental method used to probe the energy band structure of all-solid-state batteries during operation, and presents the band diagrams of each battery component (cathode, electrolyte, anode, and current collector) at different cycling stages.

***For Fig.17***: Fig.17A is a conceptual diagram illustrating the fundamental energy band relationships at a typical metal/semiconductor interface, including work function, Fermi level alignment, and band bending. This basic schematic serves as the foundation for understanding charge transfer across such interfaces. Fig.17B presents a literature-based example of a Schottky barrier, demonstrating how the barrier is influenced by the interfacial distance between the metal and semiconductor. Fig.17C extends this understanding by showing a more realistic case that incorporates crystal lattice structures at the interface.

***For Fig.18***: We also listed an example of electrocatalysis in zinc-air batteries to show the energy band at a specific interface in Fig.18A. Furthermore, another example discussing the difference between ohmic contact and Schottky contact was introduced in a methanol fuel cell.

***For Fig.19***: Similar to Fig.18, we show some examples of metal/semiconductor.

***For Fig.20***: This figure presents representative examples that emphasize the significance of incorporating the concept of the EDL into solid-state battery systems. Comparative analyses for solid-liquid and solid-solid EDLs shown in Figs.20B and 20C further highlight how EDL-related considerations impact interfacial properties and overall electrochemical performance.

***For Figs. 21 and 22***: Discussions about the potential future directions.

***For Fig.23***: Review of main conclusions.

**4)** For "In addition there are many ill-defined concepts, such as the "chemical potential interaction" (page 8), that do not exist: The chemical potential is a thermodynamic quantity, introduced to describe systems at the macroscopic scale, while interactions are occurring at the microscopic scale. There are many more examples like this that I will not detail. In conclusion, I do not recommend publication of this review in any journal."

The original sentence is "Lippmann hypothesized that the observed movement was due to the relationship between surface tension and chemical potential interactions." Sorry for the confusing point. This "interaction" refers to the interaction between "surface tension" and "chemical potential", not "chemical potential interaction".

**Change 1**: The sentence was modified to avoid any confusion: "Lippmann hypothesized that the observed movement was due to the relationship between surface tension and the chemical potential".

We have carefully reviewed all the equations and figures, and made appropriate revisions accordingly. Should you have any further comments, we would be glad to make additional improvements. Hopefully, this revised manuscript meets the publication requirements.

(When a better answer appears, this part will be updated.)

------------------------------------------------------------

**Q3**: (anonymous) This is describing a water splitting process (Page 12), yes, an e-chem process. Please expand this to make a connection/implication to batteries.

**A3**: This example was included to highlight simulation studies of hydrogen-bond networks near electrochemical interfaces. Fundamentally, electrocatalysis and electrochemical energy storage are deeply interconnected and cannot be entirely separated. In simple terms, we seek low overpotentials for electrocatalysts, while for battery electrodes, a large and stable potential window is desired. In fact, most of the materials, such as NiO, NiOOH, and related compounds, can function both as electrocatalysts and as active materials for energy storage. When the applied voltage becomes too large, energy storage materials may undergo phase transitions or surface reactions that shift their role toward electrocatalysis.

According to the classical definition, a catalyst is "a substance that increases the rate of a chemical reaction without itself undergoing any permanent chemical change." However, in recent literature, many electrocatalysts have been shown to undergo a series of structural or compositional transformations during operation. These observations suggest that the traditional definition may need refinement when applied to electrochemical systems. A more suitable definition for electrocatalysts might be: "a substance that increases the rate of an electrochemical reaction without undergoing **loss of its main active elements**." This adjusted view reflects the dynamic nature of electrocatalytic interfaces while preserving the core concept of catalytic stability. Therefore, electrocatalysis can certainly be used as a model to explain phenomena observed in battery systems.

(When a better answer appears, this part will be updated.)